\documentclass[%
reprint,
amsmath,amssymb,
aps,
prx,
fleqn,
floatfix,
]{revtex4-1}

\usepackage{amsmath,mathtools}
\usepackage{mathrsfs}
\usepackage{accents}
\usepackage{pgfplots}
\usepackage{epstopdf}
\usepackage{sublabel}
\usepackage{subfigure}
\usepackage{accents,amsfonts}
\usepackage{graphicx}
\usepackage{helvet}
\usepackage{color}
\usepackage{url}
\usepackage{bm}
\usepackage{upgreek}
\usepackage{soul}

\newcommand{\AAA}{\mathcal{A}}
\newcommand{\PPP}{\mathcal{P}}
\newcommand{\ER}{Erd\H{o}s-R\'enyi\ }
\newcommand{\Rossler}{R\"ossler\ }
\newcommand{\at}[2][]{#1|_{#2}}

\usepackage{graphicx}
\usepackage{ifluatex}
\ifluatex
  \usepackage{ifpdf}
  \ifpdf
    \expandafter\ifx\csname Gin@rule@.pdf\endcsname\relax
      \usepackage{grfext}
      \AppendGraphicsExtensions{.pdf,.pdf}
    \fi
  \fi  
\fi

\pgfplotsset{compat=newest}

\usepackage{soul}

\begin{document}

\title{Invertible generalized synchronization: A putative mechanism for implicit learning in biological and artificial neural systems}

\author{Zhixin Lu}
\author{Danielle S. Bassett}
\email{dsb@seas.upenn.edu}
\altaffiliation[Also at ]{%
Department of Physics \& Astronomy, College of Arts \& Sciences, University of Pennsylvania, Philadelphia, PA, 19104\\
Department of Electrical \& Systems Engineering, School of Engineering \& Applied Sciences, University of Pennsylvania, Philadelphia, PA, 19104\\
Department of Neurology, Perelman School of Medicine, University of Pennsylvania, Philadelphia, PA, 19104\\
Department of Psychiatry, Perelman School of Medicine, University of Pennsylvania, Philadelphia, PA, 19104\\
Santa Fe Institute, Santa Fe, NM, 87501\\ 
}%

\affiliation{Department of Bioengineering, School of Engineering \& Applied Sciences, University of Pennsylvania, Philadelphia, PA, 19104}

\begin{abstract}
Regardless of the marked differences between biological and artificial neural systems, one fundamental similarity is that they are essentially dynamical systems that can learn to imitate other dynamical systems, without knowing their governing equations. The brain is able to learn the dynamic nature of the physical world via experience; analogously, artificial neural systems can learn the long-term behavior of complex dynamical systems from data. Yet, precisely how this implicit learning occurs remains unknown. Here, we draw inspiration from human neuroscience and from reservoir computing to propose a first-principles framework explicating putative mechanisms of implicit learning. Specifically, we show that an arbitrary dynamical system implicitly learns other dynamical attractors (chaotic or non-chaotic) by embedding them into its own phase space through \emph{invertible generalized synchronization}. By sustaining the embedding through fine-tuned feedback loops, the arbitrary dynamical system can imitate the attractor dynamics it has learned. To evaluate the mechanism's relevance, we construct several distinct neural network models that adaptively learn and imitate multiple attractors. We observe and theoretically explain the emergence of five distinct phenomena reminiscent of cognitive functions: (i) imitation of a dynamical system purely from learning the time series, (ii) learning of multiple dynamics by a single system, (iii) switching among the imitations of multiple dynamical systems, either spontaneously or driven by external cues, (iv) filling-in missing variables from incomplete observations of a learned dynamical system, and (v) deciphering superimposed input from different dynamical systems. Collectively, our findings support the notion that artificial and biological neural networks can learn the dynamic nature of their environment, and systems within their environment, through the single mechanism of invertible generalized synchronization.
\end{abstract}

\maketitle

The human brain is not only a set of $1.4\times 10^{26}$ atoms, but also a dynamical system that evolves non-autonomously according to physical principles. One of its particularly notable functions is the capacity to imitate some other dynamical process solely by being exposed to the output of that process, without ever formally encoding the governing equations \cite{karuza2016local,reber1993implicit,cleeremans1998implicit,marcus1999rule,murphy2008rule,seger1994implicit}. To gain a physical intuition for this phenomenon, we can consider a few pertinent examples. Imitating a dynamical system with a fixed point or limit cycle attractor corresponds to memorizing and reciting an exemplary trajectory. Imitating a more complex dynamical system with, for example, a quasi-periodic torus or chaotic attractor corresponds to inventing new trajectories that are distinct from the original exemplary trajectory, while still obeying its underlying dynamical rules. Practically, this same phenomenon manifests in the human's ability to envision the orbits of physical objects based on previous observations of the physical world, the human's ability to speak grammatically correct sentences simply by listening to others, and the human's ability to hum a familiar or invented melody in their mind. While we may often take these capabilities for granted, they are reflective of a common mechanism that remains far from understood.

Notably, the capacity for dynamical systems to imitate other dynamical systems generalizes beyond neural systems and, in fact, even beyond biological systems. Indeed, over the last two decades intricate machine learning models and algorithms have been engineered to learn complex dynamics from data \cite{nemenman2000information,tu2013dynamic,anderson1996comparison,ly2012learning,hefny2015supervised,raissi2018multistep,raissi2018deep,altmann1999rule,schapiro2017complementary,rajan2016recurrent,rajan2016recurrent,alemi2017learning,gilra2017predicting,deneve2017brain,abbott2016building,maass2002real,jaeger2004harnessing,sussillo2009generating,pathak2017using,pathak2018model}. Such models and algorithms come in many flavors, with a particularly powerul type being reservoir computing (RC) networks, also known as echo-state networks or a liquid state machines. RC networks demonstrate a marked capacity to learn and persistently imitate various dynamical systems. For example, RC networks can learn (i) the recorded periodic movement of a human running \cite{sussillo2009generating}, (ii) low-dimensional chaotic systems such as the Lorenz or R\"ossler attractors \cite{jaeger2004harnessing,pathak2017using}, and (iii) even the Kuramoto-Sivashinsky system which is governed by partial differential equations that exhibit spatiotemporal chaos \cite{pathak2018model}. Is the correspondence between RC networks and biological neural networks in their capacity to imitate external dynamical systems more than happenstance?

We suspect there exists a general principle that allows a naive dynamical system to persistently imitate and reproduce the behavior of a target dynamical system, after implicitly assimilating only example trajectories. Without great loss of generality\footnote{Ongoing empirical work suggests that music and language follow definitive rules of composition or grammar, respectively, in a manner that is conceptually akin to distinctive trajectories on a chaotic attractor \cite{mayer1992musical,elman1995language,winters2009musical,bidlack1992Chaotic,castilho15chaoticsystems,mackenzie1995chaotic}.}, we confine our discussion to cases where the dynamics being learned is definite: the target dynamical system is autonomous and its trajectory $\mathbf{s}(t)\in\mathbb{R}^n$ on an attractor $\AAA$. We begin by describing a framework that allows a naive dynamical system to imitate other dynamical systems through a particular type of synchrony: \emph{invertible generalized synchronization}. Briefly, as the learning system is driven by $\mathbf{s}(t)$ on attractor $\AAA$, it encodes $\AAA$ into its own phase space $\mathbf{x}\in\mathbb{R}^N$ as a homeomorphism $\PPP={\bm \psi}(\AAA)$. In the process of encoding, the learning system gradually tunes its parameters such that it can reproduce the target dynamics by autonomously evolving on the attained $\PPP$ when the driving trajectory $\mathbf{s}(t)$ is later removed. By exercising this mechanism in a range of \emph{in silico} experiments, we observe and explain the emergence of several other phenomena reminiscent of human cognitive functions: (i) the ability to switch between imitating different dynamical patterns, either spontaneously \cite{kessler2009choosing} or when driven by an external stimulus \cite{koch2006cue}; (ii) the capacity to fill in missing information from sequential input \cite{constantino2017dynamic,komatsu2006neural}; and (iii) the ability to resolve distinct rules underlying mixed inputs generated from distinct structures \cite{ding2012emergence,xiang2010competing,luo2007phase,mesgarani2012selective}. The proposed learning mechanism based on invertible generalized synchronization provides a concise theory that unifies diverse learning functions.

\begin{figure}
  \center
  \includegraphics[width=8.5cm]{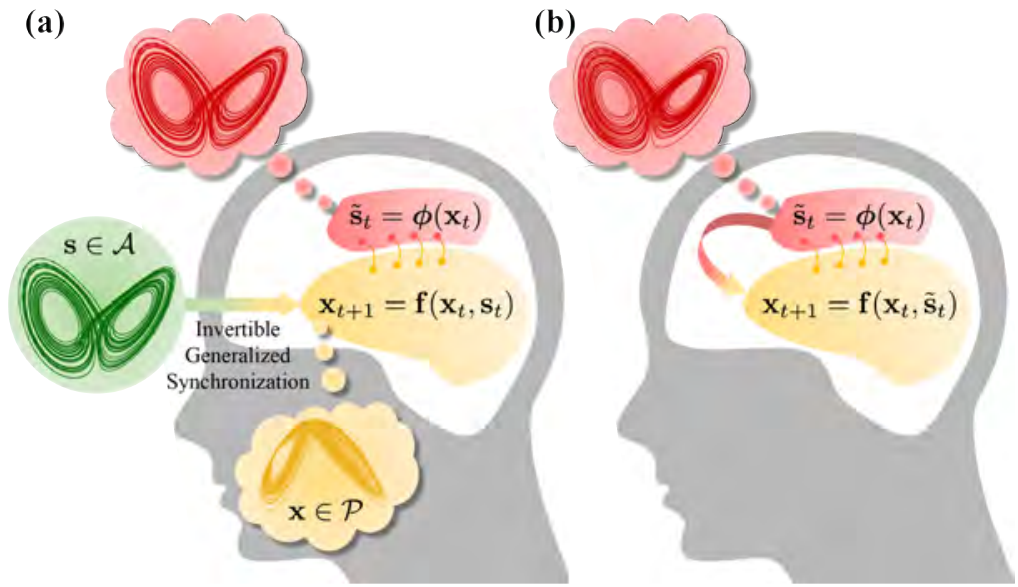}
  \caption{\textbf{Schematic depiction of our learning framework.} \emph{(a)} During the learning phase, the mind encodes the observed dynamics as it evolves on attractor $\AAA$. The mind is operationalized as a central dynamical system here shown in yellow, where sensory information is initially encoded and from which sensory information can later be retrieved. The exemplary sensory input trajectory being generated by an autonomous input dynamical system is shown in green. The internally generated output that imitates the external sensory input is shown in red. \emph{(b)} During the testing phase, the mind reproduces the learned dynamics as it evolves on attractor $\PPP$. The internally generated signal is directed into the central system through a feedback loop (shown by the curved arrow) that is recruited during the testing phase.}
  \label{fig:schm}
\end{figure}

\section{Informing the Learning Framework with Underlying Biological Features}
\label{sec:framework}

In this section, we show that a simple dynamical framework built upon invertible generalized synchronization \cite{lu2018attractor} can be informed by biological features of the human brain that have been operationalized in appropriate mathematics. We choose a generic set of features to ensure that the framework affords a systematic and analytical investigation of the underlying mechanism that is relevant to many natural and artificial learning systems.

\subsection{Attractor encoding by invertible generalized synchronization}

We begin by considering the reinstatement hypothesis, which posits that the content-specific activity in the human cortex at the time of encoding is reinstated as the encoded information is being retrieved \cite{james1890principles,tulving1973encoding}. Recent experimental evidence supports this idea, suggesting that patterns of neural activity initially encode sensory input and then those same patterns of neural activity recur when the encoded information is retrieved \cite{wheeler2000memory,kragel2017similar,nyberg2000reactivation,duzel2003human,khader2005content,ranganath2005functional,woodruff2005content,slotnick2006nature,johnson2007recollection,diana2013parahippocampal,rugg2013brain,bosch2014reinstatement}. In building our mathematical framework, we parallel this reinstatement hypothesis by recruiting a central dynamical system, $\mathbf{x}_{t+1} = \mathbf{f}(\mathbf{x}_t,\cdot)$, where sensory information is initially encoded and from which sensory information can later be retrieved (Fig.~\ref{fig:schm}).

We consider the exemplary sensory input trajectory $\mathbf{s}(t)\in\mathbb{R}^n$ being generated by an autonomous input dynamical system
\begin{equation}
\mathbf{s}(t+1) = \mathbf{g}(\mathbf{s}(t)).
\label{eqn:input_system}
\end{equation}
For input dynamical systems that are continuous in time, $\mathbf{g}(\cdot)$ can be considered as the flow map which takes $\mathbf{s}$ forward in time by a time step $\tau$. During the learning phase (Fig.~\ref{fig:schm}(a)), the central system is driven by $\mathbf{s}(t)$, and evolves following 
\begin{equation}
\mathbf{x}(t+1) = \mathbf{f}(\mathbf{x}(t),\mathbf{s}(t)),
\label{eqn:central_system}
\end{equation}
where $\mathbf{x}\in\mathbb{R}^N$, and $N\gg1$. 

In this setup, the autonomous input system (Eq.~(\ref{eqn:input_system})) and the non-autonomous central system (Eq.~(\ref{eqn:central_system})) are one-way coupled and collectively form a drive-response system. Through invertible \cite{lu2018attractor} \emph{generalized synchronization} \cite{afraimovich1986stochastic,PhysRevLett.64.821,rulkov1995generalized}, the central system becomes synchronized to the input system in a particular way. Specifically, starting from any initial state, the $\mathbf{x}(t)$, after a transient time, becomes uniquely determined by the state of the input system,
\begin{equation}
\mathbf{x}(t)={\bm \psi}(\mathbf{s}(t)),
\label{eqn:GS}
\end{equation}
and evolves onto $\PPP={\bm \psi}(\AAA)$, where ${\bm \psi}:\AAA\rightarrow\PPP$ is invertible. The attractor $\PPP$, an embedding of $\AAA$ in $\mathbb{R}^N$, plays the role of the central system's internal representation of $\AAA$.

Invertible generalized synchronization allows the central dynamical system to embed a dynamical attractor $\AAA$ into its own phase space. However, it is important to note that invertible generalized synchronization is not generically possible for any arbitrary pair of one-way coupled dynamical systems. In the context of the brain, this non-universality means that the human brain may not be able to learn any arbitrary dynamical system. To determine whether and when invertible generalized synchronization is possible between two coupled dynamical systems, we can consider the common criterion that the largest conditional Lyapunov exponent of the central system is negative (see Appendix A.1 for further details). As a consequence of the Whitney embedding theorem \cite{whitney1936differentiable}, if generalized synchronization does occur, the ${\bm \psi}(\cdot)$ in Eq.~(\ref{eqn:GS}) is likely to be invertible as long as the dimension of the central system is large ($N\gg n$)  \cite{lu2018attractor}.

\subsection{Internal imitation of the external signal}

Complementing the central system that encodes the attractor $\AAA$, we consider an internally generated output that imitates the external sensory input $\mathbf{s}$. The internal imitator operationalizes the human capacity to enhance and consolidate learned information by interacting with working memory \cite{marvel2012storage,perrone2014little}; colloquially, this phenomenon is sometimes called an \emph{inner voice} \cite{marvel2012storage,baddeley1974recent,perrone2014little,morin2011self,williams2012inner,alderson2015inner,scott2013corollary}. The external sensory signal $\mathbf{s}$ that the central system encodes is internalized, and can be recreated even when there is no external signal \cite{scott2013corollary}. We consider this process to be downstream of the central system, and formalize it as a decoding function,
\begin{equation}
\tilde{\mathbf{s}}(t) = {\bm \phi}(\mathbf{x}(t)),
\label{eqn:inner_voice}
\end{equation}
where $\tilde{\mathbf{s}}\in\mathbb{R}^n$ is the internally generated output. During the learning phase, the ${\bm \phi}(\cdot)$ is adaptively modified towards
\begin{equation}
{\bm \phi}^*(\mathbf{x})\equiv{\bm \psi}^{-1}(\mathbf{x}),\text{ for } \forall\mathbf{x}\in\PPP,
\label{eqn:goal}
\end{equation}
such that the internal imitator produces a trajectory that is concurrent with the external sensory input: $\tilde{\mathbf{s}}(t)=({\bm \psi}^{-1}\circ{\bm \psi})(\mathbf{s}(t))=\mathbf{s}(t)$.

\subsection{Retrieving and imitating the learned dynamics}

Upon the completion of the learning phase, the whole system evolves autonomously due to the absence of the external input signal $\mathbf{s}$. We refer to this period as a testing phase (Fig.~\ref{fig:schm}(b)), during which the internal output $\tilde{\mathbf{s}}$ imitates the dynamical trajectory on the learned attractor $\AAA$, as the central system retrieves and reinstates its trajectories on $\PPP$. Here, we consider a specific form of information retrieval to constitute the testing phase. In the study of human memory, extra cortical regions are engaged when humans mentally rehearse and retrieve previously learned sensory information in the absence of exogenous input \cite{addis2007remembering,kosslyn2001neural,halpern1999tune,buckner2007self,wheeler2000memory}. Motivated by these observations, we let the internally generated $\tilde{\mathbf{s}}$ be directed into the central system through a feedback loop that is recruited only during the testing phase. As $\tilde{\mathbf{s}}$ replaces the absent external $\mathbf{s}$, the central system reinstates its behavior on $\PPP$. In this setup, the central system during the testing phase evolves autonomously without external input, following 
\begin{equation}
\mathbf{x}(t+1) = \mathbf{f}(\mathbf{x}(t),\tilde{\mathbf{s}}(t))=\mathbf{f}(\mathbf{x}(t),{\bm \phi}(\mathbf{x}(t))).
\label{eqn:auto_brain}
\end{equation}

When the internal output function implemented in Eq.~(\ref{eqn:auto_brain}) is ideal, i.e., ${\bm \phi}= {\bm \psi}^{-1}$, any learning phase trajectory $\mathbf{x}(t)$ on $\PPP$ (from the driven central system governed by Eqs.~(\ref{eqn:input_system},\ref{eqn:central_system})), is also a solution of the autonomous central system equation (Eq.~(\ref{eqn:auto_brain})). Thus, we say that this $N$-dimensional autonomous central system incorporates the local dynamics of the $n$-dimensional dynamical system (Eq.~(\ref{eqn:input_system})) on the attractor $\AAA$. Moreover, to guarantee that the internal output accurately and persistently imitates the learned dynamics on $\AAA$, the autonomous central system should evolve on $\PPP$ in a manner that is transversely stable. To ensure that $\PPP$, an embedding of the low-dimensional dynamical attractor $\AAA$ in $N$-dimensional space, is indeed an attractor of the autonomous central system (Eq.~(\ref{eqn:auto_brain})), one can analyze its transverse stability by checking the sign of the corresponding transversal Lyapunov exponents. As previously shown \cite{lu2018attractor}, the rate of divergence from the learned attractor during the autonomous mode can be predicted by the value of the largest transversal Lyapunov exponent (see Appendix A.2 for further details).

\subsection{Instantiating the learning framework \emph{in silico}}
\label{sec:silico}

With this general dynamical framework, many complex learning functions naturally emerge, including learning multiple dynamics, switching between learned dynamics, filling-in missing variables, and separating superimposed signals. To study these functions, we instantiate the central system \emph{in silico} as a high dimensional dynamical system that responds consistently to external driving signals; this choice facilitates lossless encoding through invertible generalized synchronization in the learning phase. Here we let the central system be a random RNN with sparse \ER topology akin to a typical reservoir computer \cite{lu2017reservoir,PhysRevLett.120.024102,pathak2017using}. In Appendix C, we study other choices for the central system, including an RNN with a non-\ER topology and a dynamical network with random polynomial-type interactions.

\begin{figure}
  \center
  \includegraphics[width=8.5cm]{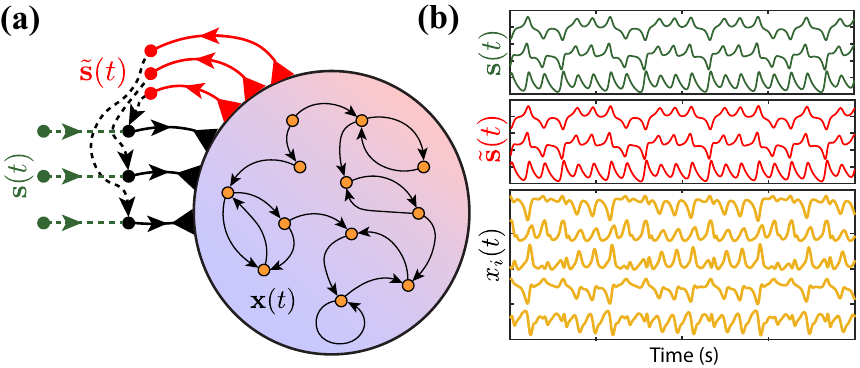}
  \caption{\textbf{Instantiating the learning framework \emph{in silico}.} \emph{(a)} We let the central system be a random RNN with network size $N=2000$ and sparse \ER topology akin to a typical reservoir computer. The output function (red) is adaptively updated following a simple delta rule that is inspired by the biological process of spike-timing-dependent plasticity. \emph{(b)} In green, we show an exemplary external input trajectory $\mathbf{s}(t)$ generated by a Lorenz system. In red, we show the successfully trained concurrent internal output $\tilde{\mathbf{s}}(t)$ that imitates $\mathbf{s}(t)$. In yellow, we show the concurrent activity of five randomly chosen nodes in the central system $x_i(t)$.}
  \label{fig:RNN_in_silico}
\end{figure}

Formally, we define the central system as
\begin{equation}
\mathbf{f}(\mathbf{x},\mathbf{s})=\tanh(\mathbf{A}\mathbf{x}+\mathbf{W}_{\text{in}}\mathbf{s}+\mathbf{c}),
\label{eqn:RNN}    
\end{equation}
where $\mathbf{A}\in\mathbb{R}^{N\times N}$ is the adjacency matrix of a sparse \ER network, $\mathbf{W}_{\text{in}}\in\mathbb{R}^{N\times n}$ is the input weight matrix, and $\mathbf{c}\in\mathbb{R}^{N\times 1}$ is a constant bias vector. While more complex approaches exist \cite{cybenko1989approximations}, we implement the internal output as a simple linear function 
\begin{equation}
{\bm \phi}(\mathbf{x}) = \mathbf{W}_{\text{out}}\mathbf{x},
\label{eqn:linear_output_function}
\end{equation}
where $\mathbf{W}_{\text{out}}$ is an $n$-by-$N$ matrix. From a random initial $\mathbf{W}_{\text{out}}$, it is adapted during the learning phase according to a delta rule,
\begin{equation}
\mathbf{W}_{\text{out}}(t+1) = \mathbf{W}_{\text{out}}(t) +\alpha {\bm \Delta}(t) \mathbf{x}^T(t),
\label{eqn:adapt}
\end{equation}
such that the discrepancy between the input and the output, $\mathbf{s}(t)-\tilde{\mathbf{s}}(t)= {\bm \Delta}(t)\in\mathbb{R}^{n\times 1}$, is gradually eliminated. This particular adaptive learning rule is inspired by the biological process of spike-time-dependent plasticity and is also used in other artificial learning systems \cite{bengio2015towards}.

\begin{figure}
  \center
  \includegraphics[width=8cm]{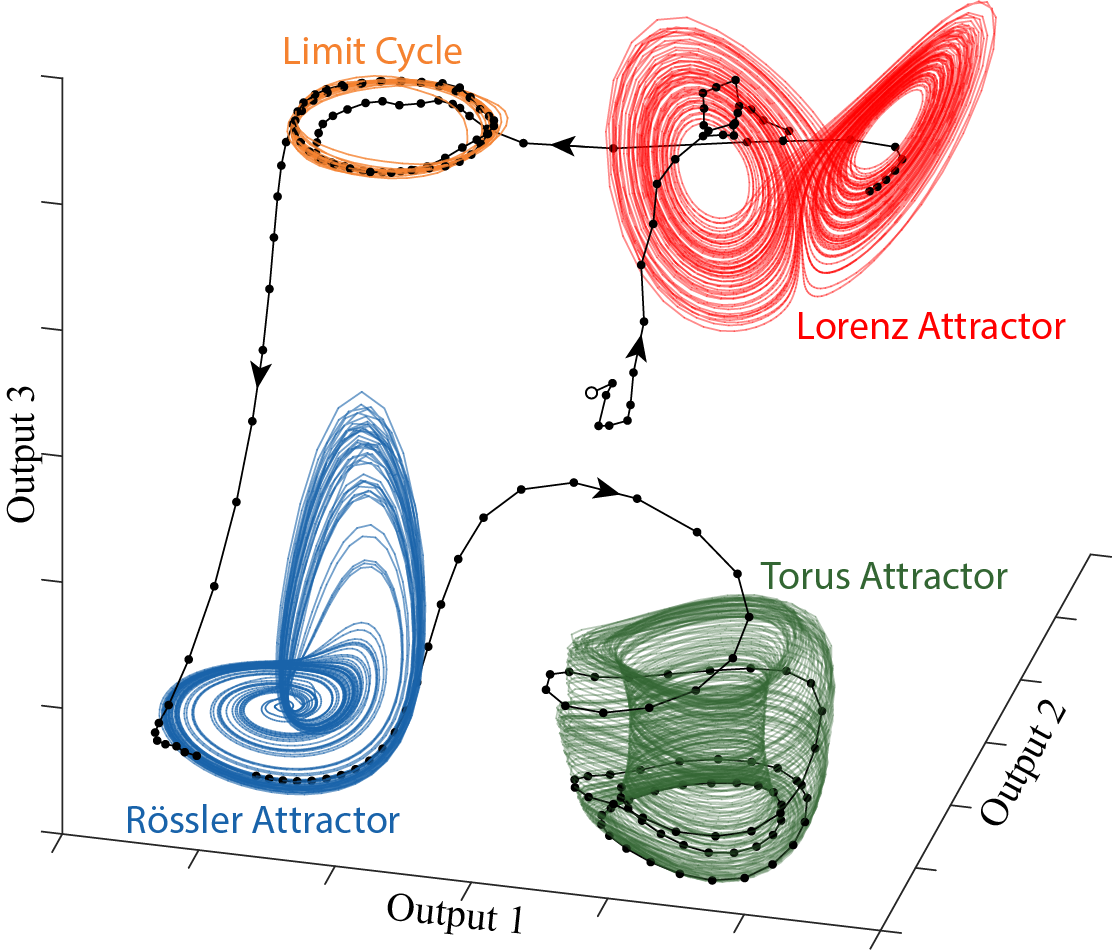}
  \caption{\textbf{Multistability in task switching between multiple learned rules.} An example learning system with a trained $\tilde{\mathbf{s}}$ that successfully imitates the dynamics of $4$ different attractors. The black dotted trajectories are the system's output signals as it switches between different tasks. This switching is induced by external cues. The instantiated central system in this example is an RNN with $N=2000$ nodes and an \ER topology.}
  \label{fig:multi_four_task}
\end{figure}

\begin{figure*}
  \center
  \includegraphics[width=16cm]{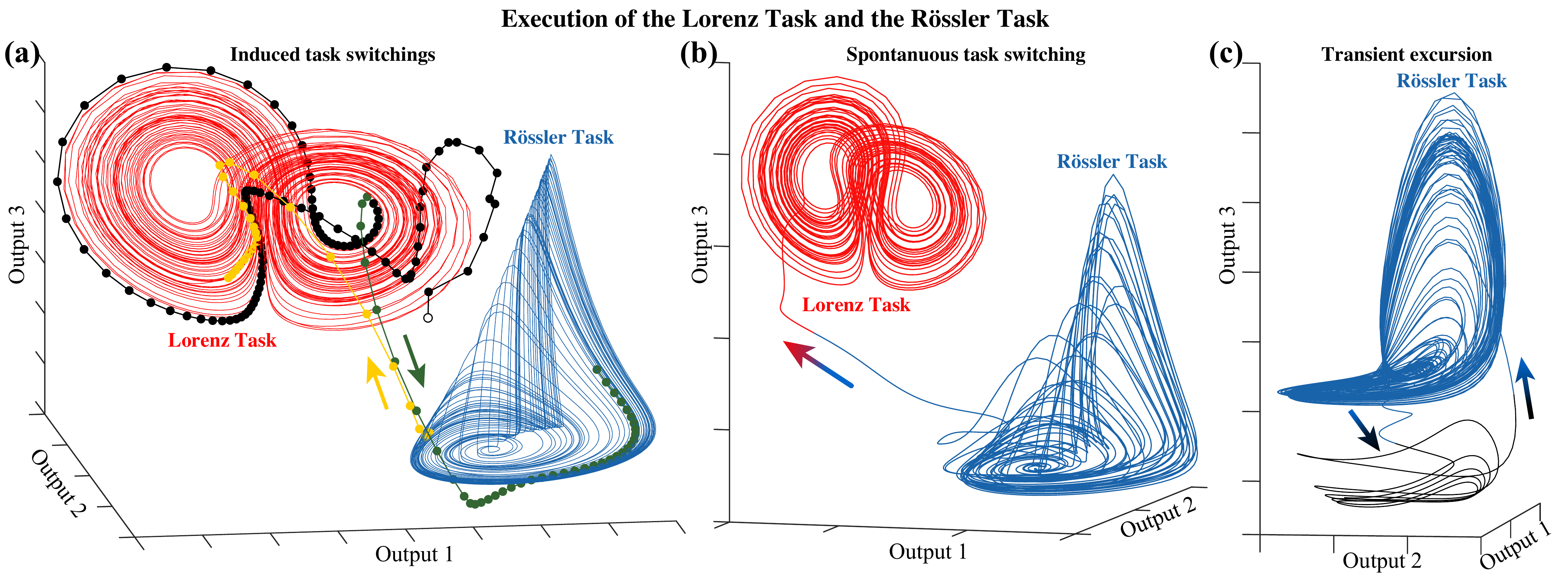}
  \caption{\textbf{Distinct types of task switching between multiple learned rules.} \emph{(a)} Output $\tilde{\mathbf{s}}$ generated by a learning system that successfully learns both the Lorenz and the R\"ossler attractors. Switching back and forth between two attractors can be induced by external cues. \emph{(b)} Spontaneous task switching from one attractor that is not stably learned to another attractor that is stably learned. \emph{(c)} A transient excursion observed when the R\"ossler attractor is not stably learned.}
  \label{fig:multi_task_1}
\end{figure*}

\section{Multistability in Task Switching between Multiple Learned Rules}

Capitalizing on the power of invertible generalized synchronization, the embedding-based learning framework that we describe can be easily extended to learn multiple attractors, which are either chaotic or non-chaotic, denoted by $\AAA_k\subset\mathbb{R}^n$ where $k=1,2,...,K$. As a proof-of-principle experiment, we show in Fig.~\ref{fig:multi_four_task} a system that learns to imitate $K=4$ attractors: one periodic, one quasi-periodic, and two chaotic (see Appendix B for further details). We refer to the imitation of each dynamical attractor as a task. The linear output weight matrix $\mathbf{W}_{\text{out}}$ is adapted following Eq.~(\ref{eqn:adapt}) as the whole system repeatedly goes through the $4$ learning phases (see Appendix D for further details regarding this multi-attractor learning schedule). While this learning framework can learn to imitate non-chaotic attractors, we are particularly interested in the learning of chaotic attractors, where the system does more than reciting a memorized exemplary trajectory. Thus, and also for simplicity and clarity of exposition, in the remainder of this paper, we only consider learning and imitating two chaotic attractors: the Lorenz and the R\"ossler attractors. 

Formally, through invertible generalized synchronization in each of the $K$ learning phases, the central system forms the corresponding internal representation by embedding $\AAA_k\subset\mathbb{R}^n$ into its own phase space $\mathbb{R}^N$. That is, 
\begin{equation}
\PPP_k={\bm \psi}_k(\AAA_k),
\label{eqn:manifold_GS}
\end{equation}
where ${\bm \psi}_k:\AAA_k\rightarrow\PPP_k$ is invertible. Given that $\PPP_k\cap\PPP_l=\emptyset$ and $\AAA_k\cap\AAA_l=\emptyset$ for any $k\neq l$, we combine the $K$ generalized synchronization functions into a single invertible function ${\bm \psi}:\cup_{k=1}^{K}\AAA_k\rightarrow\cup_{k=1}^{K}\PPP_k$. (See Appendix B.5 for a method to improve learning performance by avoiding overlap between $\PPP$s.) Then, the ideal output function is the inverse of ${\bm \psi}$ that maps each $\PPP_k$ back onto $\AAA_k$, i.e., $\AAA_k={\bm \phi}^*(\PPP_k)$ for $k=1,2,...,K$. Given that ${\bm \phi}={\bm \phi}^*$ in Eq.~(\ref{eqn:auto_brain}), if each representation $\PPP_k$ is an attractor of the testing phase central system, the central system is able to imitate any of the $K$ attractors. After implementing this setup \emph{in silico}, we observe that the central system depicted in Fig.~\ref{fig:RNN_in_silico}(a) can successfully learn to imitate both the Lorenz and the R\"ossler dynamics (Fig.~\ref{fig:multi_task_1}(a)).

Once multiple attractors $\AAA_k$ are learned, it is of interest to consider the phenomenon of task switching, which is a capacity displayed by many natural neural systems. When $\PPP_k$ becomes an attractor of Eq.~(\ref{eqn:auto_brain}), we say that the attractor $\AAA_k$ is stably learned, since the output trajectory $\tilde{\mathbf{s}}(t)$ cannot spontaneously depart from that attractor. In this case, we can consider explicitly triggering a switch from one task to another; that is, from  the scenario in which the central system imitates one attractor to the scenario in which the central system imitates another attractor. Notice that the generalized synchronization guarantees that the central system state $\mathbf{x}(t)$ converges to ${\bm \psi}(\mathbf{s}(t))$ after a short period of time in the testing phase. Thus, we can use a short external input cue, $\mathbf{s}(t)\in\AAA_k$, and lead the central system $\mathbf{x}$ from an arbitrary state to the desired $\PPP_k$. 

Instantiating these ideas \emph{in silico}, we can begin the testing phase dynamics from a random initial state $\mathbf{x}(t=0)$. In Fig.~\ref{fig:multi_task_1}(a), the output $\tilde{\mathbf{s}}(t)$ evolves following the black dotted line and converges onto the Lorenz attractor $\AAA_{\text{Lorenz}}$. Then, for a very short period of time, we provide an external input $\mathbf{s}(t)$ from a R\"ossler system. We observed that this external input successfully induced a switch from $\AAA_{\text{Lorenz}}$ to $\AAA_{\text{R\"ossler}}$ (green dotted line). Thereafter the central system evolves autonomously without any external input for a long period of time, generating the $\tilde{\mathbf{s}}(t)$ that traces out the R\"ossler attractor. We then use a short Lorenz input $\mathbf{s}(t)$ to drive the central system away from imitating the R\"ossler attractor and towards imitating the Lorenz attractor (yellow dotted line). Thereafter, the central system evolves autonomously with the testing phase dynamics, and generates a very long output trajectory that traces out the Lorenz attractor (red).

Next, it is interesting to consider the case in which not all attractors are stably learned. In this scenario, we would expect spontaneous task switching, independent of any external input. In Fig.~\ref{fig:multi_task_1}(b), we consider a case in which the Lorenz system is stably learned but the R\"ossler system is not. That is, the autonomously evolving central system is transversely stable on $\PPP_{\text{Lorenz}}$ but transversely unstable on $\PPP_{\text{R\"ossler}}$. Here we show that starting from an initial condition $\mathbf{x}\in\PPP_{\text{R\"ossler}}$, the output of the testing phase $\tilde{\mathbf{s}}$ remains on the R\"ossler attractor (blue) for an extended period of time and then spontaneously departs from the R\"ossler attractor and converges onto the Lorenz attractor (red). Relatedly, it is interesting to consider the scenario in which an attractor is not stably learned. In Fig.~\ref{fig:multi_task_1}(c), we observe that the output of the testing phase $\tilde{\mathbf{s}}(t)$ exhibits a transient excursion as it unstably imitates the R\"ossler attractor, in contrast to the dynamics observed in spontaneous task switching. 

\begin{figure*}
  \center
  \includegraphics[width=14cm]{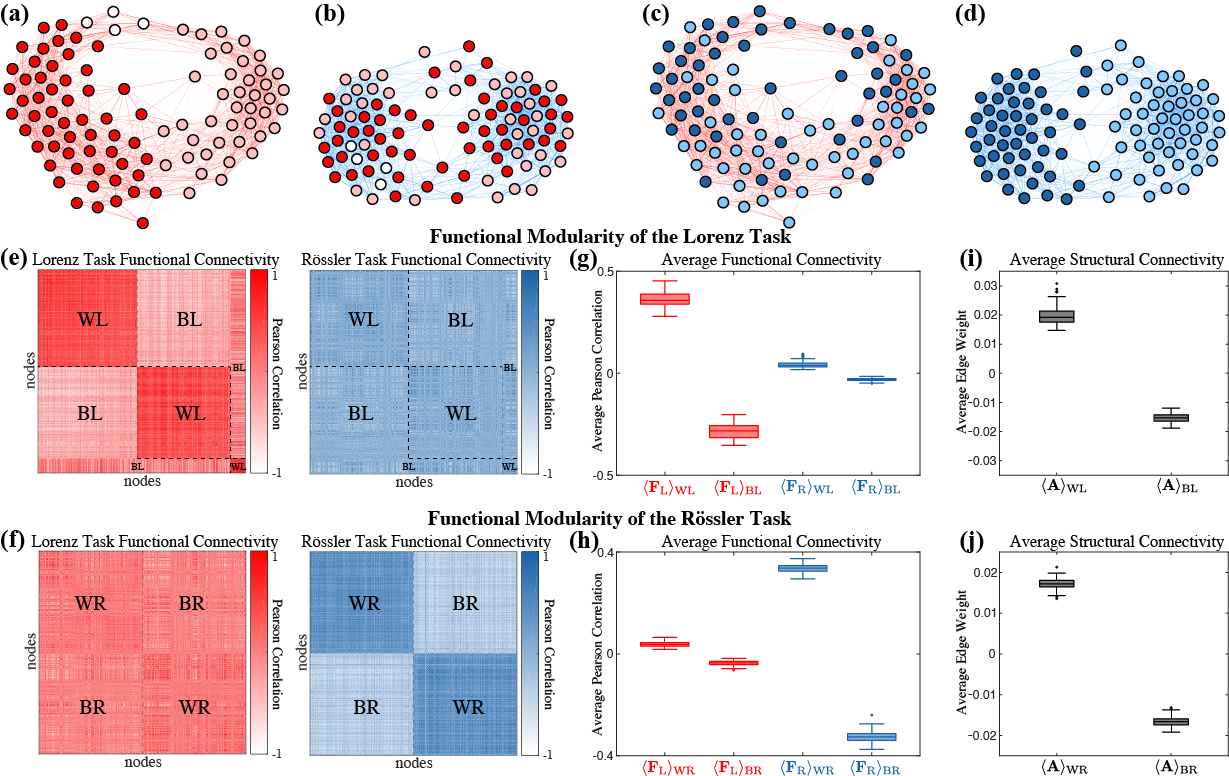}
  \caption{\textbf{The task-dependent functional connectome estimated from the statistical similarity in activity time series between neurons in the RNN.} We show the functional connectome for the Lorenz task \emph{(a,c)} and for the R\"ossler task \emph{(b,d)} for an example RNN using the Force Atlas Layout in Gephi. For the sake of visualization, we only show $100$ randomly chosen neurons out of the full $N=2000$. In panels \emph{(e-f)} we show the full functional connectivity matrices where neurons in panel \emph{(e)} (\emph{(f)}) are sorted into clusters detected from the Lorenz-task (R\"ossler-task) functional connectivity matrix using a data-driven community detection algorithm. Neurons in panels \emph{(a,b)} (\emph{(c,d)}) are colored according to the three (two) clusters detected from the Lorenz-task (R\"ossler-task) functional connectome. \emph{(g-h)} The average functional connectivity within and between communities as estimated in an ensemble of $100$ RNNs. \emph{(i-j)} The average structural connectivity (synaptic weights) within and between communities as estimated in the same ensemble of $100$ RNNs.}
  \label{fig:multi_task_2}
\end{figure*}

\section{Network Mechanics Underlying Multiple Learned Attractors}

Critically, because the learning of multiple attractors is achieved by embedding and incorporating multiple attractors into a single system (Eq.~(\ref{eqn:auto_brain})), different imitation tasks can be executed simply by letting the central system state $\mathbf{x}(t)$ revisit the corresponding $\PPP_k$. An important corollary of this fact is that no hyperparameters such as synaptic strengths are altered. We refer to the physical synapses as structural connections, and note that they are given by $\mathbf{A}$. While we cannot explain the learning of multiple attractors with the pattern of structural connections, it is possible that there is explanatory content in the emergent pattern of functional connections, which are defined as statistical similarities in neuronal time series. 

To investigate this possibility, we consider the structural connectivity given by the random adjacency matrix $\mathbf{A}\in\mathbb{R}^{N\times N}$, and the functional connectivity given by the Pearson correlation matrix of the recorded activity $\mathbf{x}(t)\in\mathbb{R}^{N\times 1}$ for $t$ during the Lorenz (R\"ossler) task, $\mathbf{F}_{\text{L}}\in\mathbb{R}^{N\times N}$ ($\mathbf{F}_{\text{R}}\in\mathbb{R}^{N\times N}$). To summarize the emergent patterns of functional connectivity, we apply a commonly used community detection technique \cite{porter2009communities} known as modularity maximization \cite{newman2006modularity} enacted with a particularly effective heuristic \cite{blondel2008fast} to identify groups of neurons that show similar time series. We found strong but distinct community structure in $\mathbf{F}_{\text{L}}$ and $\mathbf{F}_{\text{R}}$. In Fig.~\ref{fig:multi_task_2}(e), we show $\mathbf{F}_{\text{L}}$ (red) and $\mathbf{F}_{\text{R}}$(blue) with all nodes sorted by the community structure identified in $\mathbf{F}_{\text{L}}$. Similarly, in Fig.~\ref{fig:multi_task_2}(f), we show $\mathbf{F}_{\text{L}}$ (red) and $\mathbf{F}_{\text{R}}$(blue) with all nodes sorted by the community structure identified in $\mathbf{F}_{\text{R}}$. These observations indicate that while neurons remain identically structurally connected in both tasks, their emergent collective dynamics differ. 

To quantify these observations more fully, we constructed $100$ randomly organized and trained neural networks, and for each we calculated the average functional connectivity among pairs of neurons that are within \emph{versus} between communities identified from either $\mathbf{F}_{\text{L}}$ or $\mathbf{F}_{\text{R}}$ (labeled ``WL'', ``BL'', ``WR'', and ``BR'', respectively, in Fig.~\ref{fig:multi_task_2}(e,f)). In Figs.~\ref{fig:multi_task_2}(g,h)), we observe that the functional connectivity estimated from the Lorenz task and averaged \emph{within} the communities identified from the Lorenz task data, $\langle\mathbf{F}_{\text{L}}\rangle_{\text{WL}}$, is significantly larger than the functional connectivity estimated from the Lorenz task and averaged \emph{between} the communities identified from the Lorenz task data, $\langle\mathbf{F}_{\text{L}}\rangle_{\text{BL}}$. Similarly, for the R\"ossler task, $\langle\mathbf{F}_{\text{R}}\rangle_{\text{WR}}$ is significantly larger than $\langle\mathbf{F}_{\text{R}}\rangle_{\text{BR}}$. In contrast, the fact that $\langle\mathbf{F}_{\text{L}}\rangle_{\text{WR}}$ is not significantly larger than $\langle\mathbf{F}_{\text{L}}\rangle_{\text{BR}}$, and similarly $\langle\mathbf{F}_{\text{R}}\rangle_{\text{WL}}$ is not significantly larger than $\langle\mathbf{F}_{\text{R}}\rangle_{\text{BL}}$, supports our qualitative observation that the community structure in the functional connectivity matrix of the Lorenz task is distinct from the community structure in the functional connectivity matrix of the R\"ossler task. 

Lastly, we asked whether there existed any structural basis for the observed emergent functional communities. Importantly, random networks are far from homogeneous, and can display locally dense areas as well as locally sparse areas that occur simply by chance. It is therefore intuitively possible that the random network $\mathbf{A}$ contains degenerate weak community structure that supports the distinct community structure present in the functional connectivity when the system generates different patterns of emergent dynamics. To investigate this possibility, we calculated the average structural connectivity within the Lorenz communities, $\langle\mathbf{A}\rangle_{\text{WL}}$, within the R\"ossler communities, $\langle\mathbf{A}\rangle_{\text{WR}}$, between the Lorenz communities, $\langle\mathbf{A}\rangle_{\text{BL}}$, and between the R\"ossler communities, $\langle\mathbf{A}\rangle_{\text{BR}}$. We observed greater average structural connectivity within communities than between communities: that is, $\langle\mathbf{A}\rangle_{\text{WL}}$ and $\langle\mathbf{A}\rangle_{\text{WR}}$ were significantly larger than $\langle\mathbf{A}\rangle_{\text{BL}}$ and $\langle\mathbf{A}\rangle_{\text{BR}}$ (Fig.~\ref{fig:multi_task_2}(g,h)). This observation motivates two open questions: (i) whether some structural networks more easily (or less easily) support diverse functional community structures, and thus the learning of multiple systems, and (ii) for a given structural network, can one predict the number of possible emergent community structures, and therefore the number of systems that can be learned. While both questions remain to be answered, in Appendix E we provide additional quantitative analysis to assess differences in the strength and topology of structural support for functional communities elicited by the two tasks.

\begin{figure}
  \center
  \includegraphics[width=8.5cm]{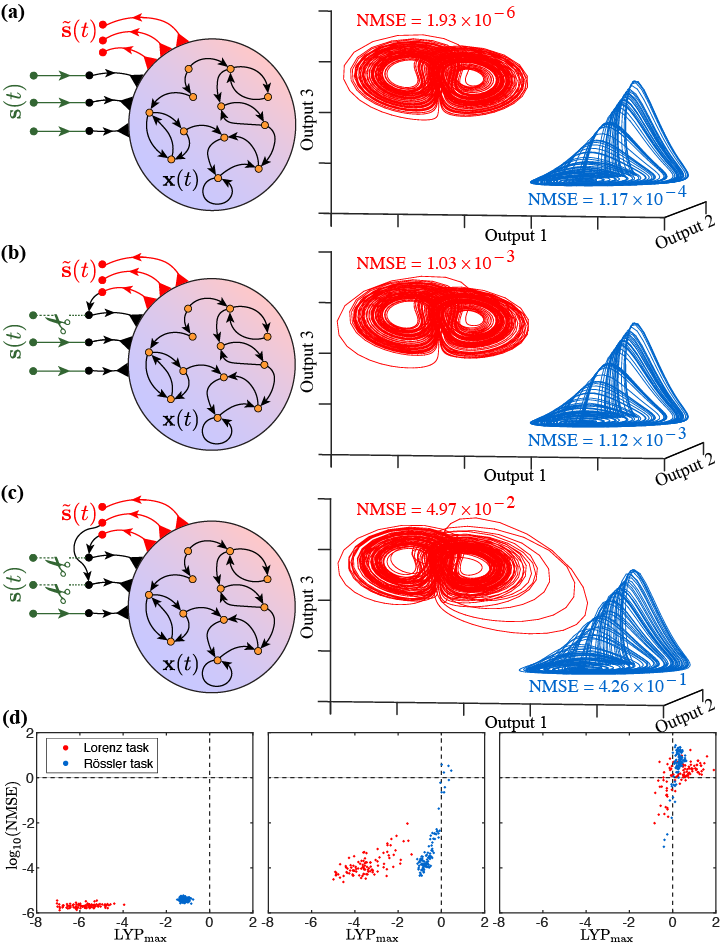}
  \caption{\textbf{Inferring the task for both the Lorenz and R\"ossler systems when full or partial input variables are provided.} \emph{(a)} The inferred output $\tilde{\mathbf{s}}$ from the central system as well as the normalized mean squared error for the Lorenz (red) and the R\"ossler (blue) tasks when $3$ out of $3$ input variables are provided. \emph{(b)} The inferred output $\tilde{\mathbf{s}}$ from the central system as well as the normalized mean squared error for the Lorenz (red) and the R\"ossler (blue) tasks when $2$ out of $3$ input variables are provided. \emph{(c)} The inferred output $\tilde{\mathbf{s}}$ from the central system as well as the normalized mean squared error for the Lorenz (red) and the R\"ossler (blue) tasks when $1$ out of $3$ input variables is provided. \emph{(d)} The log of the normalized mean squared error in the inference task with respect to the largest conditional Lyapunov exponents, from $100$ randomly constructed neural networks.}
  \label{fig:observer}
\end{figure}

\section{Infer Missing Variables from Learned Dynamical Systems}

Here we consider the problem of filling in missing variables of a dynamical system that is previously learned. Given that dynamical attractors $\AAA_k$ are successfully learned, we consider the case in which a new trajectory $\mathbf{s}(t)$ on attractor $\AAA_k$ is given but with some of the variables $[\mathbf{s}]_i$ missing. The goal is to use the acquired knowledge of $\AAA_k$, together with the remaining variables $[\mathbf{s}]_j$, to infer values of the missing variables $[\mathbf{s}]_i$, where $i\neq j$. To perform this inference, we evolve the central system following the testing phase dynamics shown in Eq.~(\ref{eqn:central_system}), and we replace the missing variables $[\mathbf{s}]_i$ in $\mathbf{s}$ by the corresponding output variables $[\tilde{\mathbf{s}}]_i$ obtained from Eq.~(\ref{eqn:inner_voice}) (Fig.~\ref{fig:observer}(a--c)). If the central system driven by available variables $[\mathbf{s}]_j$ and its own output variables $[\tilde{\mathbf{s}}]_i$ maintains generalized synchronization with the input dynamical system that generates $\mathbf{s}$, then the inference is expected to be successful. 

To instantiate this problem \emph{in silico}, we train an RNN with $N=2000$ neurons to learn both the Lorenz and the R\"ossler systems. Then, we test the inference of this RNN in three scenarios, where zero, one, and two variables $[\mathbf{s}]_i$ from $\mathbf{s}(t)$ are missing (Fig.~\ref{fig:observer}(a-c)). The inferred trajectories are the central system outputs $\tilde{\mathbf{s}}$ for the Lorenz and the R\"ossler tasks, respectively. In agreement with intuition, more missing variables leads to poorer inference quality, where the quality is quantified by the normalized mean squared error $\text{NMSE}=\text{mean}(\Vert\tilde{\mathbf{s}}(t)-\mathbf{s}(t)\Vert^2)/\text{var}(\Vert\mathbf{s}(t)\Vert^2)$. This NMSE assessing the encoding error is also referred to as the ``learning phase error'' in the Appendix A. To show that the inference quality is related to the quality of the generalized synchronization, we train an ensemble of $100$ RNNs, and we calculate both their inference error as well as the largest conditional Lyapunov exponent $\text{LYP}_{\text{max}}$. A more negative $\text{LYP}_{\text{max}}$ suggests a stronger generalized synchronization. In Fig.~\ref{fig:observer}(d), we observe that, for the three scenarios, the inference error $\log_{10}(\text{NMSE})$ is indeed higher whenever the generalized synchronization is weaker; that is, when the largest conditional Lyapunov exponent of the central system $\text{LYP}_{\text{max}}$ is less negative.

\begin{figure}
  \center
  \includegraphics[width=8cm]{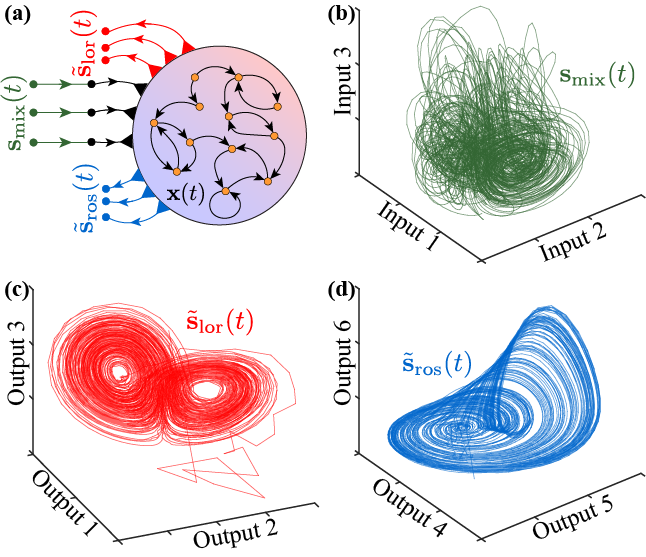}
  \caption{\textbf{Separating superimposed input from different dynamical system sources.} \emph{(a)} Schematic of a learning system being trained to separate superimposed input. The central system is driven by $3$-dimensional superimposed input $\mathbf{s}_{\text{mix}}(t)=\mathbf{s}_{\text{lor}}(t)+\mathbf{s}_{\text{ros}}(t)$. During the learning phase, the output weight matrices for the two $3$-dimensional outputs $\tilde{\mathbf{s}}_{\text{lor}}$ and $\tilde{\mathbf{s}}_{\text{ros}}$ approximate the concurrent Lorenz trajectory $\mathbf{s}_{\text{lor}}$ and R\"ossler trajectory $\mathbf{s}_{\text{ros}}(t)$. \emph{(b)} The testing phase input trajectory to the central system, $\mathbf{s}_{\text{mix}}$, which is a superposition of a new Lorenz trajectory and a new R\"ossler trajectory, both different from the exemplary trajectories used in the learning phase. \emph{(c)} The $\tilde{\mathbf{s}}_{\text{lor}}$ deciphered by this model during the testing phase. After a transient excursion from the initial state, the deciphered output converges to a good approximation of the actual Lorenz trajectory. \emph{(d)} The $\tilde{\mathbf{s}}_{\text{ros}}$ deciphered by this model during the testing phase. After a transient excursion from the initial state, the deciphered output converges to a good approximation of the actual R\"ossler trajectory.}
  \label{fig:decipher}
\end{figure}

\section{Separating Superimposed Input from Different Sources}

Thus far, we have considered cases in which the external input comprises a trajectory generated by one chaotic system at a time. However, in real environments, complex adaptive systems -- including the human brain -- often process mixed sensory input given by a superposition of multiple input systems \cite{ding2012emergence}. It is therefore natural to ask whether the encoding mechanism (invertible generalized synchronization) allows our learning system to execute such a function. To address this question, we employ a $2000$-node RNN with \ER topology as the central system (Fig.~\ref{fig:decipher}), and we construct a $3$-dimensional mixed input signal $\mathbf{s}_{\text{mix}}(t)=\mathbf{s}_{\text{lor}}(t)+\mathbf{s}_{\text{ros}}(t)$, where $\mathbf{s}_{\text{lor}}(t)$ and $\mathbf{s}_{\text{ros}}(t)$ are independent trajectories from the Lorenz attractor $\AAA_{\text{Lorenz}}$ and the R\"ossler attractor $\AAA_{\text{R\"ossler}}$, respectively. As the central system encodes the mixed input into its own state variable $\mathbf{x}(t)$, we seek two output weight matrices, $\mathbf{W}_{\text{lor}}$ and $\mathbf{W}_{\text{ros}}$, that can successfully reconstruct the segregated signals, i.e., $\mathbf{W}_{\text{lor}}\mathbf{x}(t)\approx \mathbf{s}_{\text{lor}}(t)$ and $\mathbf{W}_{\text{ros}}\mathbf{x}(t)\approx \mathbf{s}_{\text{ros}}(t)$. We find that such output matrices do exist, because the delta-rule adaptation of both $\mathbf{W}_{\text{lor}}$ and $\mathbf{W}_{\text{ros}}$ converges after many repetitions of the learning phase.

Next we test the decoding power of the trained central system by constructing a new mixed input signal $\mathbf{s}_{\text{mix}}(t)$ (Fig.~\ref{fig:decipher}(b)), which differs markedly from the mixed signal used in the learning phase. Starting from a random initial state $\mathbf{x}(t)$, we drive the central system with this new mixed input $\mathbf{s}_{\text{mix}}(t)$ and monitor the two $3$-dimensional outputs $\tilde{\mathbf{s}}_{\text{lor}}(t)=\mathbf{W}_{\text{lor}}\mathbf{x}(t)$ and $\tilde{\mathbf{s}}_{\text{ros}}(t)=\mathbf{W}_{\text{ros}}\mathbf{x}(t)$ (Fig.~\ref{fig:decipher}(c,d)). We observe that the segregated signals are indeed reconstructed from the central system by the two output matrices. Taken together, these simulations demonstrate that segregated signals can be retrieved and reconstructed from the state variable of the central system in our learning framework when the central system receives mixed signals. 

In order to build a deeper intuition, it is helpful to consider the relevant underlying mechanism within the dynamical systems framework. Formally, we consider the input signal $\mathbf{s}_{\text{mix}}=\mathbf{s}_{\text{lor}}+\mathbf{s}_{\text{ros}}$ a superposition of any Lorenz trajectory and any \Rossler trajectory, and we formulate a $6$-dimensional dynamical system that combines the Lorenz system and the \Rossler system. The state variable of this combined system is the Cartesian product of the state variables of the Lorenz system and the \Rossler system,
\begin{equation}
\mathbf{s}_{\text{com}}(t)=\begin{bmatrix}
    \mathbf{s}_{\text{lor}}(t) \\
    \mathbf{s}_{\text{ros}}(t)
\end{bmatrix}\in\mathbb{R}^6.
\label{eqn:s_cob}
\end{equation}
As $\mathbf{s}_{\text{lor}}(t)$ and $\mathbf{s}_{\text{ros}}(t)$ evolve on the Lorenz attractor $\AAA_{\text{Lorenz}}$ and the R\"ossler attractor $\AAA_{\text{R\"ossler}}$, respectively, the $\mathbf{s}_{\text{com}}(t)$ evolves onto an attractor that is the Cartesian product of the two attractors, $\AAA_{\text{cob}}=\AAA_{\text{Lorenz}}\times\AAA_{\text{R\"ossler}}\subset\mathbb{R}^6$. Then, we consider the central system as a response system that is one-way coupled to the combined dynamical system through an incomplete measurement function $h(\cdot)$,
\begin{equation}
\mathbf{s}_{\text{mix}} = h(\mathbf{s}_{\text{com}}) = \begin{bmatrix}
    1 & 0 & 0 & 1 & 0 & 0 \\
    0 & 1 & 0 & 0 &1 & 0 \\
    0 & 0 & 1 & 0 & 0 & 1
\end{bmatrix}
\begin{bmatrix}
    \mathbf{s}_{\text{lor}} \\ \mathbf{s}_{\text{ros}}
\end{bmatrix}.
\label{eqn:measurement}
\end{equation}
Although $h(\cdot)$ only provides a $3$-dimensional projection $\mathbf{s}_{\text{mix}}$ of the $6$-dimensional full measurement of the combined system, $\mathbf{s}_{\text{com}}$, the invertible generalized synchronization between the central system and the combined system may still occur \cite{lu2018attractor}. Once synchronized, the state of the central system $\mathbf{x}(t)\rightarrow{\bm \psi}(\mathbf{s}_{\text{com}}(t))$, where ${\bm \psi}(\cdot)$ is invertible on $\AAA_{\text{com}}$. The $\AAA_{\text{com}}$ is then encoded into the central system in the form of an internal representation $\PPP_{\text{com}} = {\bm \psi}(\AAA_{\text{com}})$. Thus, it becomes obvious that one can reconstruct both $\mathbf{s}_{\text{lor}}$ and $\mathbf{s}_{\text{ros}}$ by output weight matrices that approximate the ${\bm \psi}^{-1}(\cdot)$. That is,
\begin{equation}
\begin{bmatrix}
    \mathbf{W}_{\text{lor}} \\ \mathbf{W}_{\text{ros}}
\end{bmatrix}\mathbf{x} \approx {\bm \psi}^{-1}(\mathbf{x}),
\label{eqn:inverse_demix_function}
\end{equation}
for $\mathbf{x}\in\PPP_{\text{com}}$. Thus we see that it is the nature of the information encoding that allows for the existence of output matrices that can reconstruct segregated signals from $\mathbf{x}(t)$. The mechanism of information encoding -- invertible generalized synchronization -- does not only encode the mixed input $\mathbf{s}_{\text{mix}}$, but also the two segregated signals $\mathbf{s}_{\text{lor}}$ and $\mathbf{s}_{\text{lor}}$ into its state space, even though the segregated signals are not explicitly given to the central system. 

\section{Conclusion}

Complementing probabilistic models for learning hidden properties from experience \cite{lake2015human,battaglia2013simulation,tenenbaum2011grow,chater2010bayesian,griffiths2008bayesian,kemp2008discovery,chater2006probabilistic,tenenbaum2006theory}, we propose a general dynamical systems model derived from first principles. Surprisingly, this simple dynamical model produces successful learning of attractor dynamics from exemplary trajectories on these attractors. Our model also supports many other complex learning functions, such as task-switching, filling-in missing variables, and separating mixed inputs. We believe that the model serves as a promising preliminary framework that can be extended in future studies to include additional biological details supporting better performance. But perhaps most importantly, the framework sheds light on the possible dynamical origin of the emergence of many learning functions in biological and artificial systems.

\section*{Acknowledgements}
We thank Elisabeth A. Karuza, Christopher W. Lynn, Jason Z. Kim, Lia Papadopoulos, and Arian Ashourvan for helpful comments on earlier versions of this manuscript. ZL and DSB acknowledge support from the John D. and Catherine T. MacArthur Foundation, the Alfred P. Sloan Foundation, the ISI Foundation, the Paul Allen Foundation, the Army Research Laboratory (W911NF-10-2-0022), the Army Research Office (Bassett-W911NF-14-1-0679, Grafton-W911NF-16-1-0474, DCIST-W911NF-17-2-0181), the Office of Naval Research, the National Institute of Mental Health (2-R01-DC-009209-11, R01-MH112847, R01-MH107235, R21-M MH-106799), the National Institute of Child Health and Human Development (1R01HD086888-01), National Institute of Neurological Disorders and Stroke (R01 NS099348), and the National Science Foundation (BCS-1441502, BCS-1430087, NSF PHY-1554488 and BCS-1631550). The content is solely the responsibility of the authors and does not necessarily represent the official views of any of the funding agencies.

\section*{Appendix A: Lyapunov exponents and the stability of encoding and retrieval}

\subsection*{1. Largest conditional Lyapunov exponent and generalized synchronization}

One common criterion of generalized synchronization is the sign of the largest conditional Lyapunov exponent \cite{PhysRevLett.64.821}. By conditional, we mean that --- in this directionally coupled drive-response system (Eqs.~(\ref{eqn:input_system},\ref{eqn:central_system})) --- the stability of the response system (Eq.~(\ref{eqn:central_system})) is measured under the condition that it is driven by an unperturbed input system (Eq.~(\ref{eqn:input_system})). Specifically, we drive the learning phase central system (Eq.~(\ref{eqn:central_system})) with the same input trajectory $\mathbf{s}(t)$ that is on $\AAA$, yet from two nearby random initial states: $\mathbf{x}(0)$ and $\mathbf{x}'(0)=\mathbf{x}(0)+{\bm \delta}_0$, where $\Vert{\bm \delta}_0\Vert_2\ll 1$. The largest conditional Lyapunov exponent $\lambda_{\text{max}}$ is then the exponential convergence or divergence rate of the distance $\Vert{\bm \delta}_t\Vert_2=\Vert\mathbf{x}'(t)-\mathbf{x}(t)\Vert_2$ between these two trajectories,
\begin{equation}\tag{A1}
\lambda_{\text{max}}=\lim_{T\rightarrow\infty}\frac{1}{T}\log(\frac{\Vert{\bm \delta}_t\Vert_2}{\Vert{\bm \delta}_0\Vert_2}).
\label{eqn:lyp}
\end{equation}
Notice that ${\bm \delta}_t$ evolves following,
\begin{equation}\tag{A2}
{\bm \delta}_{t+1}=\mathbf{J}(t){\bm \delta}_{t},
\label{eqn:delta}
\end{equation}
where $\mathbf{J}(t)\in\mathbb{R}^{N\times N}$ is the Jacobian matrix of the externally driven response system,
\begin{equation}\tag{A3}
\mathbf{J}(t)=\frac{\partial\mathbf{f}(\mathbf{x},\mathbf{s})}{\partial\mathbf{x}}\at[\Big]{\mathbf{x}=\mathbf{\hat{x}}(t),\mathbf{s}=\mathbf{\hat{s}}(t)}.
\label{eqn:jacobian}
\end{equation}
Here, $\mathbf{\hat{x}}(t)$ is the trajectory of the central system (Eq.~(\ref{eqn:central_system})), and $\mathbf{\hat{s}}(t)$ the driving trajectory from the input system (Eq.~(\ref{eqn:input_system})). With Eqs.~(\ref{eqn:lyp}-\ref{eqn:jacobian}), we can numerically estimate the largest conditional Lyapunov exponent $\lambda_{\text{max}}$. If $\lambda_{\text{max}}<0$, we say that the central system exhibits generalized synchronization to the input system, and the input information is encoded into the central system in the form of $\mathbf{x}={\bm \psi}(\mathbf{s})$. Conversely, when $\lambda_{\text{max}}\geq 0$, the state of the central system can not be uniquely determined by the input $\mathbf{s}$.

\begin{figure}
  \center
  \includegraphics[width=0.6\linewidth]{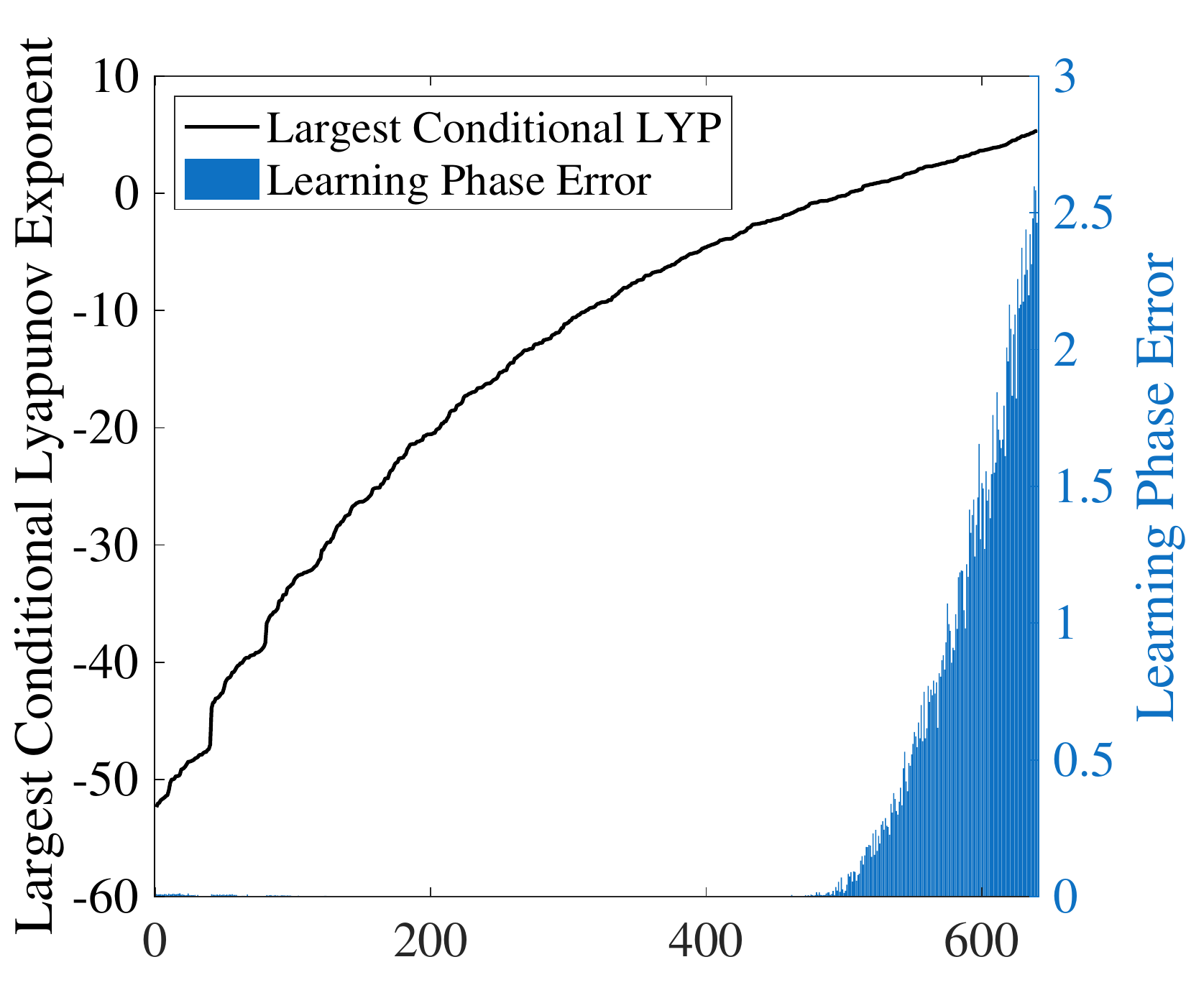}
  \includegraphics[width=0.6\linewidth]{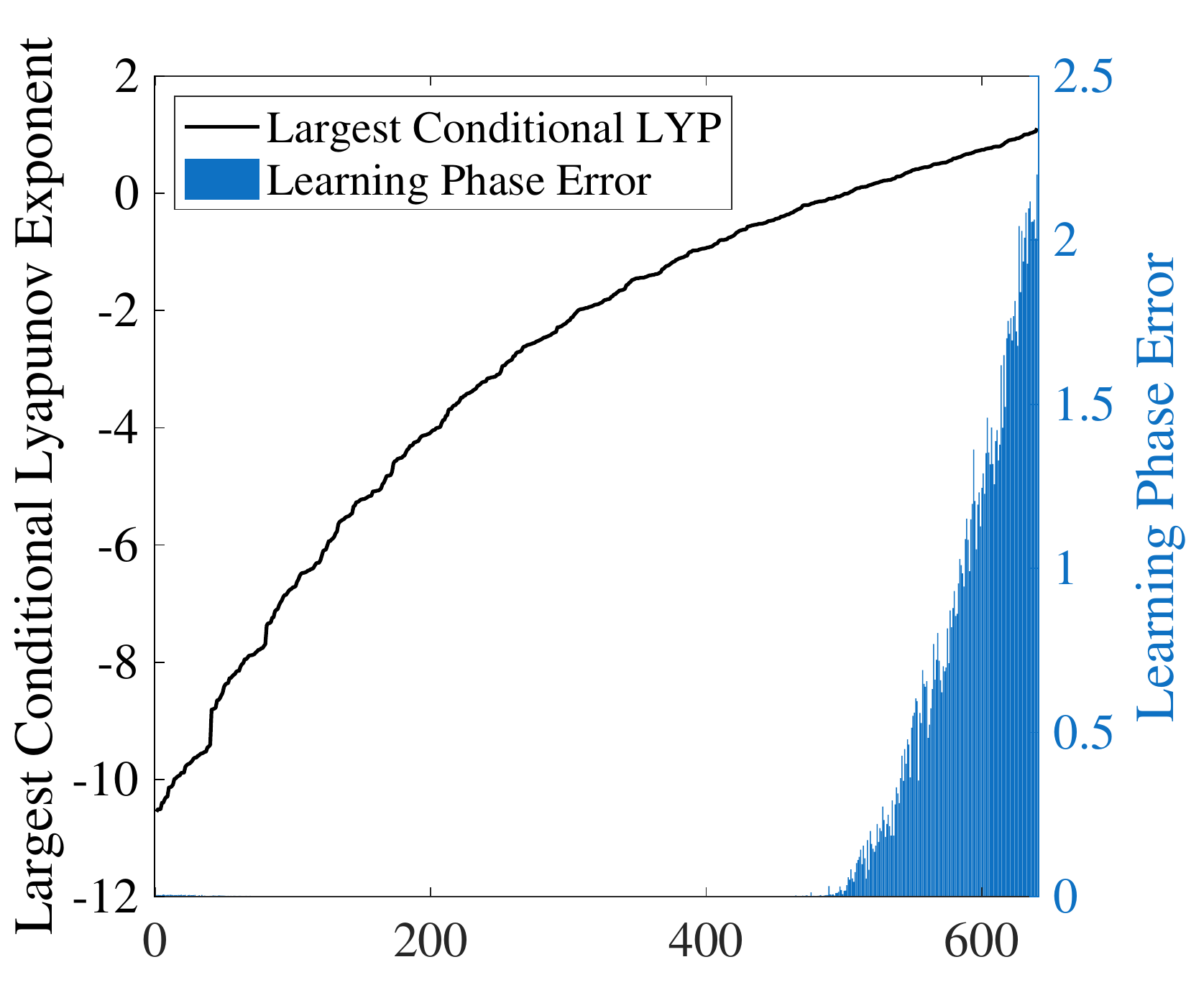}
  \linespread{1.0}
  \caption{\textbf{Relation between the largest conditional Lyapunov exponent and the learning phase error.} These two panels show (\emph{left y-axis}) the largest conditional Lyapunov exponent in the learning phase and (\emph{right y-axis}) the corresponding learning phase error, for the Lorenz task (\emph{left panel}) and the \Rossler task (\emph{right panel}) for $640$ trained RNNs with \ER topology. These $640$ RNNs were constructed using $10$ random realizations for each of the $64$ parameter pairs $(\rho,\xi)$ spanning the two-dimensional space defined by the spectral radius $\rho=0.5,\ 0.6,\ ...,\ 2.0$ of the \ER graph and the separation parameter $\xi=5,\ 10,\ 15,\ 20$ between the two attractors. Collectively, these results show that -- for either task -- the learning phase error is particularly great when the largest conditional Lyapunov exponent is positive.}
  \label{fig:Learning_LYP}
\end{figure}

To provide further intuition, we perform a series of numerical experiments. Using the learning schedule discussed in Appendix D, we train the output functions of $640$ randomly constructed RNNs with \ER topology on both the Lorenz and the \Rossler tasks. Once the learning schedules are finished, regardless of the convergence of $\mathbf{W}_\text{out}$, we calculate the learning phase error for each task; that is, we calculate the mean squared error between the actual sensory input $\mathbf{s}(t)$ and the predicted sensory input $\mathbf{\tilde{s}}(t)$. These encoding errors are plotted together with the largest conditional Lyapunov exponents in Fig.~\ref{fig:Learning_LYP}. The results are sorted by the largest conditional Lyapunov exponent in an increasing order. Consistent with prior theoretical work \cite{PhysRevLett.64.821}, the learning phase errors are particularly large when the largest conditional Lyapunov exponents become positive.

\subsection*{2. Largest transversal Lyapunov exponent and the persistence of the imitation}

To generate an output signal that reliably imitates the learned dynamics, the testing phase autonomous central system (Eq.~(\ref{eqn:auto_brain})) should remain on $\PPP$. Although the generalized synchronization guarantees that the central system remains on $\PPP$ during the learning phase, it is not necessarily the case when the central system becomes autonomous during the testing phase. Thus, it is worth analyzing the testing phase stability by calculating the Lyapunov exponents of this autonomous central system as it evolves on $\PPP$. Specifically, we calculate the Lyapunov exponents of the autonomous central system using the Jacobian of the testing phase central system
\begin{equation}\tag{A4}
\mathbf{J}(t)=\frac{d\mathbf{f}(\mathbf{x},\bm \phi(\mathbf{x}))}{d\mathbf{x}}\at[\Big]{\mathbf{x}=\mathbf{\hat{x}}(t)},
\label{eqn:jacobian_auto}
\end{equation} 
evaluated along a trajectory $\mathbf{\hat{x}}(t)$ from the learning phase central system (Eq.~(\ref{eqn:central_system})), rather than the testing phase central system (Eq.~(\ref{eqn:auto_brain})). This choice is motivated by our goal to evaluate the stability of a trajectory that remains on $\PPP$ even when the autonomous central system is transversely unstable and cannot be sustained on $\PPP$.

As the high-dimensional autonomous central system incorporates the local dynamics of the low-dimensional input system on $\AAA$, it inherits all the non-negative Lyapunov exponents from the input dynamical system \cite{lu2018attractor,pathak2017using}. Thus, $\PPP$ is an attractor of Eq.~(\ref{eqn:auto_brain}) if its $m+1$-th Lyapunov exponent is negative, where $m$ is the number of non-negative Lyapunov exponents from Eq.~(\ref{eqn:input_system}) on $\AAA$. For example, $m=2$ for both the Lorenz and the R\"ossler systems, as they both contain one positive and one zero Lyapunov exponent. For the $2$-D torus attractor, $m=2$ because it contains two zero Lyapunov exponents and zero positive Lyapunov exponents. For the limit cycle attractor, $m=1$ because it has only one non-negative Lyapunov exponent that is zero. For a fixed point attractor, $m=0$ because it does not have any non-negative Lyapunov exponents. Thus, for either the Lorenz or the \Rossler attractors, if the central system is transversally stable on the corresponding internal representation $\PPP$, the $3$rd largest Lyapunov exponent of the central system given by Eq.~(\ref{eqn:auto_brain}) should be negative.

\subsection*{3. Quantifying learning performance using the testing phase error}
\begin{figure}
  \center
  \includegraphics[width=0.6\linewidth]{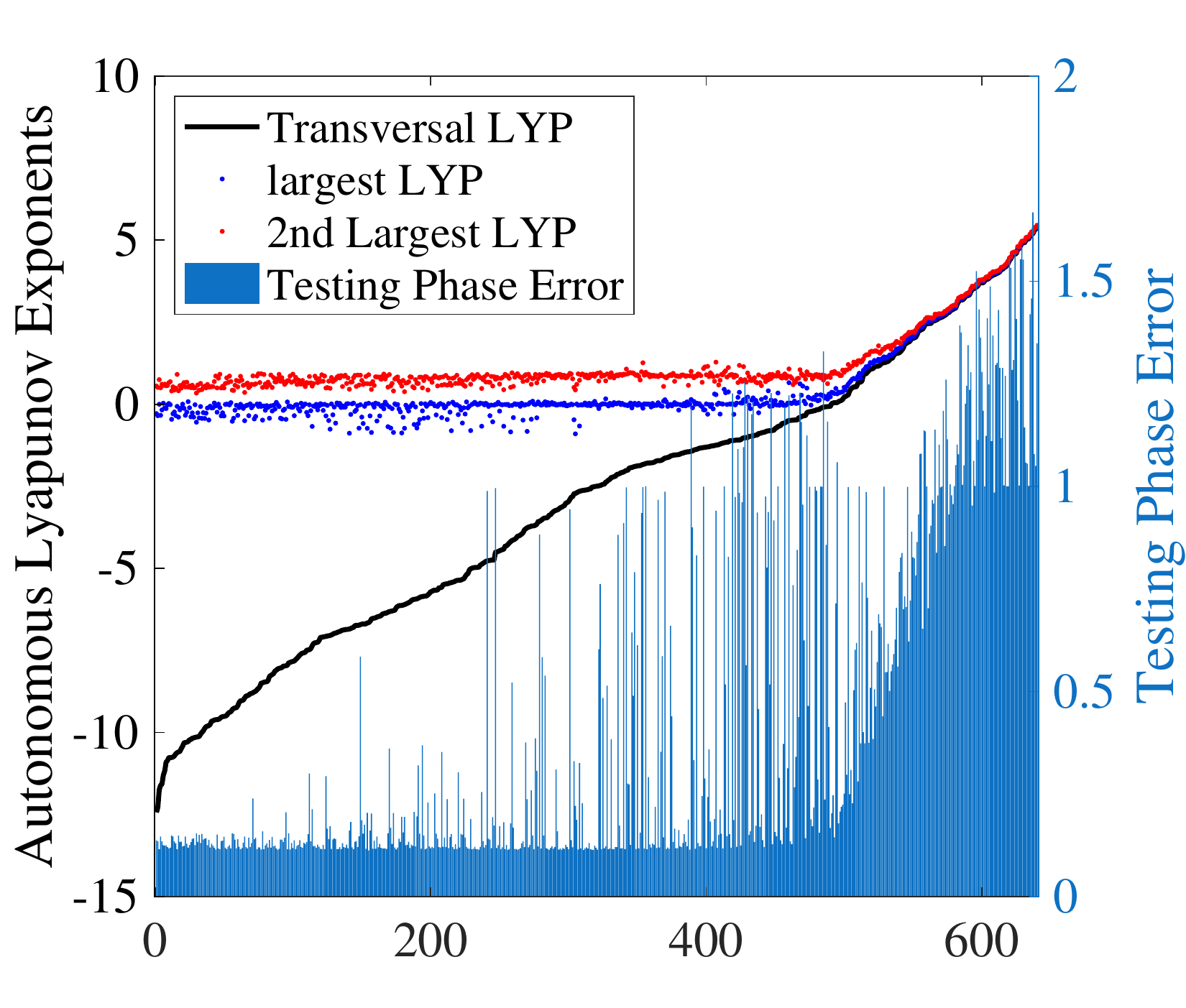}
  \includegraphics[width=0.6\linewidth]{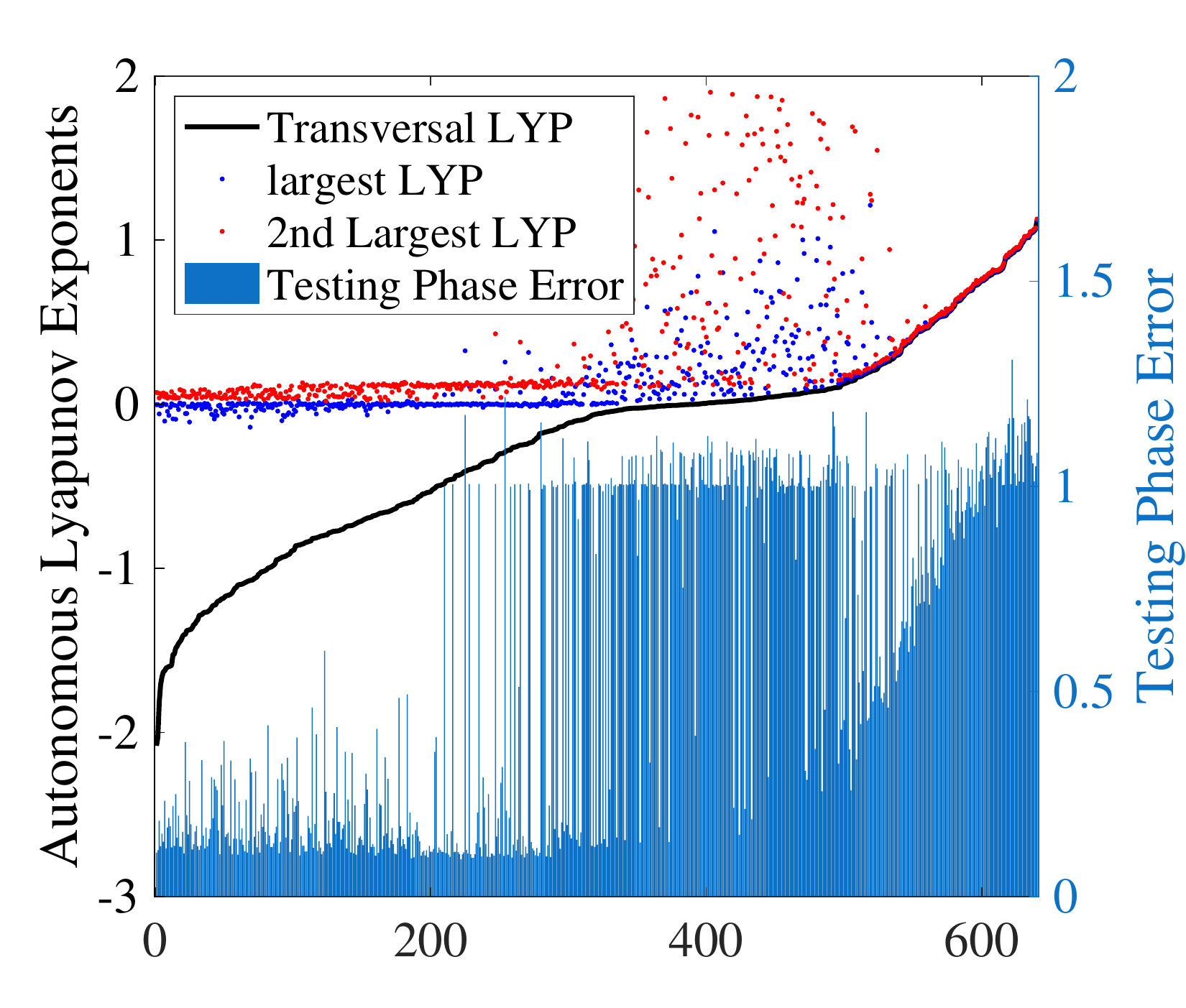}
  \linespread{1.0}
  \caption{\textbf{Relation between the largest Lyapunov exponents of the testing phase autonomous system and the testing phase error.} These two panels show (\emph{left y-axis}) the three largest Lyapunov exponents of the testing phase autonomous system as well as (\emph{right y-axis}) the testing phase error, on the Lorenz task (\emph{left panel}) and the \Rossler task (\emph{right panel}) for $640$ trained RNNs with \ER topology. These $640$ RNNs are sorted from least to greatest third largest Lyapunov exponent. When the third largest Lyapunov exponent is non-negative, the transversal Lyapunov exponent is greater than zero, and thus the internal representations are not transversely stable, resulting in notably worse performance. For those instances that have relatively low error, the first two Lyapunov exponents are close to the positive and zero Lyapunov exponents from the Lorenz and \Rossler systems.}
  \label{fig:Testing_LYP}
\end{figure}

During the testing phase, the system autonomously generates new streams of signals using the implicitly acquired dynamical rules. Since the new streams of signals are chaotic and are not intended to be identical to any of the exemplary signals used in the learning phase, it is impossible to quantify the error in the same manner as we did in the learning phase. Thus, we propose a different quantitative performance measure that we call the ``testing phase error,'' which describes the difference between the implicit dynamical rules acquired and the dynamical rules underlying the system that generated the exemplary sensory inputs.

Along an extended trajectory $\tilde{\mathbf{s}}(t)$ generated by the testing phase central system, we denote the system's movement during one unit of time as $\delta \tilde{\mathbf{s}}(t) = \tilde{\mathbf{s}}(t+1)-\tilde{\mathbf{s}}(t)$. Based on the actual dynamics of the input system (Eq.~(\ref{eqn:input_system})), we also calculate the ideal movement vector $\delta {\mathbf{s}}(t) = \mathbf{g}(\tilde{\mathbf{s}}(t))-\tilde{\mathbf{s}}(t)$ from each point $\tilde{\mathbf{s}}(t)$. The testing phase error is then defined as 
\begin{equation}\tag{A5}
\text{testing phase error}=\langle\frac{\Vert\delta {\tilde{\mathbf{s}}}(t) - \delta {\mathbf{s}}(t)\Vert_2}{\Vert\delta {\mathbf{s}}(t)\Vert_2}\rangle_t,
\label{eqn:testing_phase_error}
\end{equation}
where $\langle\cdot\rangle_t$ denote the time average along the trajectory. When the autonomously generated trajectory $\tilde{\mathbf{s}}(t)$ evolves on $\AAA$ with the correct underlying dynamics, the testing phase error is small ($\sim0.1$). When the generated trajectory escapes from $\AAA$, the testing phase error can be very large. 

In the results shown in Fig.~\ref{fig:Testing_LYP}, the testing phase error is small when the central system is transversely stable on the internal representation $\PPP$. However, due to the bubbling phenomenon \cite{venkataramani1996transitions}, we also note that -- while the largest transversal Lyapunov exponent describes the overall stability on $\PPP$ -- it is possible for the generated trajectory to escape from $\AAA$ even when the transversal Lyapunov exponent is negative. From the results reported in Fig.~\ref{fig:Testing_LYP}, we observe that there are many instances where the testing phase error is close to $1$. These instances may correspond to the cases where the generated trajectory escapes (via bubbling) from $\AAA$ and lands on a fixed point, resulting in $\delta {\tilde{\mathbf{s}}}(t)=0$, and thereby leading to a testing phase error of approximately $1$.

\section*{Appendix B: Preparing input trajectories from chaotic and non-chaotic systems}

In the main text, we demonstrate the capacity to learn the dynamics of multiple attractors with a proof-of-principle study considering $4$ attractors: one periodic, one quasi-periodic, and two chaotic. We also demonstrate the ability to successfully induce task switching among these four attractors. Outside of this specific study, the remainder of this paper predominantly focuses on the learning of the two widely studied $3$-dimensional chaotic dynamical systems: the Lorenz system and the \Rossler system. In the following subsections, we describe the $4$ dynamical systems in detail, and we also describe the construction of exemplary trajectories from each.

\subsection*{1. Input trajectory from the Lorenz system}

The input trajectory $\mathbf{s}_{\text{lor}}(t)$ from the Lorenz system is generated as follows. We integrate the differential equations of the Lorenz system 
\begin{gather}
  \frac{d}{d t}X_{\text{lor}} = 10Y_{\text{lor}}-10X_{\text{lor}}, \nonumber\\
  \frac{d}{d t}Y_{\text{lor}} = -X_{\text{lor}}Z_{\text{lor}} + 28X_{\text{lor}} -Y_{\text{lor}},\tag{B1}\\
  \frac{d}{d t}Z_{\text{lor}} = X_{\text{lor}}Y_{\text{lor}} - 8Z_{\text{lor}}/3,\nonumber
\label{eqn:Lorenz}
\end{gather}
using a $4$-th order Runge-Kutta integrator with time step $\delta t=10^{-3}$ from a random initial state. Each of the three variables $X_{\text{lor}}$, $Y_{\text{lor}}$, and $Z_{\text{lor}}$ of the trajectory is then normalized to have a mean of zero and a variance of one. This normalized trajectory is then saved as the Lorenz-task input trajectory $\mathbf{s}_{\text{lor}}(t)$ with time resolution $\tau=0.02$. 

\subsection*{2. Input trajectory from the R\"ossler system} 

Similar to the Lorenz system, the input trajectory $\mathbf{s}_{\text{ros}}(t)$ from the \Rossler system is generated as follows. We integrate the differential equations of the \Rossler system
\begin{gather}
  \frac{d}{d t}X_{\text{ros}} = -5Y_{\text{ros}} - 5Z_{\text{ros}} ,\nonumber\\
  \frac{d}{d t}Y_{\text{ros}} = 5X_{\text{ros}} + 5Y_{\text{ros}}/2, \tag{B2}\label{eqn:Rossler}\\
  \frac{d}{d t}Z_{\text{ros}} = 10 + 5Z_{\text{ros}}(X_{\text{ros}} - 4), \nonumber
\end{gather}
using a $4$-th order Runge-Kutta integrator with time step $\delta t=10^{-3}$ from a random initial state. The coefficients of the \Rossler system in Eq.~(\ref{eqn:Rossler}) are chosen such that the system has a similar time scale to that of the Lorenz system. We then normalize each of the three variables, $X_{\text{ros}}$, $Y_{\text{ros}}$, and $Z_{\text{ros}}$, such that they all have a mean of zero and a variance of one. We then save the normalized trajectory as the R\"ossler-task input trajectory $\mathbf{s}_{\text{ros}}(t)$ with time resolution $\tau=0.02$. 

\subsection*{3. Input trajectory from a system with a stable limit cycle}

In addition to the two chaotic systems, we consider a dynamical system with less complex dynamics, where the attractor is a limit cycle. In this case, orbits in the phase space of such dynamical systems spiral into a closed periodic orbit. We engineer a $3$-dimensional dynamical system with such limit cycle attractor dynamics. We integrate the differential equations of the following system
\begin{gather}
  \frac{d}{d t}X_{\text{lmc}} = 10X_{\text{lmc}}*(2-X_{\text{lmc}}^2-Y_{\text{lmc}}^2) - 10Y_{\text{lmc}},\nonumber\\
  \frac{d}{d t}Y_{\text{lmc}} = 10Y_{\text{lmc}}*(2-X_{\text{lmc}}^2-Y_{\text{lmc}}^2) + 10X_{\text{lmc}}, \tag{B3}\label{eqn:LimitCycle}\\
  \frac{d}{d t}Z_{\text{lmc}} = -10Z_{\text{lmc}}. \nonumber
\end{gather}
using a $4$-th order Runge-Kutta integrator with time step $\delta t=10^{-3}$ from a random initial state. We thereby obtain an orbit that approaches a periodic orbit. We then normalize each of the three variables, $X_{\text{lmc}}$, $Y_{\text{lmc}}$, and $Z_{\text{lmc}}$, such that they all have a mean of zero and a variance of one, except for $Z_{\text{lmc}}$. We then save the normalized trajectory as the Limit-Cycle-task input trajectory $\mathbf{s}_{\text{lmc}}(t)$ with time resolution $\tau=0.02$. 

\subsection*{4. Input trajectory from a system with a quasi-periodic torus attractor} 

Another non-trivial attractor, which is neither chaotic nor periodic, is the quasi-periodic torus attractor. The orbit of such a system flows along the surface of a $2$-dimensional torus. The motion is quasi-periodic because it is composed of oscillations of two incommensurate frequencies. The orbit on the torus attractor will densely cover the torus without collapsing into a periodic orbit. Following prior work \cite{kuznetsov2010simple}, we engineer a $3$-dimensional dynamical system with such quasi-periodic torus attractor dynamics. We integrate the differential equations of the following system
\begin{gather}
  \frac{d}{d t}X_{\text{trs}} = 2\pi Y_{\text{trs}},\nonumber\\
  \frac{d}{d t}Y_{\text{trs}} = 2\pi Y_{\text{trs}}(X_{\text{trs}}^2-0.5X_{\text{trs}}^4) - 4\pi^2 X_{\text{trs}}, \tag{B4}\label{eqn:Torus}\\
  \frac{d}{d t}Z_{\text{trs}} = 0.9-X_{\text{trs}}^2. \nonumber
\end{gather}
using a $4$-th order Runge-Kutta integrator with time step $\delta t=10^{-3}$ from a random initial state. We thereby obtain an orbit on the quasi-periodic torus attractor. We then normalize each of the three variables, $X_{\text{trs}}$, $Y_{\text{trs}}$, and $Z_{\text{trs}}$, such that they all have a mean of zero and a variance of one. We then save the normalized trajectory as the Torus-task input trajectory $\mathbf{s}_{\text{trs}}(t)$ with time resolution $\tau=0.02$. 

\begin{figure*}
  \centering
  \includegraphics[width=0.28\linewidth]{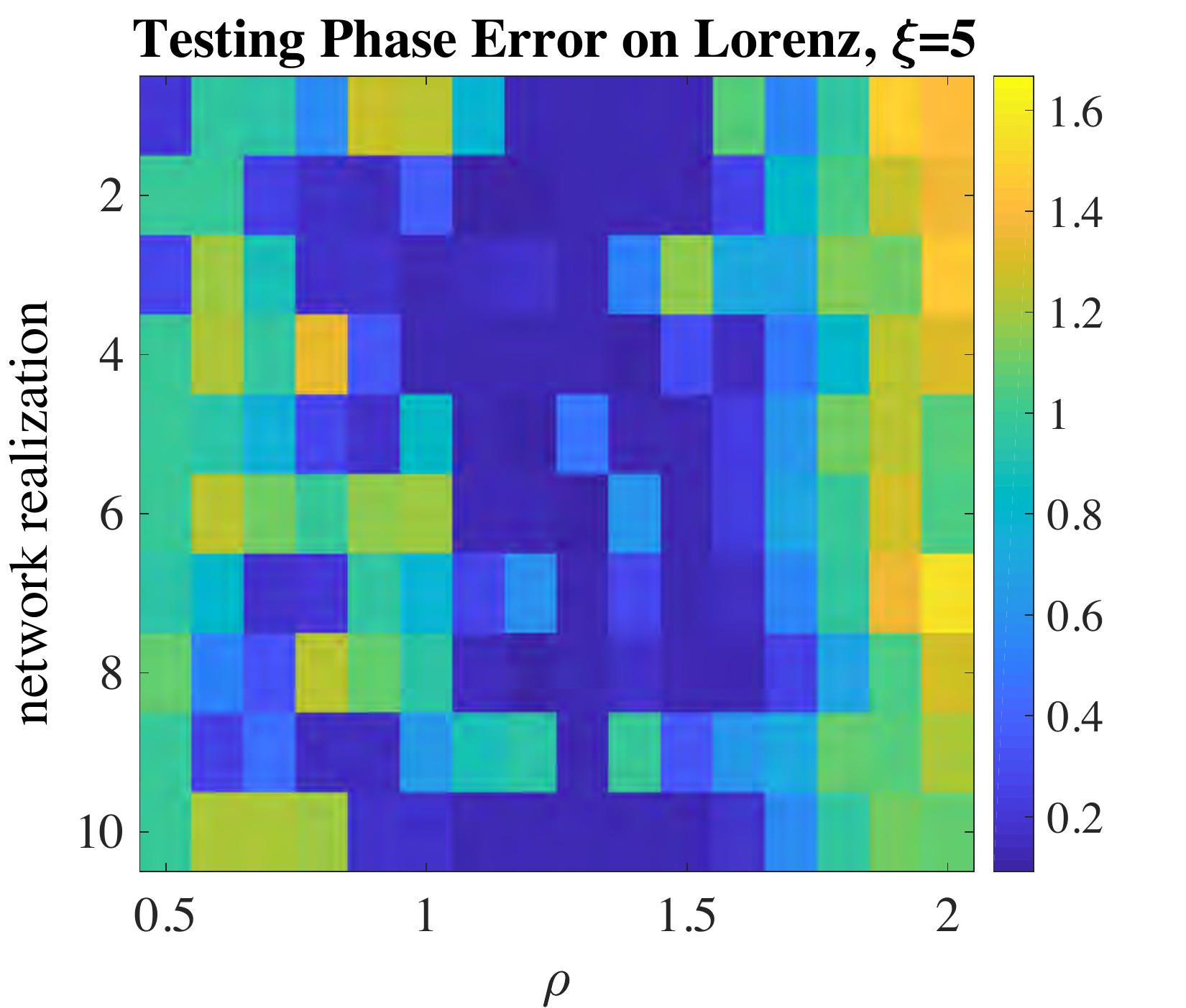}
  \includegraphics[width=0.28\linewidth]{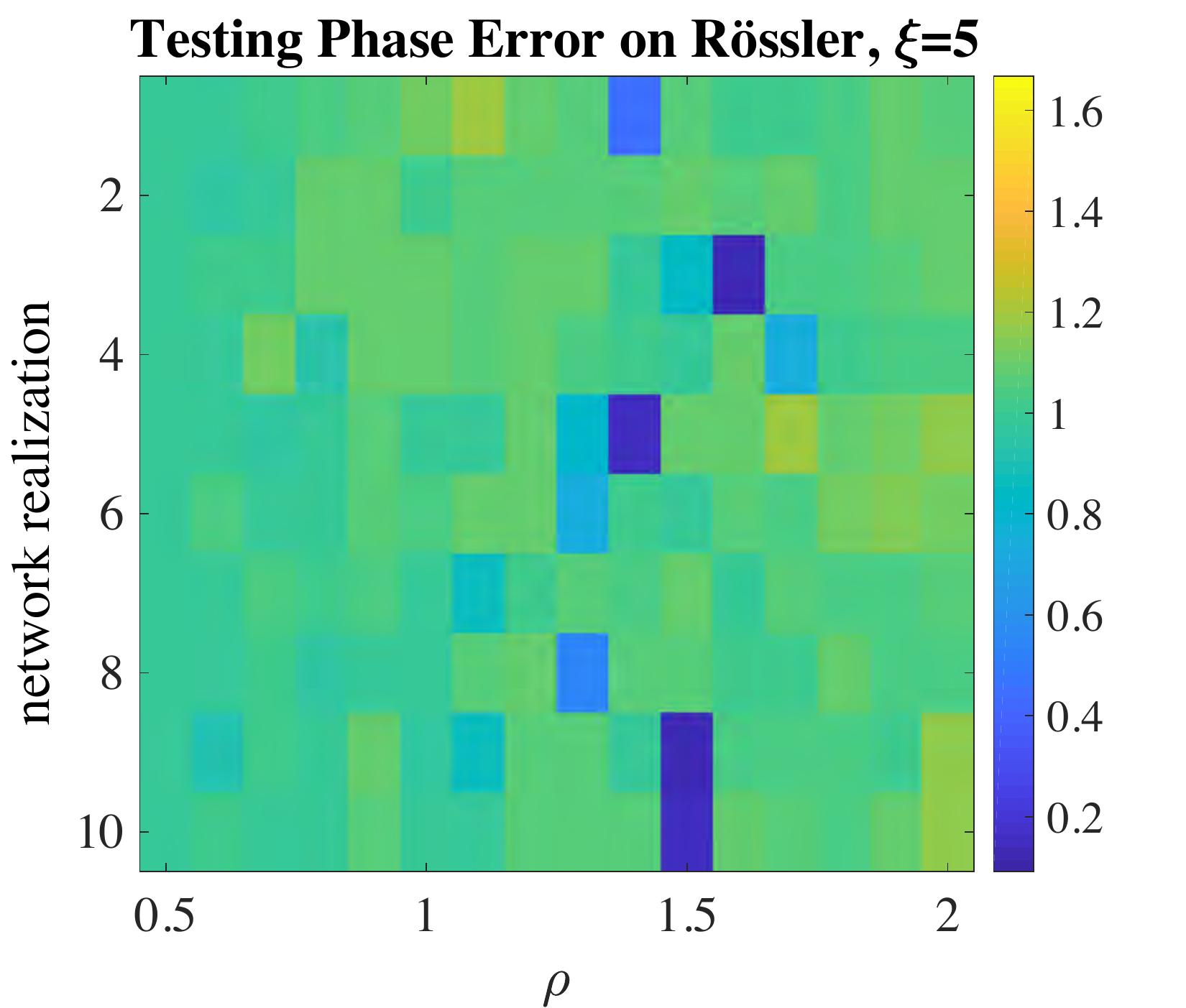}
  \includegraphics[width=0.28\linewidth]{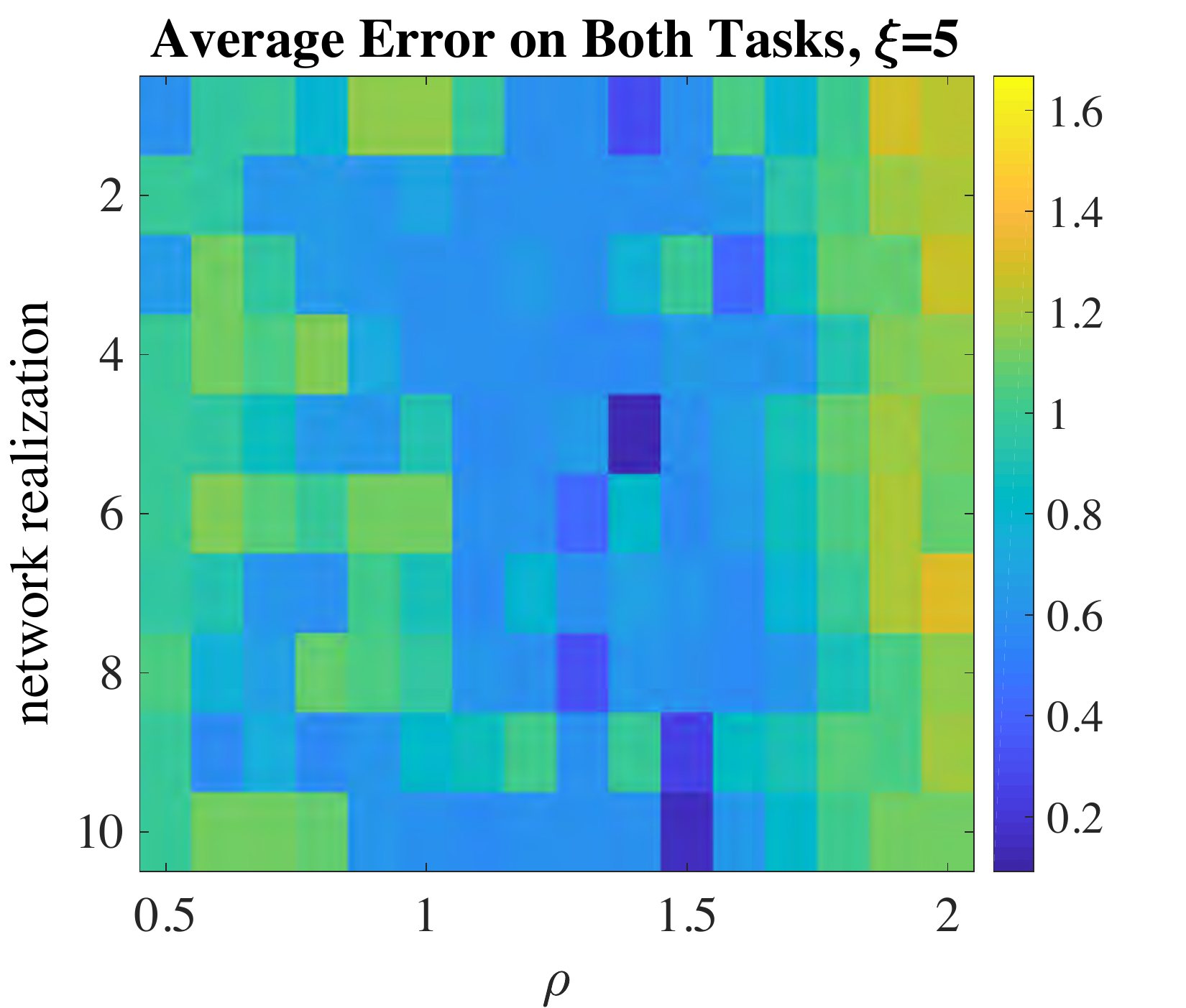}
  \includegraphics[width=0.28\linewidth]{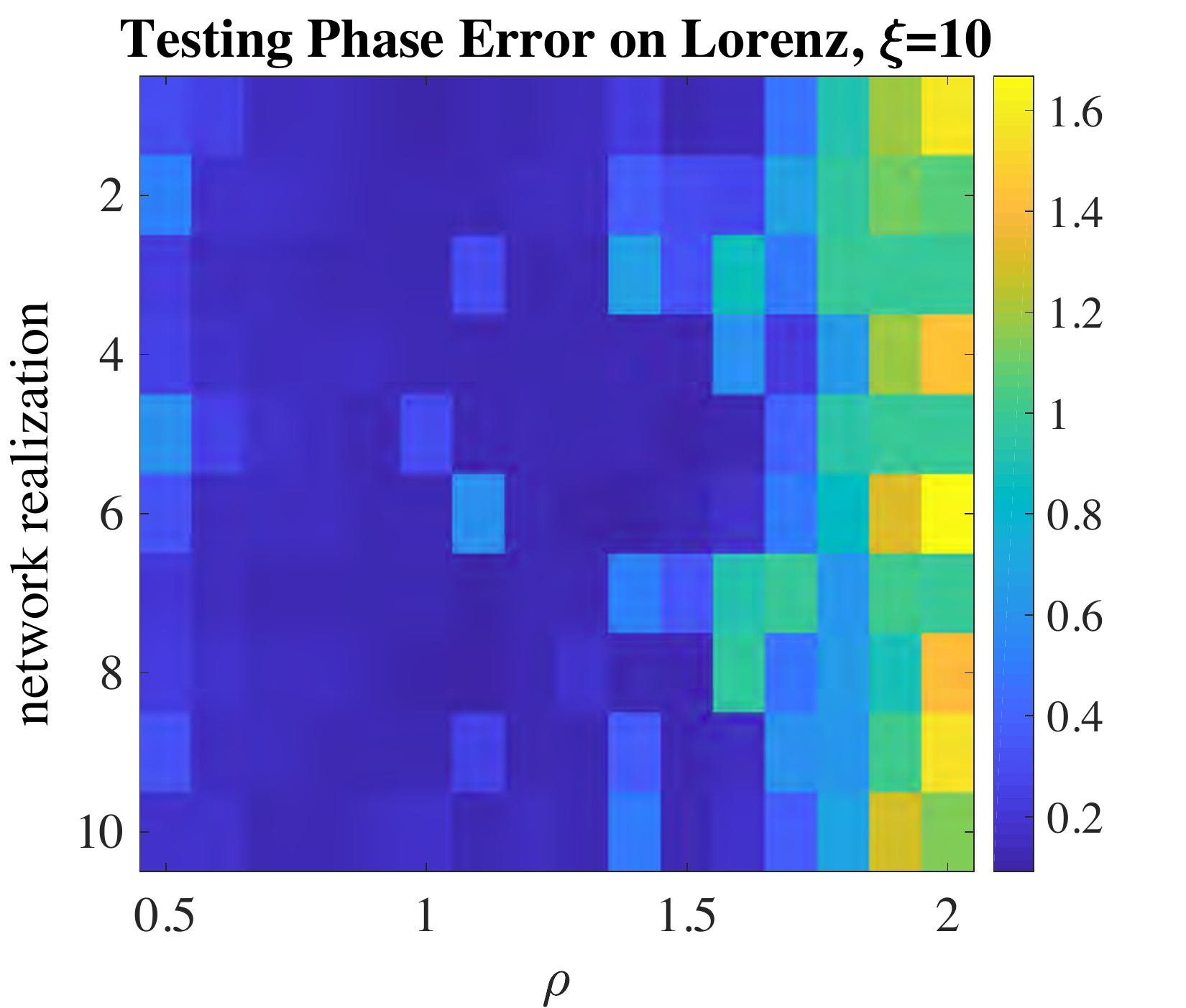}
  \includegraphics[width=0.28\linewidth]{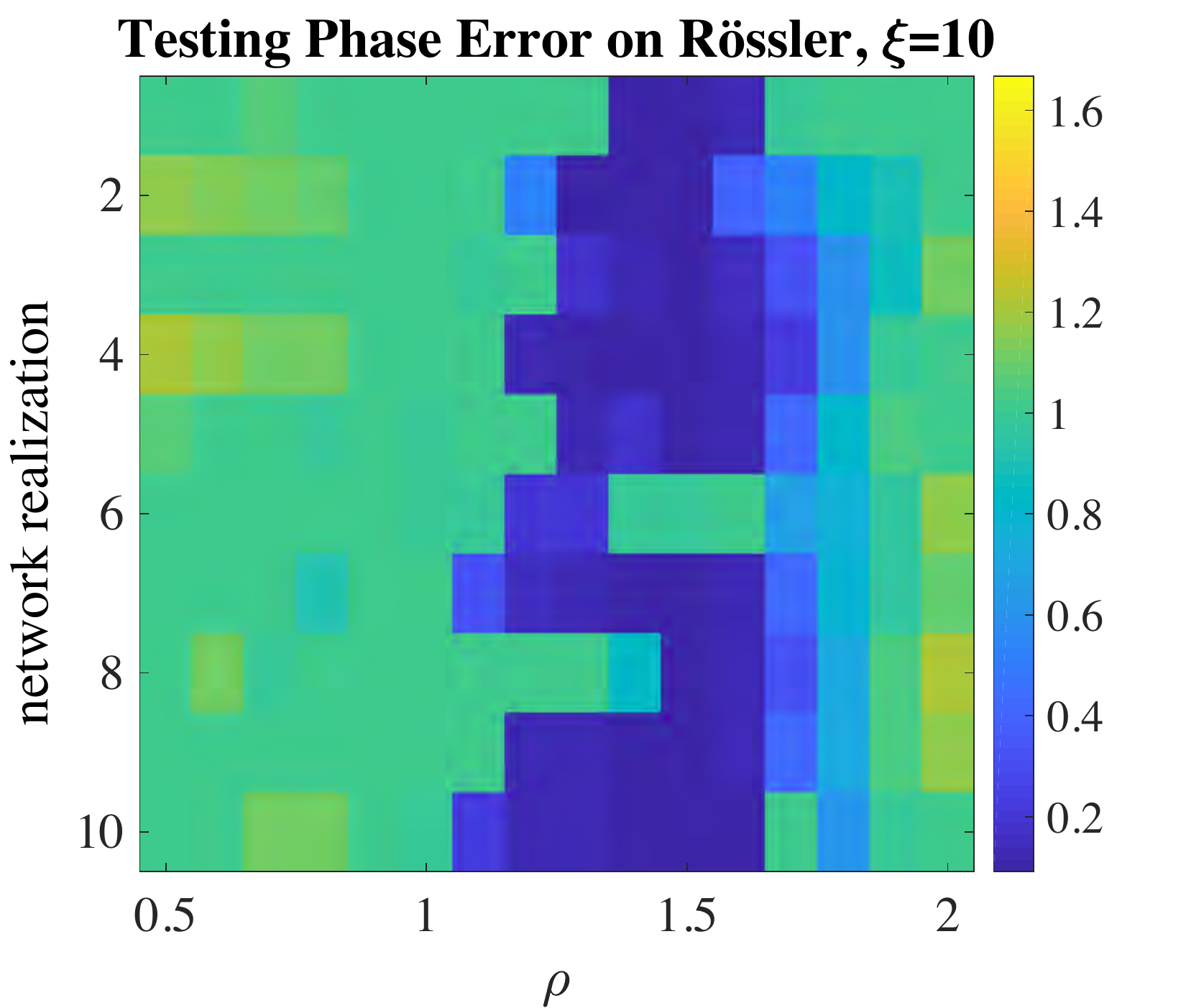}
  \includegraphics[width=0.28\linewidth]{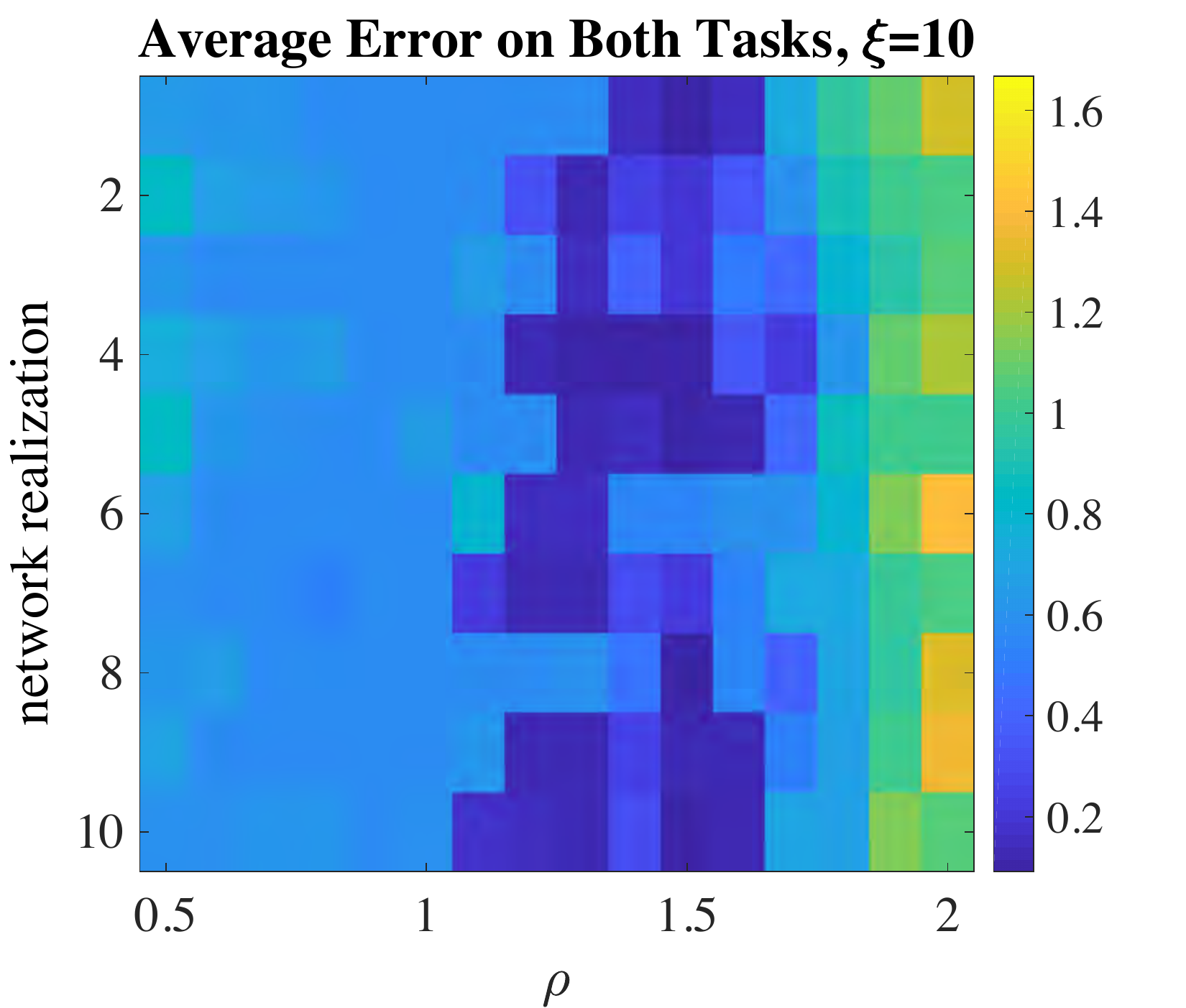}
  \includegraphics[width=0.28\linewidth]{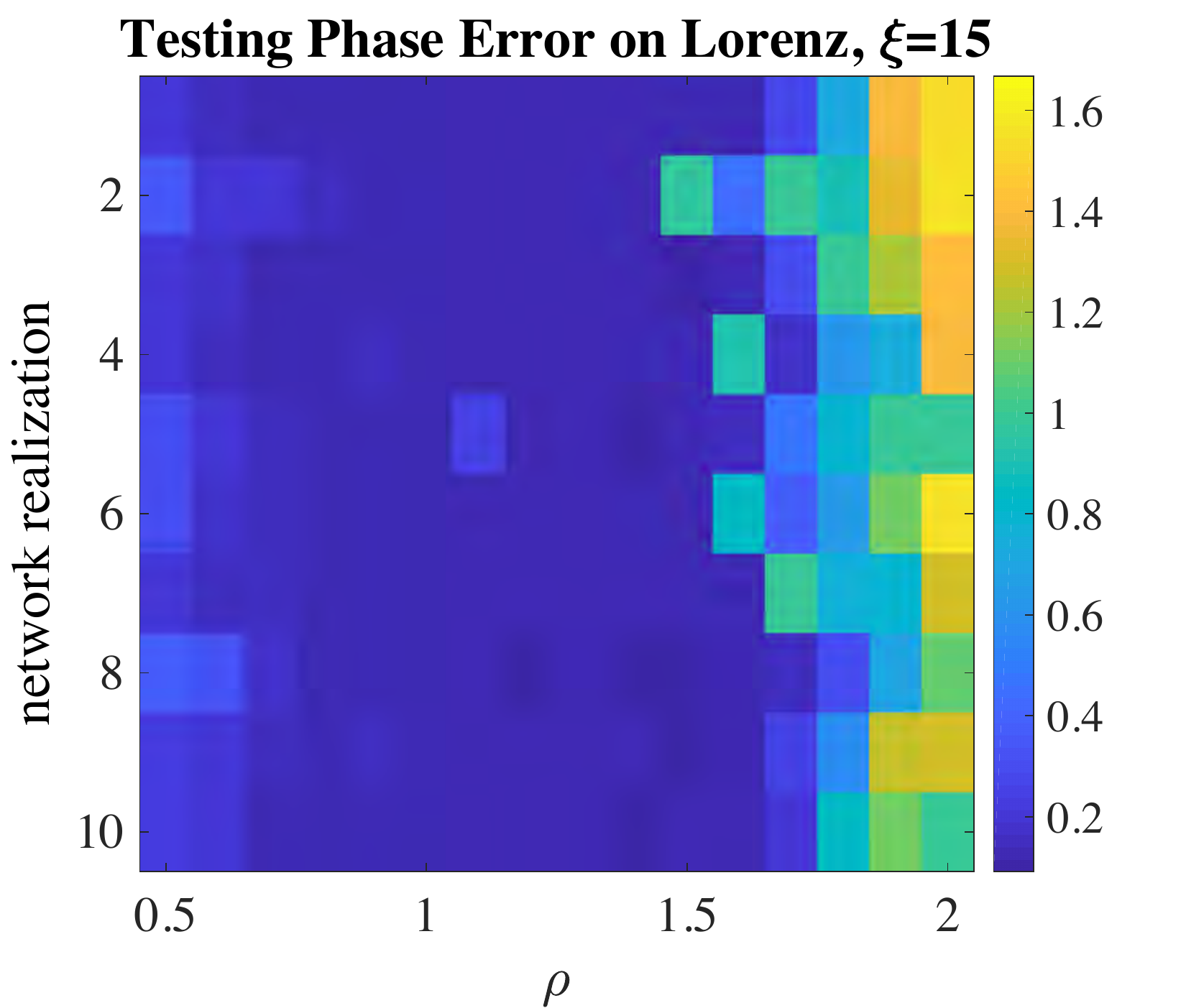}
  \includegraphics[width=0.28\linewidth]{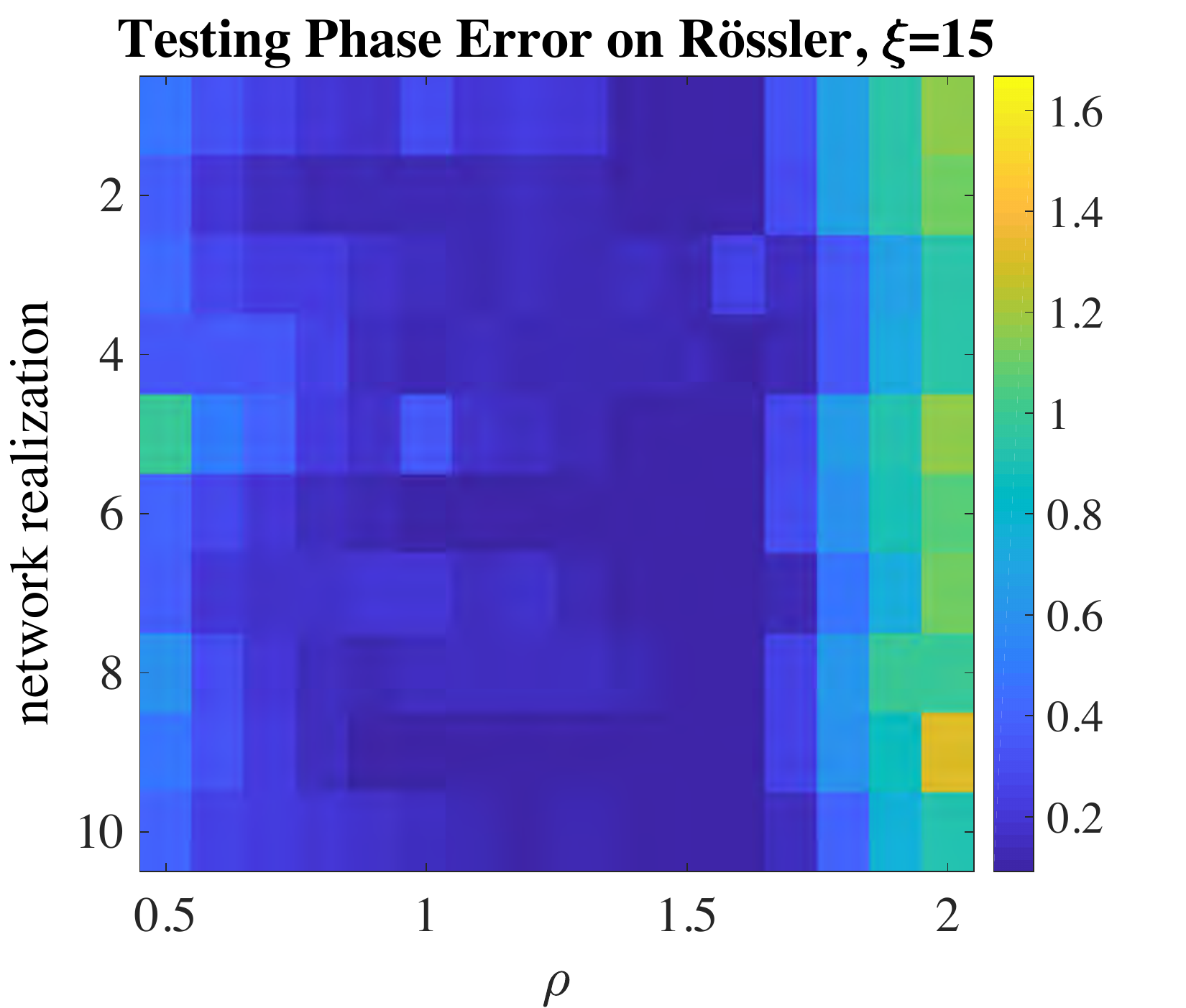}
  \includegraphics[width=0.28\linewidth]{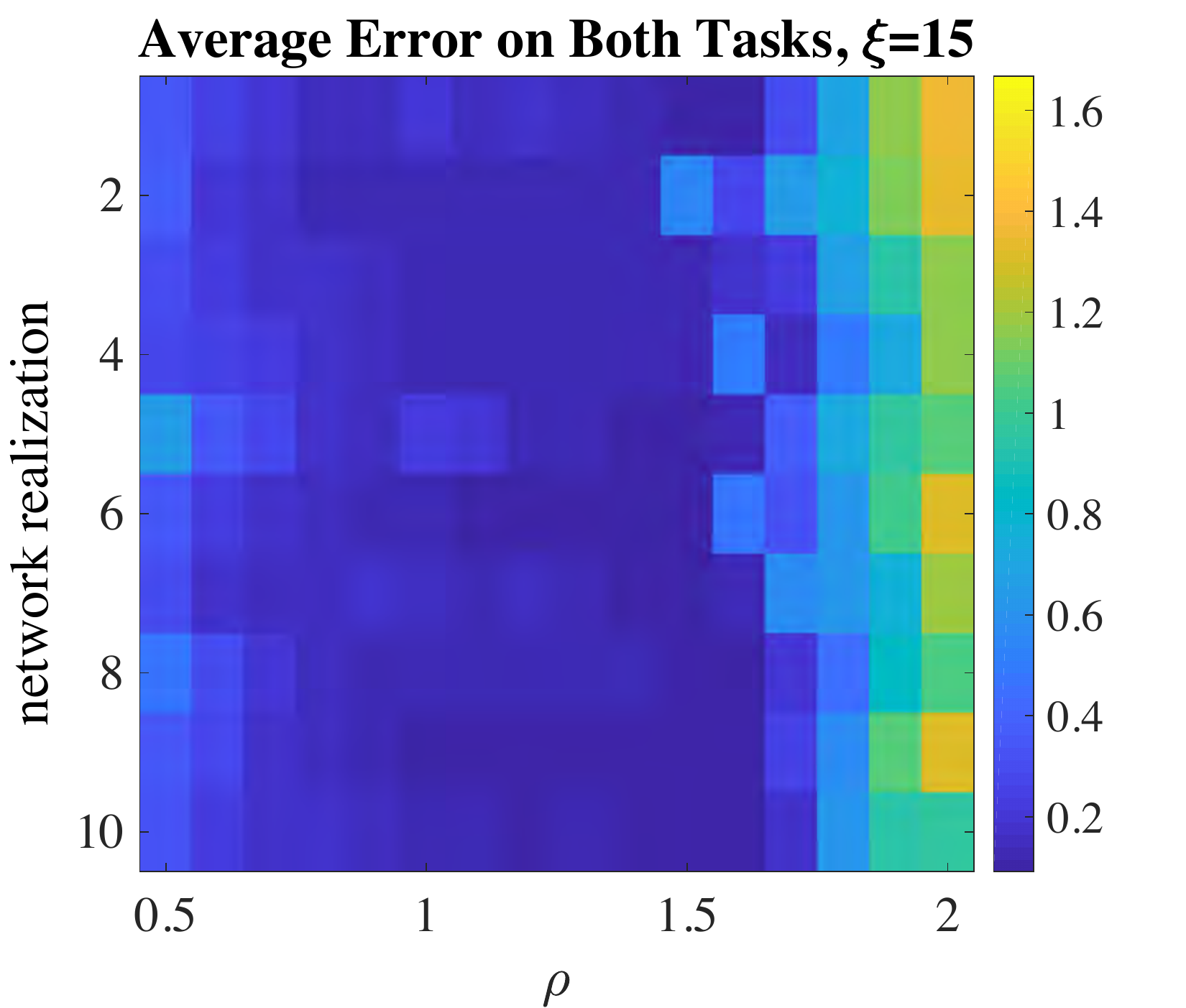}
  \includegraphics[width=0.28\linewidth]{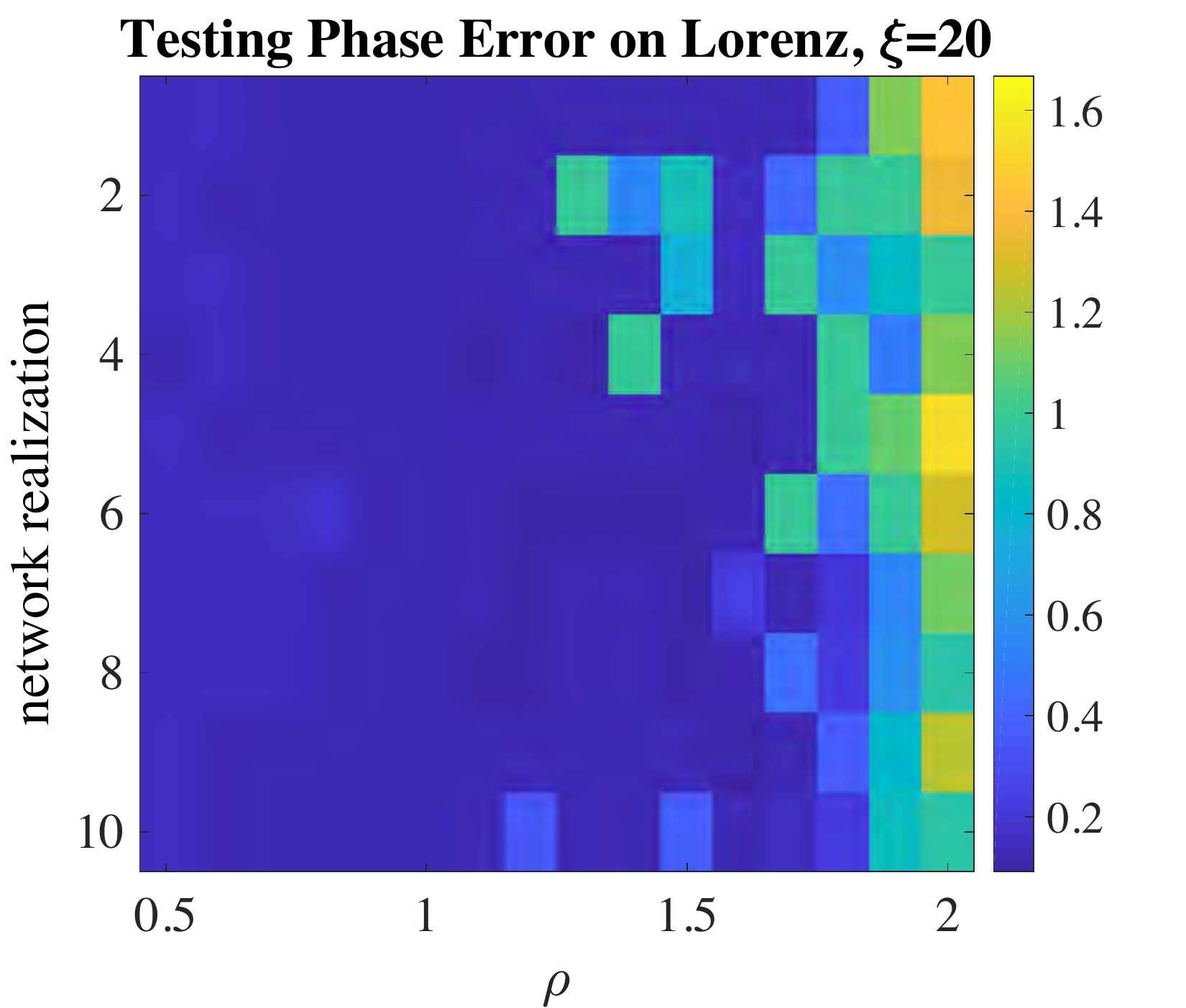}
  \includegraphics[width=0.28\linewidth]{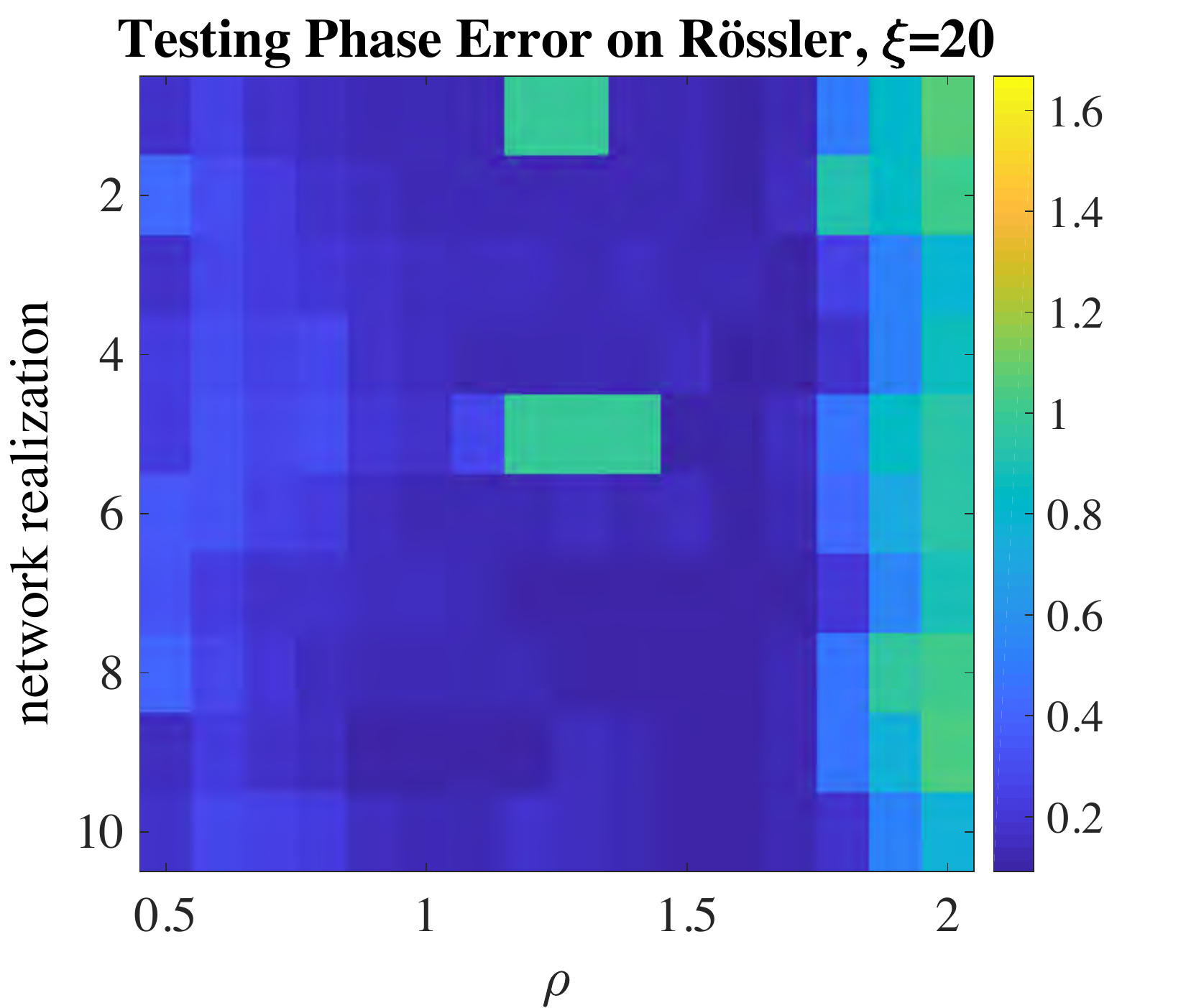}
  \includegraphics[width=0.28\linewidth]{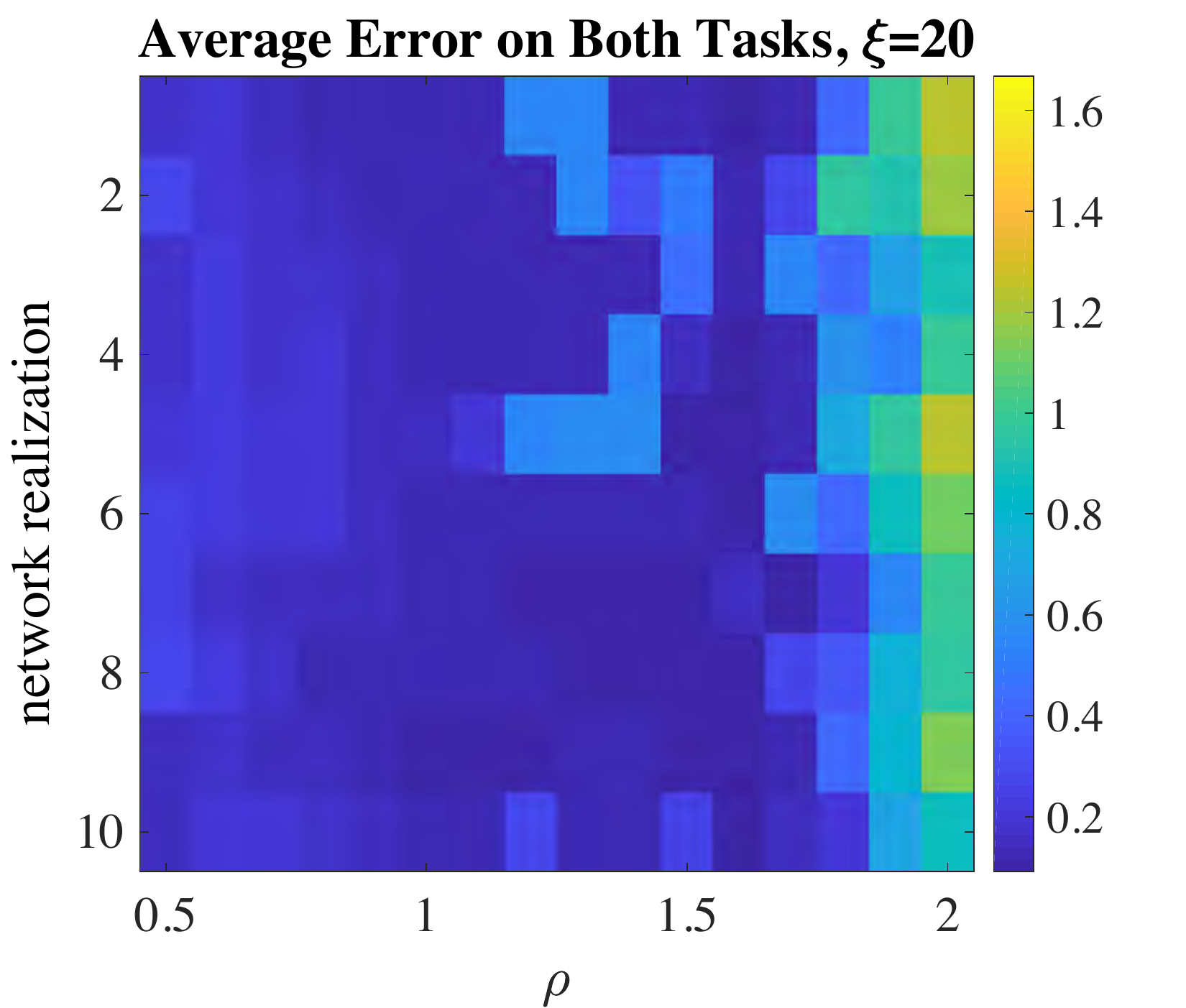}
  \linespread{1.0}
  \caption{\textbf{Relation between the testing phase error and the separation between attractors, as well as the spectral radius of the structural graph on which the RNN is instantiated.} In these heatmaps, color indicates the testing phase error on the Lorenz task (\emph{left}) and on the \Rossler task (\emph{center}). We also show the mean testing phase error averaged across the two tasks (\emph{right}). Each row of panels in the figure corresponds to a different value of the separation parameter $\xi=5,\ 10,\ 15,\ 20$, with the top row of panels corresponding to $\xi=5$ and the bottom row of panels corresponding to $\xi=20$. All panels reflect results obtained from RNNs with \ER topology, where the spectral radius $\rho$ takes on values ranging from $0.5,\ 0.6,\ ...,\ 2.0$ (\emph{x-axis})). For each $(\rho,\xi)$ pair, we construct $10$ realizations of the \ER topology (\emph{y-axis}). From these data, we can conclude that when $\xi$ is larger, more RNNs learn both the Lorenz and the \Rossler task well.}
  \label{fig:Testing_Error_on_shift}
\end{figure*}

\subsection*{5. Avoiding overlap between internal representations in multitask learning} 

In the multitask learning scenario, the central system (Eq.~(\ref{eqn:central_system})) learns attractor dynamics by forming an internal representation during the learning phase. During the testing phase, a successful central system will evolve autonomously on the internal representations of the Lorentz task $\PPP_{\text{Lorenz}}$ and of the \Rossler task $\PPP_{\text{R\"ossler}}$. However, if there exists any overlap between the two internal representations, at least one of the tasks will not be accurately learned. The reason for this inaccuracy is that the output mapping function ${\bm \phi}:\PPP\rightarrow \AAA$, where $\PPP = \PPP_{\text{Lorenz}}\cup\PPP_{\text{R\"ossler}}$ and $\AAA=\AAA_{\text{Lorenz}}\cup\AAA_{\text{R\"ossler}}$, loses its one-to-one nature on the intersected region $\mathbf{x}\in\PPP_{\text{Lorenz}}\cap\PPP_{\text{R\"ossler}}$. As a consequence, when $\mathbf{W}_\text{out}$ is sequentially updated following Eq.~(\ref{eqn:adapt}) on multiple attractors whose internal representations intersect each other, $\mathbf{W}_\text{out}$ does not converge.

To explicitly prevent such overlap, we modify the input trajectories by increasing the mean of the Lorenz-task input $\mathbf{s}_{\text{lor}}(t)$ to $[\xi,\xi,\xi]$, and by decreasing the mean of the R\"ossler-task input $\mathbf{s}_{\text{ros}}(t)$ to $[-\xi,-\xi,-\xi]$, where $\xi\geq0$. For the sake of visualization, we adopt $\xi=10$ for the simulated examples shown in the main text. Results shown in Figs.~\ref{fig:Testing_Error_on_shift} demonstrate that a larger value of $\xi$ results in lower testing phase errors on both tasks.

\section*{Appendix C: Instantiating the learning framework with different central systems}

Our goal in building this general theoretical framework was to provide a simplistic and general explanation of learning dynamics observed in many neurobiological systems. To support its generalizability, we seek to demonstrate that this dynamical learning framework specified by Eqs.~(\ref{eqn:input_system},\ref{eqn:central_system},\ref{eqn:auto_brain}) is robust to many different choices of the central system. To this end, we perform a range of numerical simulations to test several different central systems. In this section, we report on the nature of those experiments and also on the performance of these diverse central systems in learning both the Lorenz and \Rossler tasks. Specifically, in Appendix C.1-3, we let the central system be an RNN with several different topologies, while in Appendix C.4, we let the central system be a random polynomial system with polynomial degree equal to $5$, wrapped in the sigmoidal function $\tanh(\cdot)$ to prevent divergence.

\subsection*{1. RNNs with \ER topology}

In the numerical simulations whose results we have reported thus far, we model the central system as an RNN of $N=2000$ neurons with \ER topology. The $\tanh(\cdot)$ in Eq.~(\ref{eqn:RNN}) operates on a vector and returns a vector with the same shape that satisfies $[\tanh(\mathbf{x})]_i=\tanh([\mathbf{x}]_i)$. The adjacency matrix $\mathbf{A}\in\mathbb{R}^{2000\times 2000}$ is the weighted adjacency matrix of the RNN. The input weight matrix $\mathbf{W}_{\text{in}}\in\mathbb{R}^{2000\times 3}$ propagates a $3$-dimensional input $\mathbf{s}$ (either externally or internally generated) to the $2000$ neurons. The vector $\mathbf{c}\in\mathbb{R}^{2000\times 1}$ is a random vector whose elements are drawn uniformly from $[-1,1]$. We choose a simple topology of the recurrent neural network where the adjacency matrix $\mathbf{A}$ is a sparse \ER random matrix. The sparseness of the \ER-type random matrix $\mathbf{A}$ (the fraction of non-zero elements) is set to be $0.02$. All non-zero elements in $\mathbf{A}$ are drawn uniformly from $[-\sigma,\sigma]$. The value of $\sigma$ is determined by having the spectral radius of $\mathbf{A}$ (the magnitude of the largest eigenvalue of $\mathbf{A}$) be equal to $\rho=1.4$. We construct the input weight matrix $\mathbf{W}_{\text{in}}$ in such a way as to ensure that each neuron receives one and only one input variable from the $3$-dimensional input. The input connection strength (each nonzero element in $\mathbf{W}_{\text{in}}$) is drawn uniformly at random from the interval $[-0.05,0.05]$.

\subsection*{2. RNNs with a balanced Watts--Strogatz topology}
\begin{figure*}
  \centering
  \includegraphics[width=0.32\linewidth]{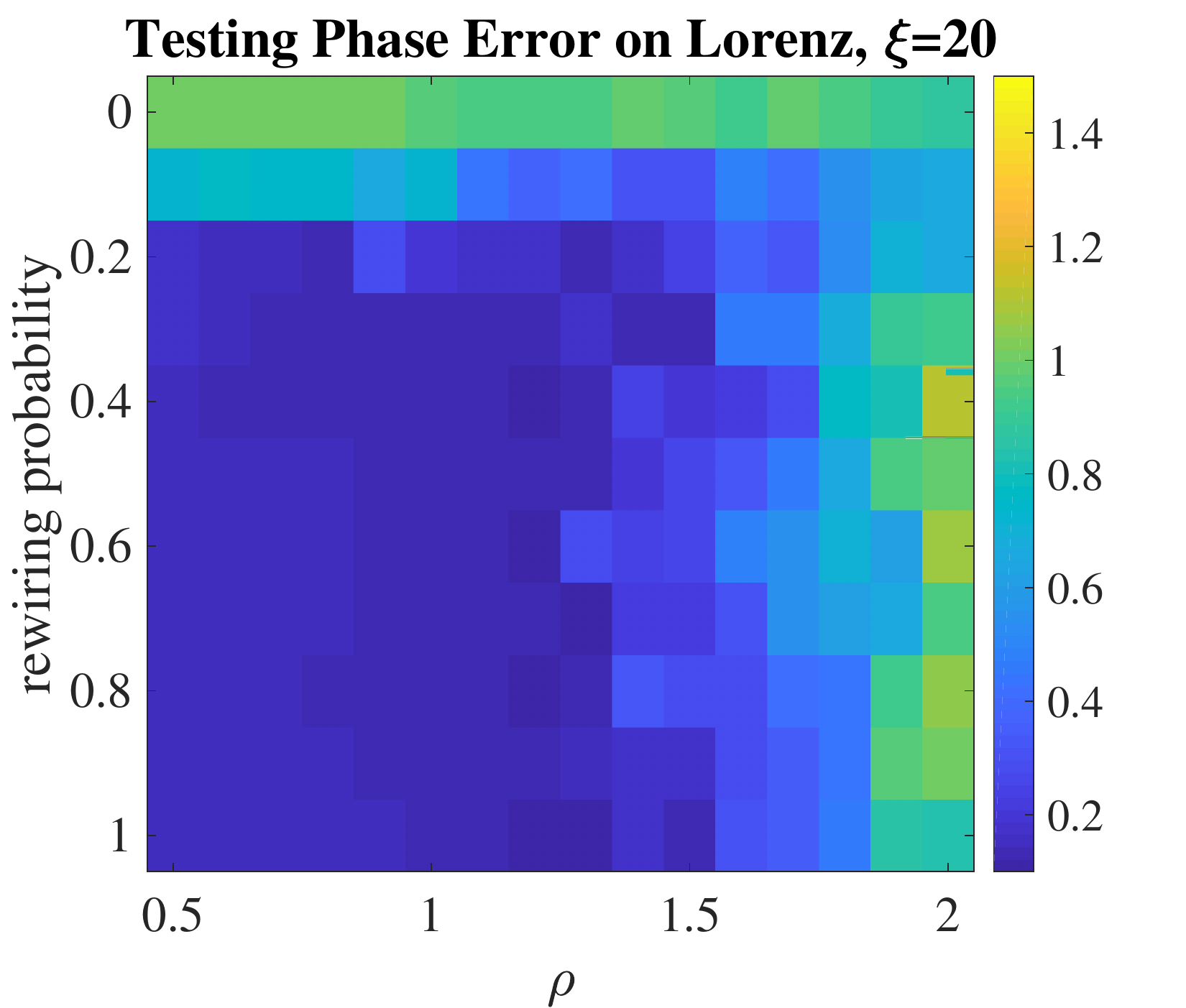}
  \includegraphics[width=0.32\linewidth]{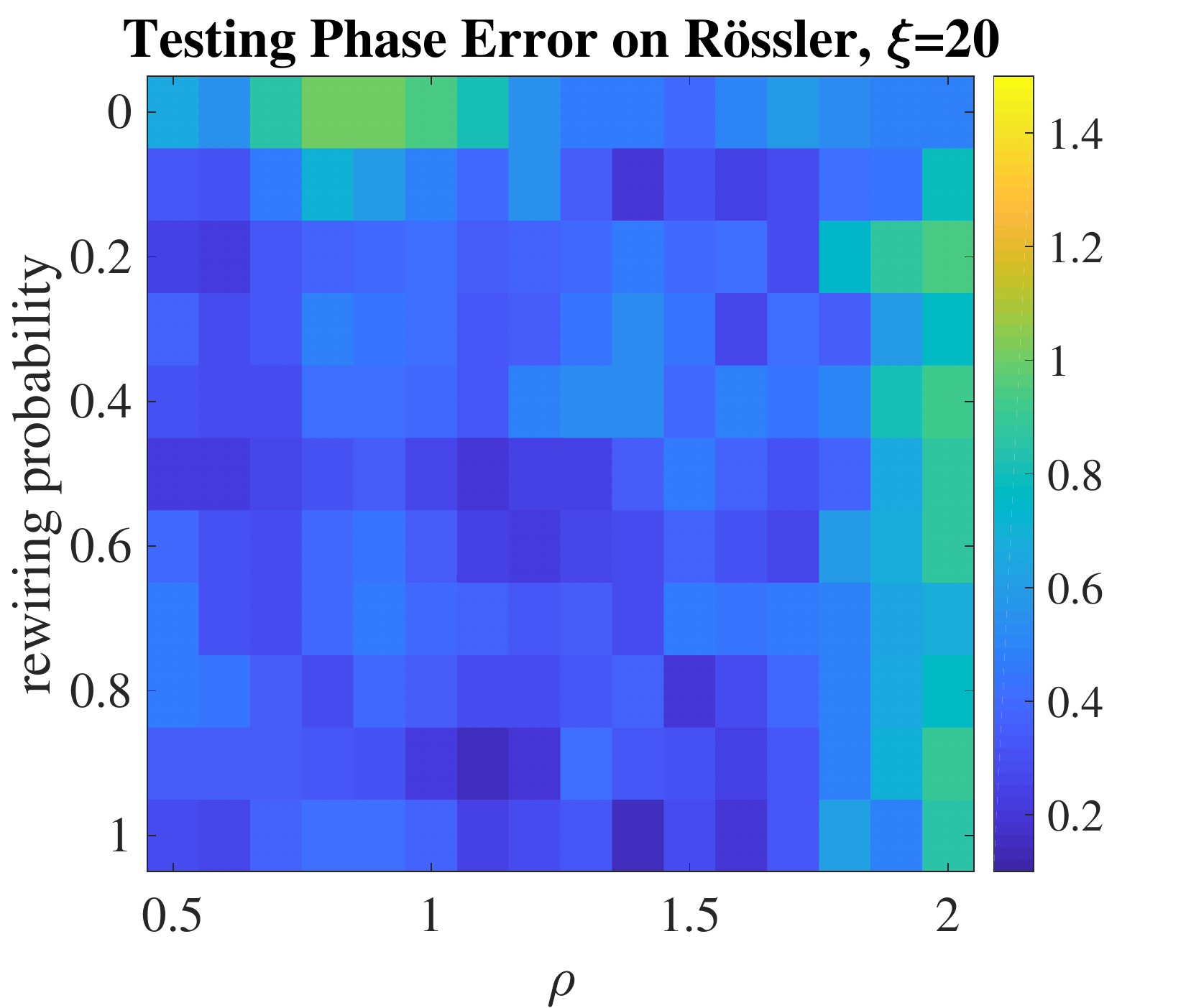}
  \includegraphics[width=0.32\linewidth]{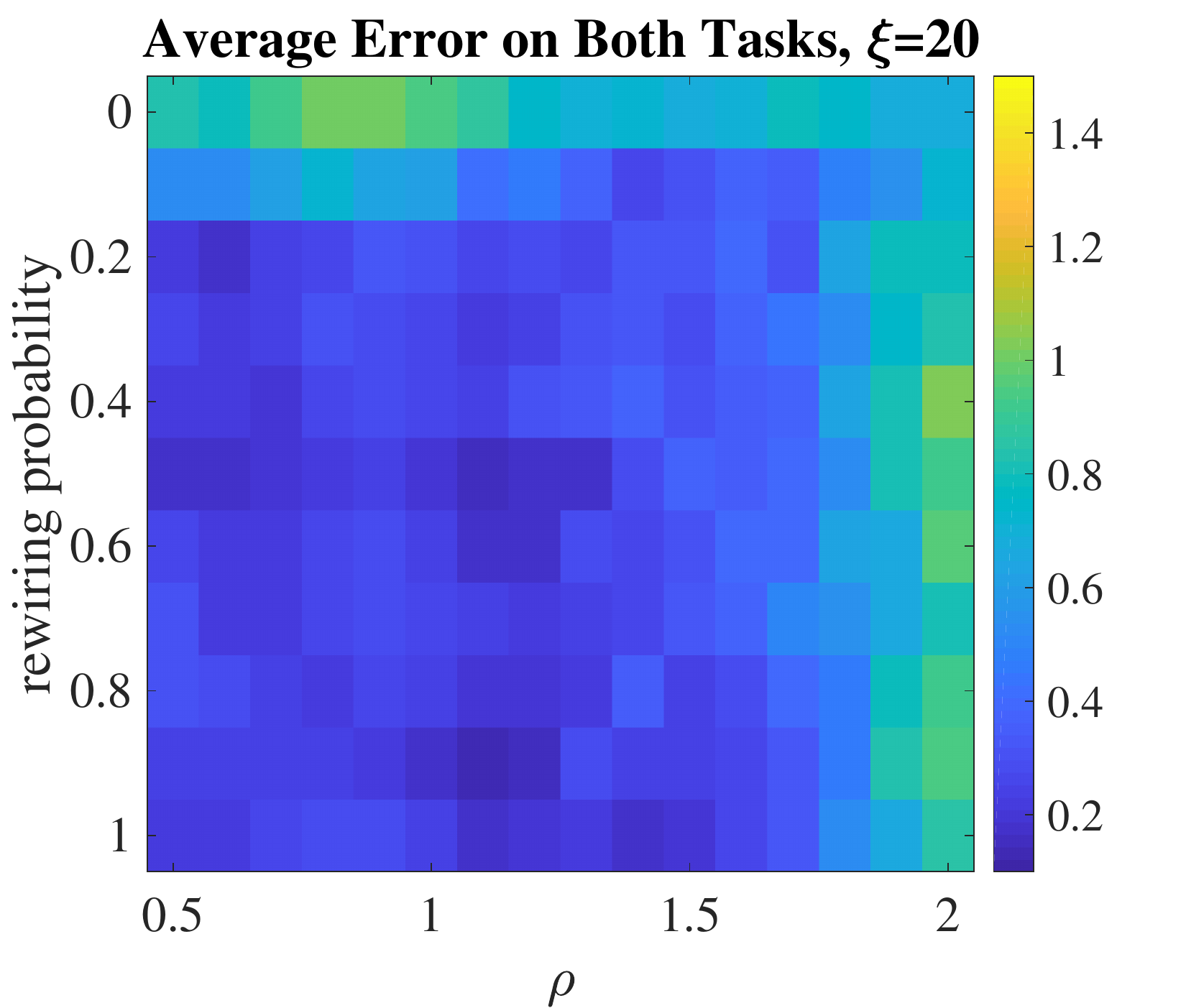}
  \linespread{1.0}
  \caption{\textbf{Relation between the testing phase error and the spectral radius of a balanced Watts--Strogatz graph on whose structure the RNN is instantiated.} In these heatmaps, color indicates the testing phase error on the Lorenz task (\emph{left}) and on the \Rossler task (\emph{center}). We also show the mean testing phase error averaged across the two tasks (\emph{right}). This set of panels corresponds to a separation parameter of $\xi= 20$. All panels reflect results obtained from RNNs with a balanced and randomly rewired Watts--Strogatz topology, where the spectral radius $\rho$ takes on values ranging from $0.5,\ 0.6,\ ...,\ 2.0$ (\emph{x-axis})). For each $(\rho,\xi)$ pair, we construct $10$ realizations of the balanced Watts--Strogatz topology for a rewiring probability between $0$ and $1$ (\emph{y-axis}) and we show the mean testing phase error across these $10$ realizations.}
  \label{fig:Testing_Error_on_WS}
\end{figure*}

In addition to modeling the central system as an RNN with \ER topology, we sought to determine whether similar learning success could be attained by a central system modeled as an RNN with non-random topology. Specifically, we considered a commonly studied small-world topology as instantiated in so-called Watts--Strogatz networks \cite{watts1998collective} that are balanced with inhibitory synapses. The setup of the central system is identical to that of the \ER examples (Eq.~(\ref{eqn:RNN})) as previously discussed, except for a different adjacency matrix $\mathbf{A}$. We construct the Watts--Strogatz adjacency matrix following the protocol described in \cite{muldoon2016small}, with network size $N=2000$ and degree $\kappa=40$. When the rewiring probability equals $0$, the network is a regular ring lattice. Before the random rewiring process, the adjacency matrix is symmetric and satisfies
\begin{equation}\tag{C1}
\mathbf{A}_{i,j>i} = \begin{cases}
    \frac{1}{d_{i,j}},& \text{if } d_{i,j}\leq \kappa/2\\
    0,              & \text{otherwise}
\end{cases},
\label{eqn:ring_lattice}
\end{equation}
where $d_{i,j}=j-i \mod N$. Then, with the given rewiring probability, each existing unidirectional connection between node $i$ and node $j$ is replaced with another unidirectional connection between node $i$ and a randomly chosen node $j'\neq i$, to which node $i$ was not already connected. To balance the network with inhibitory synapses, we independently assign a sign to each element in the adjacency matrix $\mathbf{A}$ uniformly at random and with a chance of one half. Then, we achieve a desired spectral radius $\rho$ (the largest absolute value of its eigenvalues) by multiplying the matrix $\mathbf{A}$ by a scalar. As shown in Fig.~\ref{fig:Testing_Error_on_WS}, this system can consistently learn both the Lorenz and \Rossler tasks, and its performance rises as the rewiring probability increases and the spectral radius decreases. Thus, the greatest performance is reached for the most randomly wired balanced Watts--Strogatz networks. However, we note that the networks obtained when the rewiring probability is equal to $1$ do not display the same topology as \ER networks, because synapse strengths are not uniformly distributed over the range $[-\sigma,\sigma]$.

\subsection*{3. RNNs with the Pseudo Watts--Strogatz random networks}
\begin{figure*}
  \centering
  \includegraphics[width=0.32\linewidth]{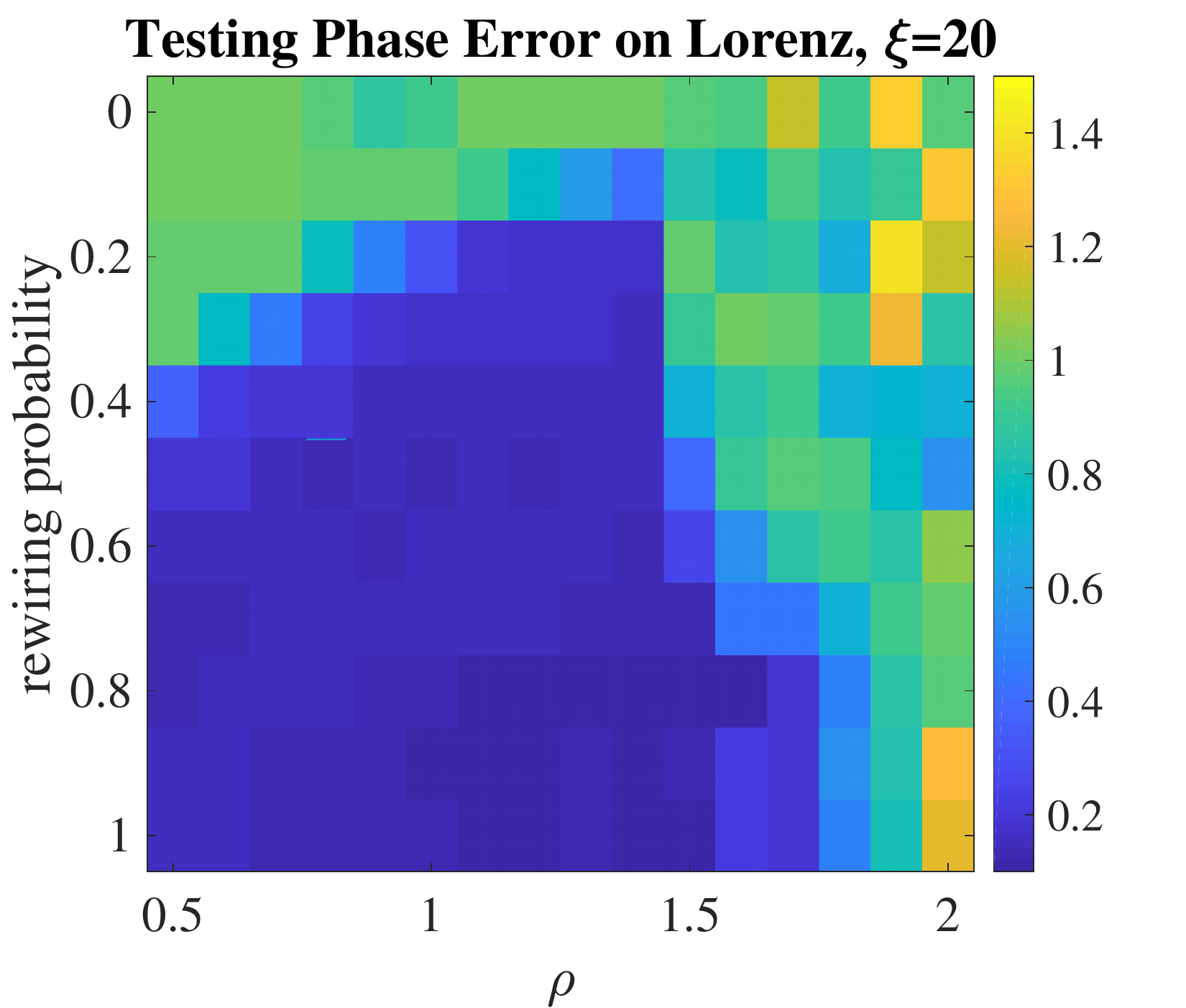}
  \includegraphics[width=0.32\linewidth]{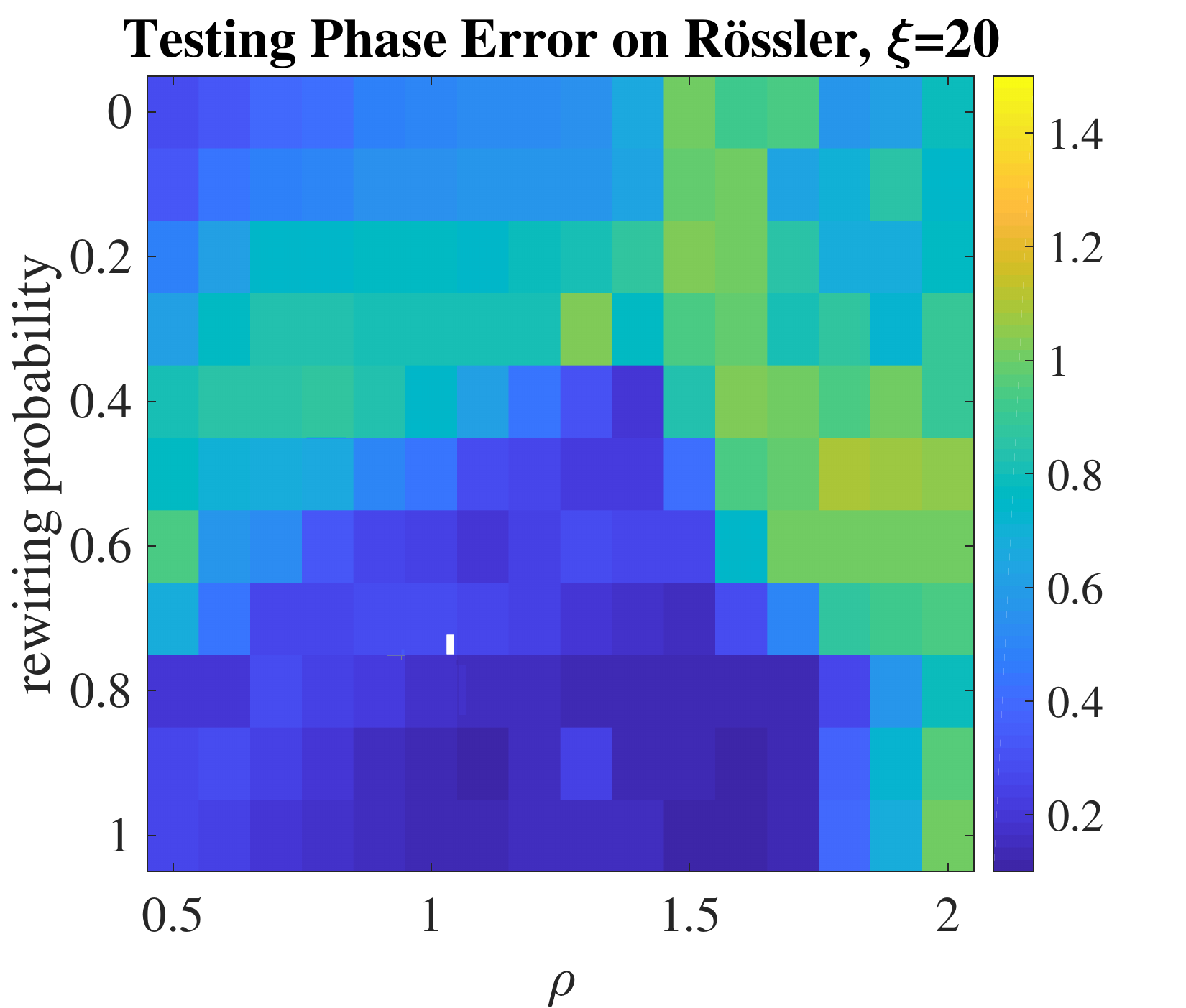}
  \includegraphics[width=0.32\linewidth]{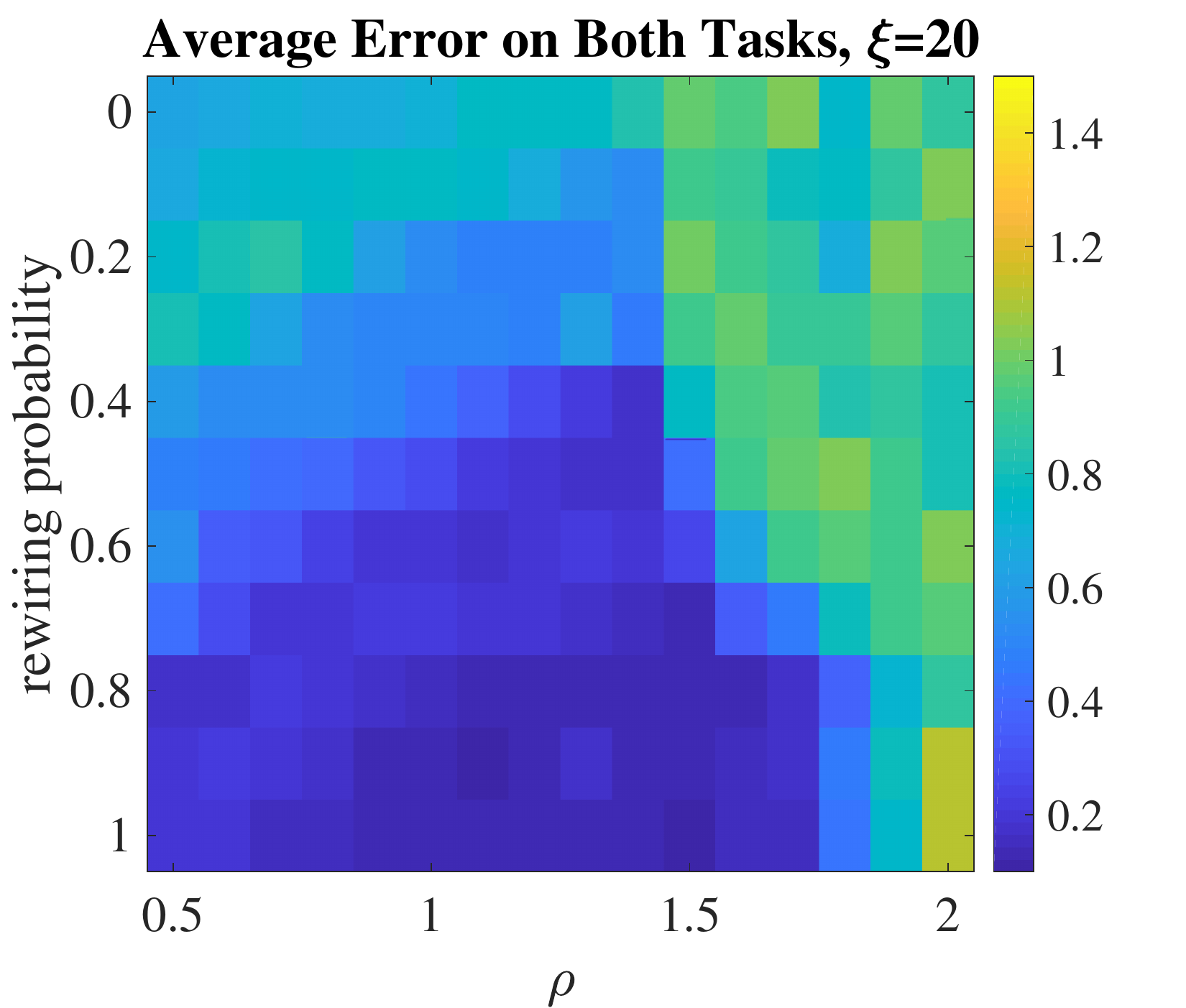}
  \linespread{1.0}
  \caption{\textbf{Relation between the testing phase error and the spectral radius of a pseudo Watts--Strogatz graph on whose structure the RNN is instantiated.} In these heatmaps, color indicates the testing phase error on the Lorenz task (\emph{left}) and on the R\"ossler task (\emph{center}). We also show the mean testing phase error averaged across the two tasks (\emph{right}). This set of panels corresponds to a separation parameter of $\xi= 20$. All panels reflect results obtained from RNNs with a pseudo and randomly rewired Watts--Strogatz topology, where the spectral radius $\rho$ takes on values ranging from $0.5,\ 0.6,\ ...,\ 2.0$ (\emph{x-axis})). For each $(\rho,\xi)$ pair, we construct $10$ realizations of the balanced Watts--Strogatz topology for a rewiring probability between $0$ and $1$ (\emph{y-axis}) and we show the mean testing phase error across these $10$ realizations.}
  \label{fig:Testing_Error_on_Pseudo_WS}
\end{figure*}

Continuing in our study of the robustness of our learning framework to the topology of the network, we further consider a slightly modified Watt--Strogatz model. Specifically, we modify the protocol used in the previous example by first constructing a ring lattice, 
\begin{equation}\tag{C2}
\mathbf{A}_{i,j>i} = \begin{cases}
    \frac{2d_{i,j}}{k},& \text{if } d_{i,j}\leq \kappa/2\\
    0,              & \text{otherwise}
\end{cases},
\label{eqn:uniform_ring_lattice}
\end{equation}
that is antisymmetric (i.e., $\mathbf{A}=-\mathbf{A}^{T}$). Then, given the rewiring probability, each directional connection from node $i$ to node $j$ is replaced by a connection to a randomly chosen node $j'\neq i$, to which node $i$ was not already connected. Then, we multiply the matrix $\mathbf{A}$ by a scalar to obtain the desired spectral radius $\rho$. With this pseudo Watt--Strogatz model, the connection strengths are approximately uniformly distributed over $[-\sigma,\sigma]$. Thus, using this model we can test how the learning performance varies as the network changes from a ring lattice structure to an \ER network. When the rewiring probability is $0$, each node is connected to only $\kappa=40$ neighboring nodes on the ring lattice, with $\kappa/2=20$ on each side. When the rewiring probability is $1$, all connections are randomly rewired, and thus the topology approximates that of an \ER network. The performance of this pseudo Watt--Strogatz network is shown in Fig.~\ref{fig:Testing_Error_on_Pseudo_WS}. Again, we find that this system can consistently learn both the Lorenz and \Rossler tasks, and its performance rises as the rewiring probability increases and the spectral radius decreases.

\subsection*{4. Random polynomial network model}
\begin{figure*}
  \centering
  \includegraphics[width=0.32\linewidth]{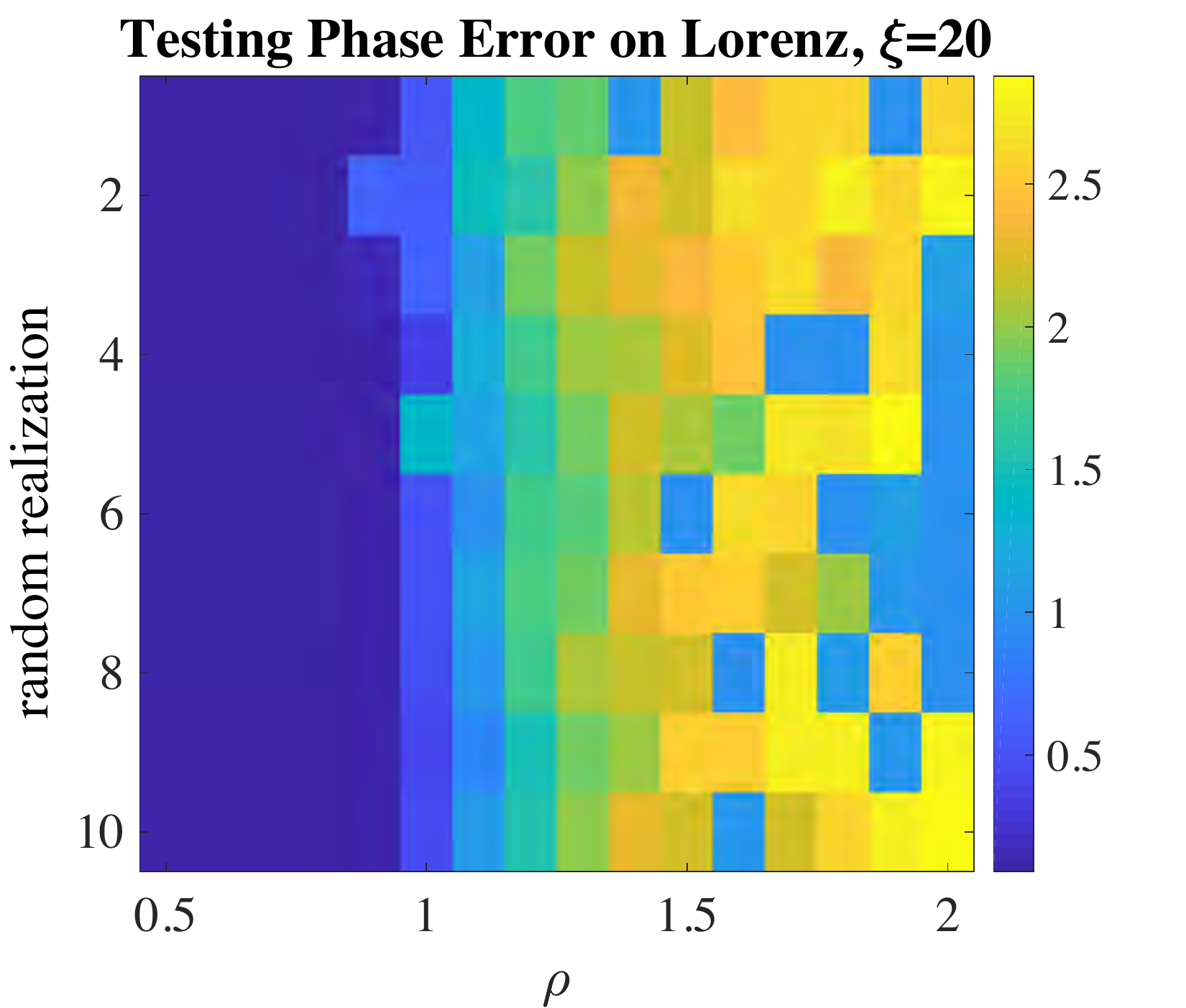}
  \includegraphics[width=0.32\linewidth]{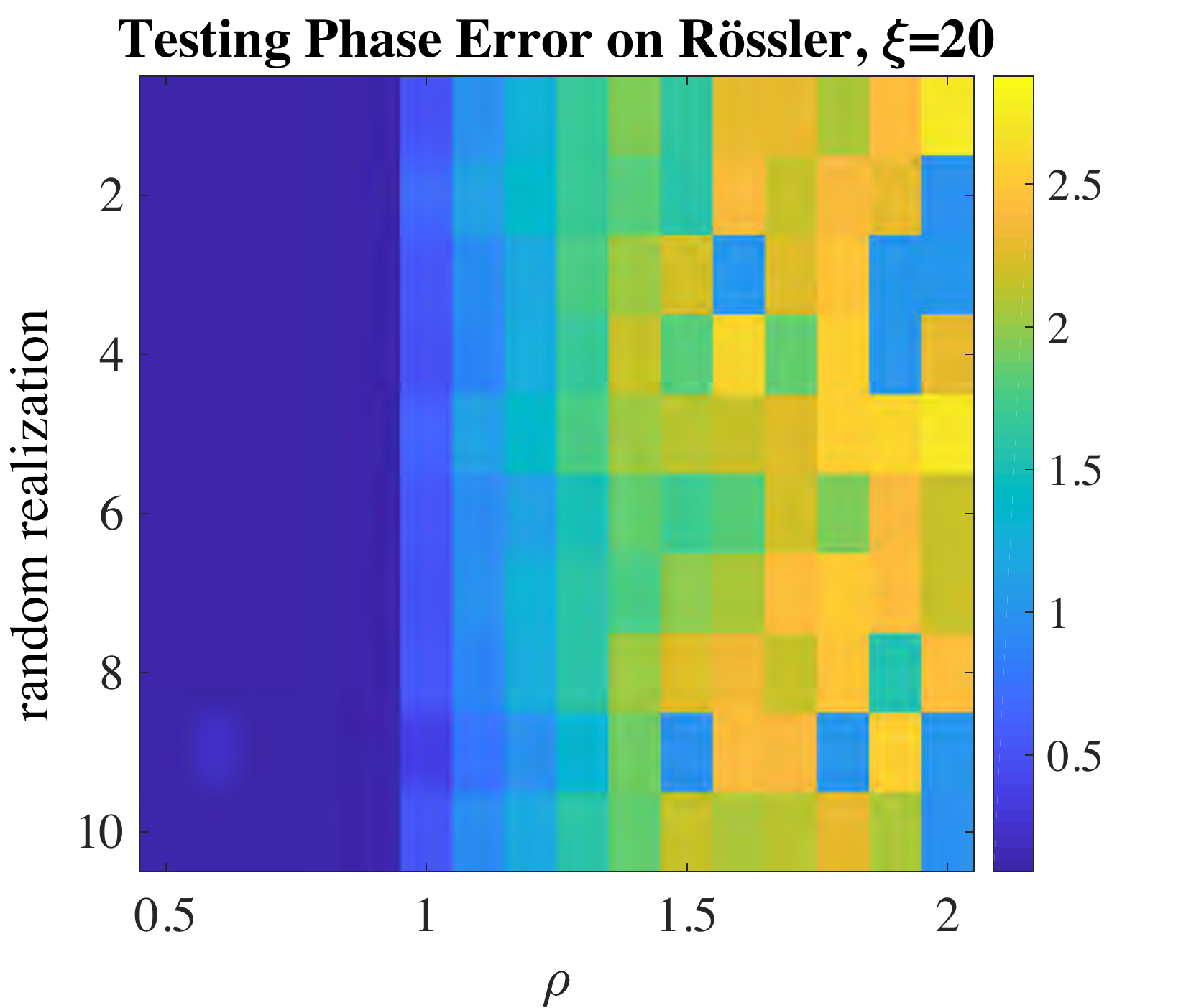}
  \includegraphics[width=0.32\linewidth]{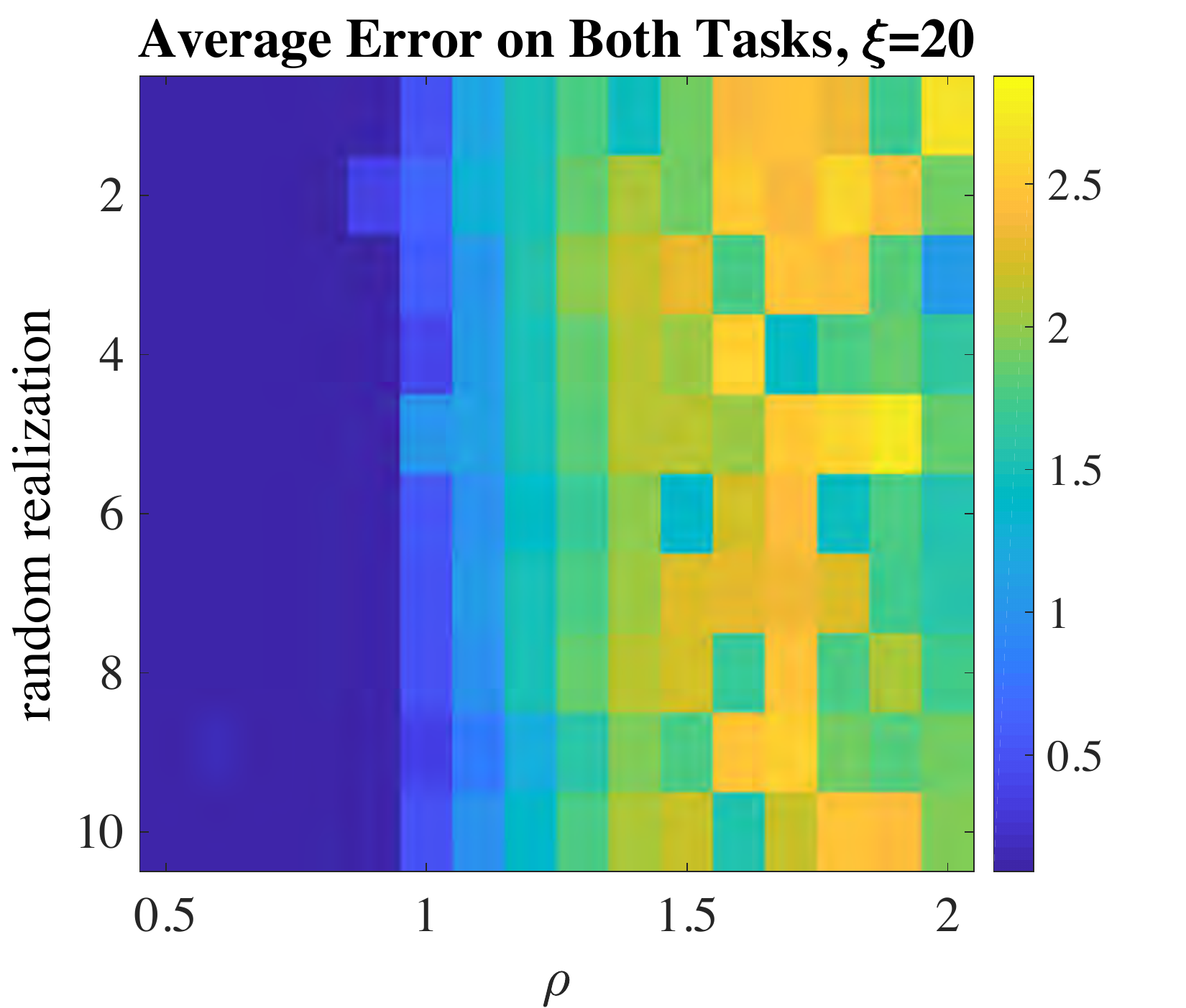}
  \linespread{1.0}
  \caption{\textbf{Relation between the testing phase error and the spectral radius of the structural graph on which the polynomial system is instantiated.} In these heatmaps, color indicates the testing phase error on the Lorenz task (\emph{left}) and on the \Rossler task (\emph{center}). We also show the mean testing phase error averaged across the two tasks (\emph{right}). This set of panels corresponds to a separation parameter of $\xi=20$. Panels reflect results obtained from $160$ randomly constructed polynomial systems, where the sparse random coefficient matrices $\mathbf{A}_{1,2,...,5}$ have been scaled to obtain a spectral radius of $\rho=0.5,\ 0.6,\ ...,\ 2.0$ (\emph{x-axis}). For each of the $16$ values of $\rho$, we construct $10$ realizations of the topology (\emph{y-axis}).}
  \label{fig:Testing_Error_for_polynomial}
\end{figure*}

To further demonstrate the generalizability of this learning framework (Eqs.~(\ref{eqn:input_system},\ref{eqn:central_system},\ref{eqn:auto_brain})), we consider another high-dimensional dynamical system that differs markedly from an RNN. Specifically, we construct the following high-dimensional random polynomial-type central system with polynomial degree $5$. In this model, the state of the central system $\mathbf{x}\in\mathbb{R}^{N=2000}$, driven by a $3$-dimensional input signal $\mathbf{s}\in\mathbb{R}^{n=3}$ (either externally or internally generated), evolves following
\begin{equation}\tag{C3}
\begin{aligned}
x_i(t+1) = \tanh(\sum_{j=1}^{2000}\sum_{k=1}^{5}[\mathbf{A}_k]_{i,j}\cdot x_j^k(t) \\ 
+\sum_{j=1}^3 [\mathbf{W}_{\text{in}}]_{i,j}\cdot s_j(t)+[\mathbf{c}]_i),
\end{aligned}
\label{eqn:polynomial}
\end{equation}
where the input weight matrix $\mathbf{W}_{\text{in}}$ and the bias term $\mathbf{c}$ are defined similarly to those in the previous models. To prevent divergence of the central system, we wrap the random polynomial with the sigmoidal function $\tanh(\cdot)$. The coefficients of these polynomials are denoted by five random matrices $\mathbf{A}_k\in\mathbb{R}^{N\times N}$, each with a fraction of non-zero elements set to $0.02$. The non-zero elements are randomly and independently drawn from the uniform distribution $[-\sigma,\sigma]$, where $\sigma>0$ is chosen such that the spectral radius of each $\mathbf{A}_k$ equals $\rho$. In Fig.~\ref{fig:Testing_Error_for_polynomial}, we show the performance of the polynomial-type central system. We observe that these systems can robustly learn both tasks with very small testing phase error when the spectral radius $\rho$ is less than one.

\section*{Appendix D: Learning by adapting the output weight matrices}

Throughout our numerical simulations, all exemplary Lorenz and \Rossler input trajectories have length $1000$ time units, spanned by $5\times10^4$ data points obtained with a sampling rate of $\tau = 0.02$. During each learning phase, the output matrix $\mathbf{W}_{\text{out}}$ adapts following Eq.~(\ref{eqn:adapt}) where $\alpha>0$ is the learning rate.

Notice that by starting the central system from an arbitrary initial state $\mathbf{x}(0)$, the generalized synchronization occurs after a short transient period. As a result, we consider that the input signal is not being encoded into the state of the central system during this transient period. We therefore freeze the $\mathbf{W}_{\text{out}}(t)$ for the first $5\times 10^3$ time points (chosen to be longer than the transient period) and only allow $\mathbf{W}_{\text{out}}(t)$ to adapt for the remaining $4.5\times 10^4$ time points during the learning phase. We repeat this process $1000$ times to allow $\mathbf{W}_{\text{out}}(t)$ to converge. For cases where two or more attractors are learned, we let the system repetitively learn all tasks in an alternate manner. Specifically, we continually adapt the output connection strengths as the system repetitively engages in the learning phases of each task $200$ times. For the remaining $800$ repetitions, in order to enhance the convergence of the multitask learning, we break down the learning phases into even shorter pieces during which we continue to allow $\mathbf{W}_{\text{out}}(t)$ to adapt. To ensure that $\mathbf{W}_{\text{out}}$ converges, we empirically choose the learning rate $\alpha=10^{-3}$ for the first $300$ repetitions and then we decrease the learning rate thereafter so that the later learning periods serve to fine-tune $\mathbf{W}_{\text{out}}$.

At this point in the exposition, we wish to emphasize that the above learning scheme is a somewhat biologically plausible scheme that we chose from a set of many other possible schemes. The convergence of this particular adaptive learning scheme depends on many heuristic factors, such as the value of the learning rate $\alpha$. However, it is of particular interest to address the general convergence for different adaptive learning schemes. Based on our previous discussion of the information encoding mechanism, we are guaranteed that the internal representations $\PPP$s are solely determined by the generalized synchronization function ($\PPP={\bm \Psi}(\AAA)$), and are thus not affected by updating $\mathbf{W}_\text{out}$. Therefore, regardless of the particular learning scheme that one adopts, the learning of $\mathbf{W}_\text{out}$ only converges if there exists an output weight matrix such that $\mathbf{s}=\mathbf{W}_\text{out}\mathbf{x}$ where $\mathbf{x}\in\PPP$ and $\mathbf{s}\in\AAA$. Consequently, from our theory we obtain two necessary but not sufficient conditions for learning convergence. One necessary condition is an invertible generalized synchronization between the drive system and the central system, which occurs when the largest conditional Lyapunov exponent is negative. The other necessary condition specific to learning multiple attractors is that the internal representations of different attractors cannot overlap.

\section*{Appendix E: Difference in structural connectivity matrix within different functional connectivity clusters}

\begin{figure*}
  \center
  \includegraphics[width=13cm]{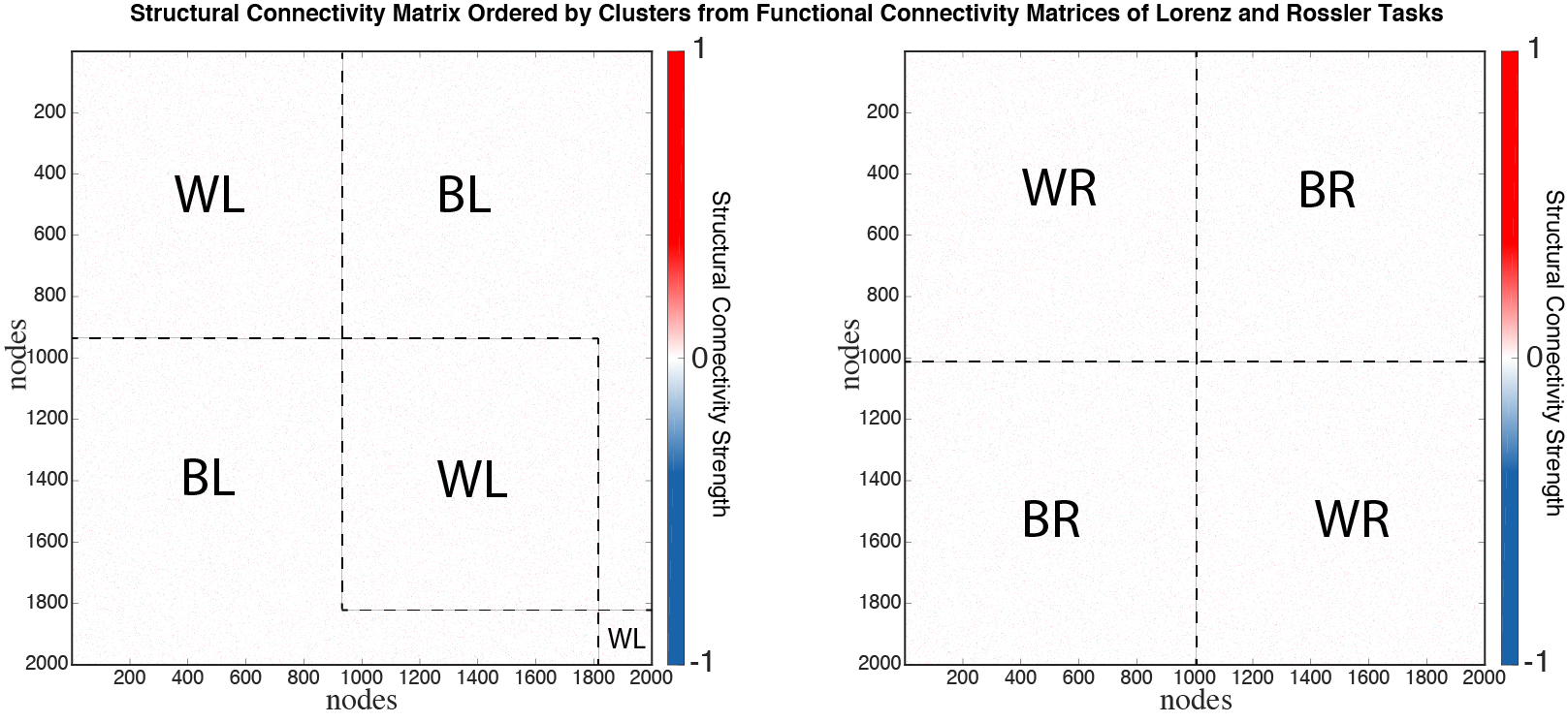}
  \caption{\textbf{Structural connectivity matrix ordered based on the community structure from functional connectivity matrices.} \emph{Left} The structural connectivity matrix $\mathbf{A}$ ordered based on the functional community structure elicited by the Lorentz task. \emph{Right} The structural connectivity matrix $\mathbf{A}$ ordered based on the functional community structure elicited by the \Rossler task. We conclude that the marked sparsity of the structural matrix hampers any explicit inferences from this visualization.}
  \label{fig:structural_matrix}
\end{figure*}

In this section, we examine the intra-community structural connectivity for functional communities elicited by either the \Rossler or Lorentz tasks. First, we visualize the structural connectivity matrices, ordered either by the functional communities elicited by the Lorentz task (left panel in Fig.~\ref{fig:structural_matrix}) or by the functional communities elicited by the \Rossler task (right panel in Fig.~\ref{fig:structural_matrix}). Because of the marked sparsity of the structural matrix, it is difficult to make any explicit inferences from this visualization. 

\begin{figure*}
  \center
  \includegraphics[width=5.5cm]{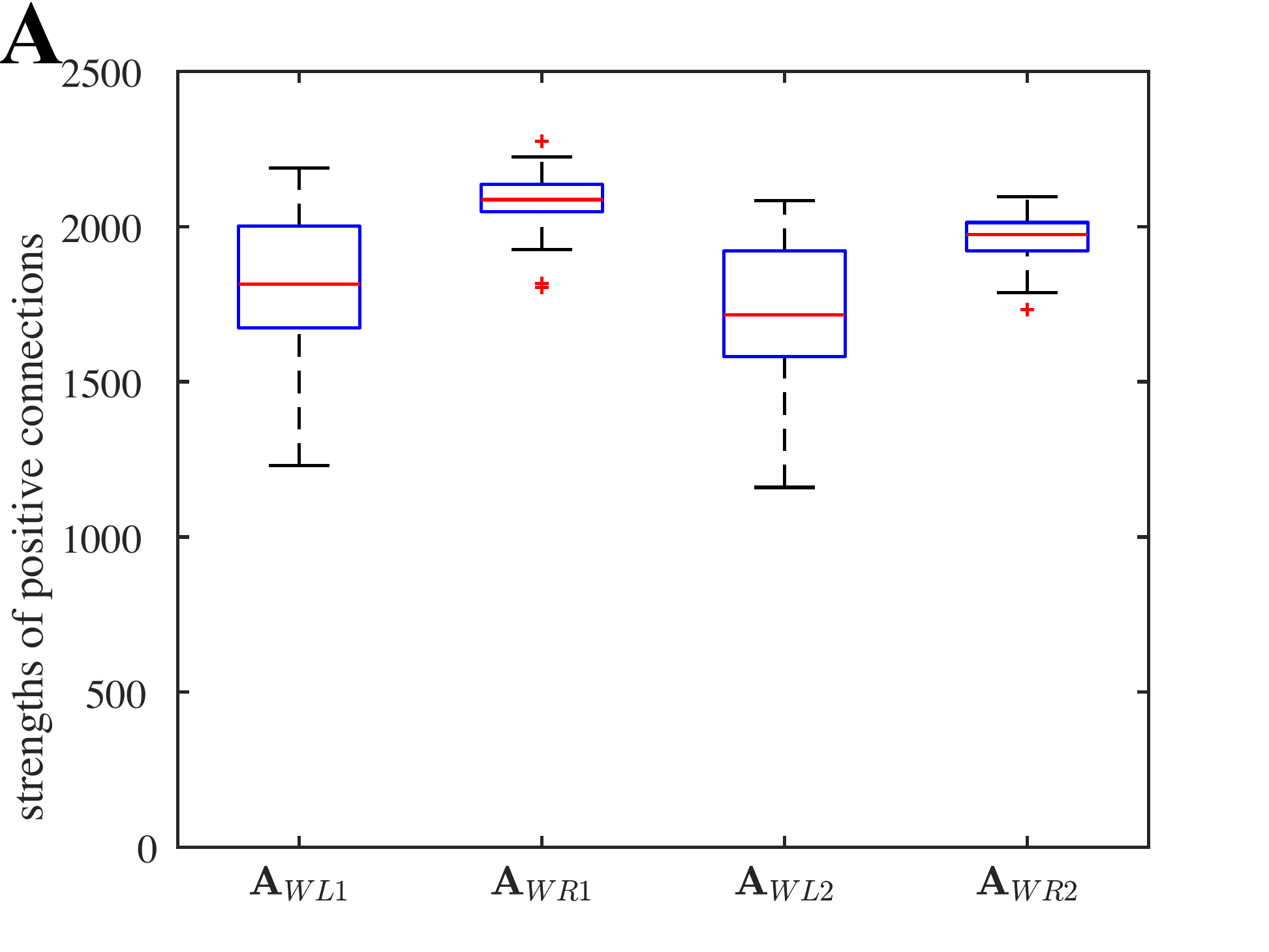}
  \includegraphics[width=5.5cm]{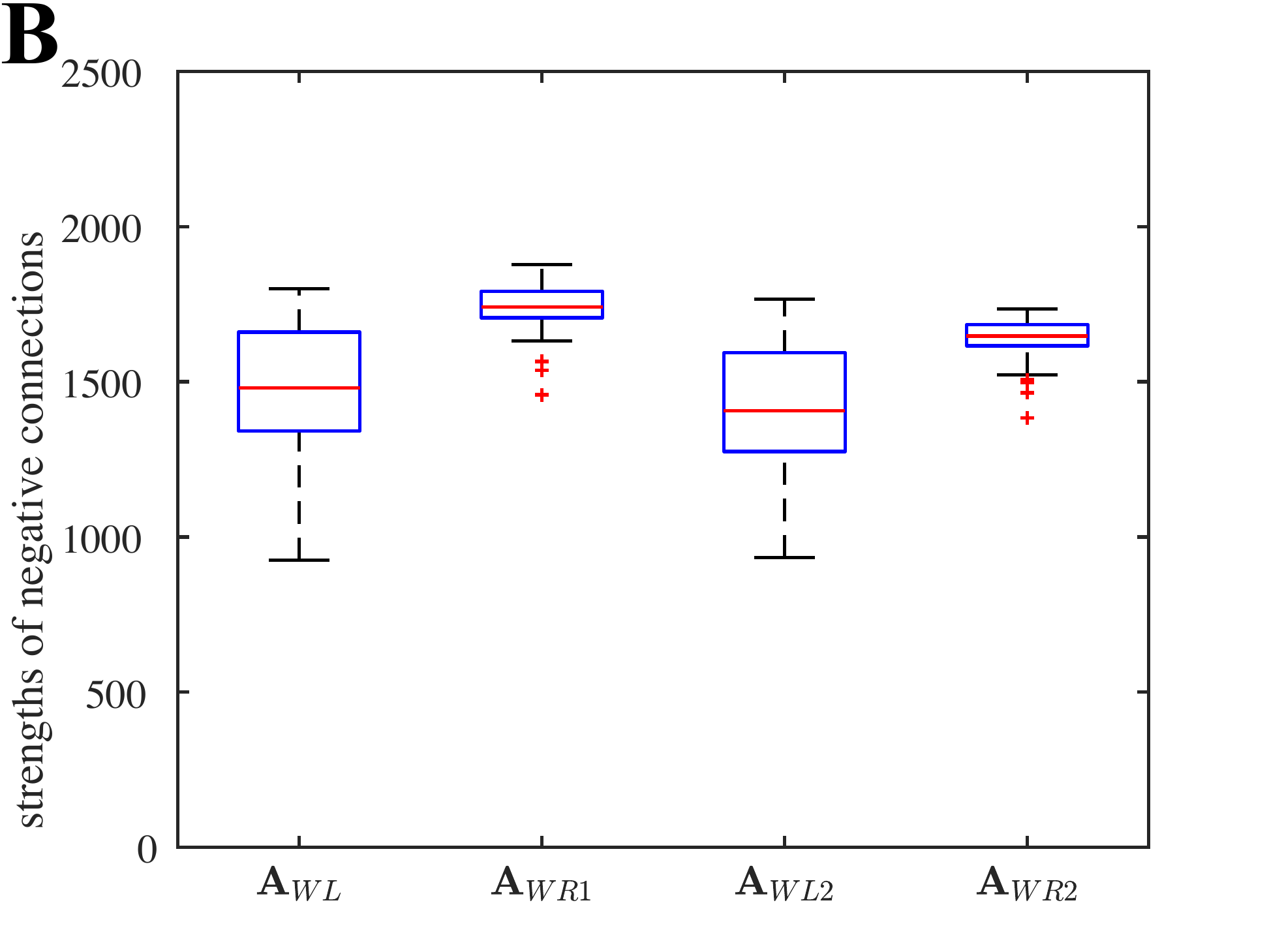}
  \includegraphics[width=5.5cm]{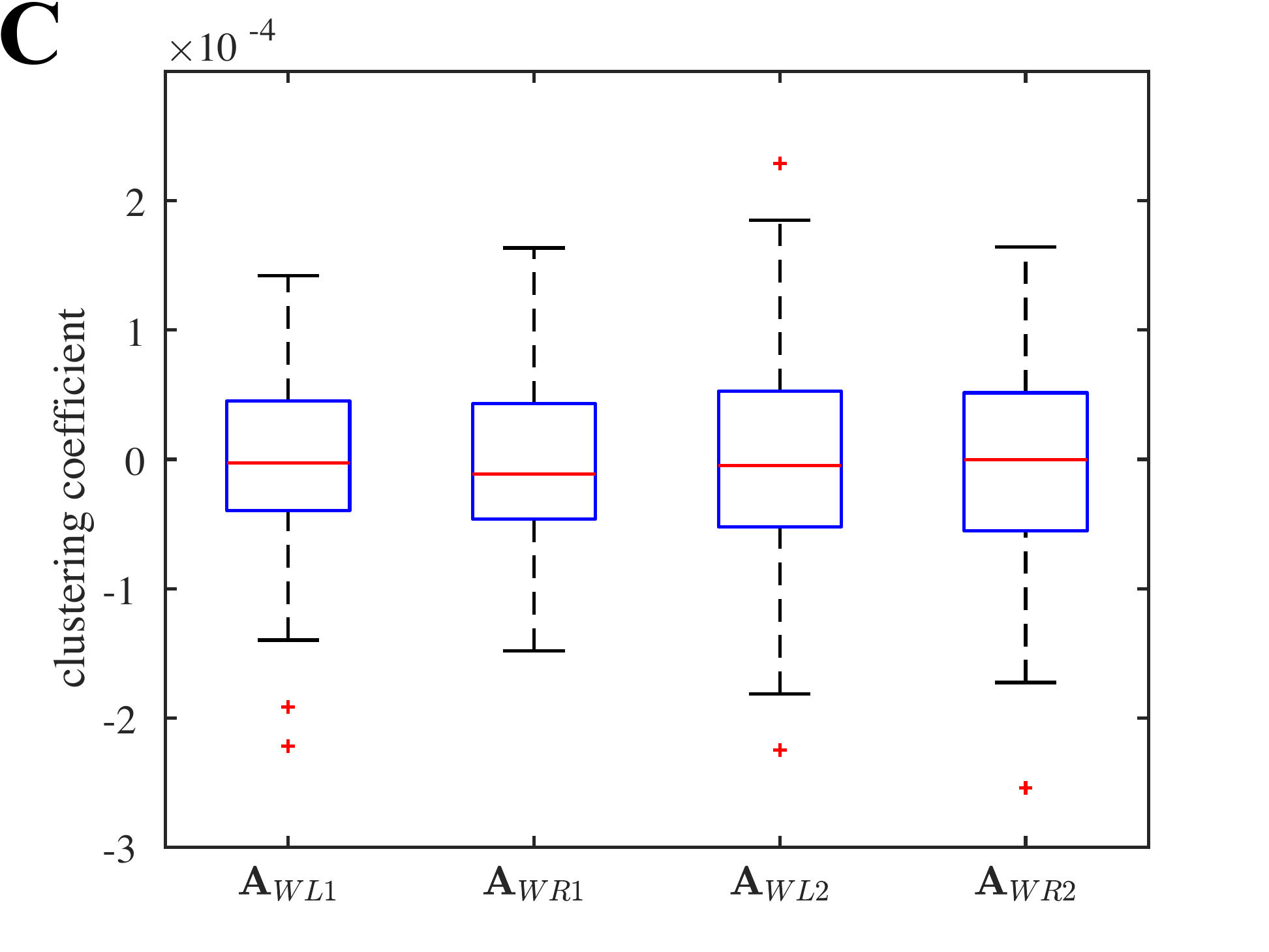}
  \includegraphics[width=5.5cm]{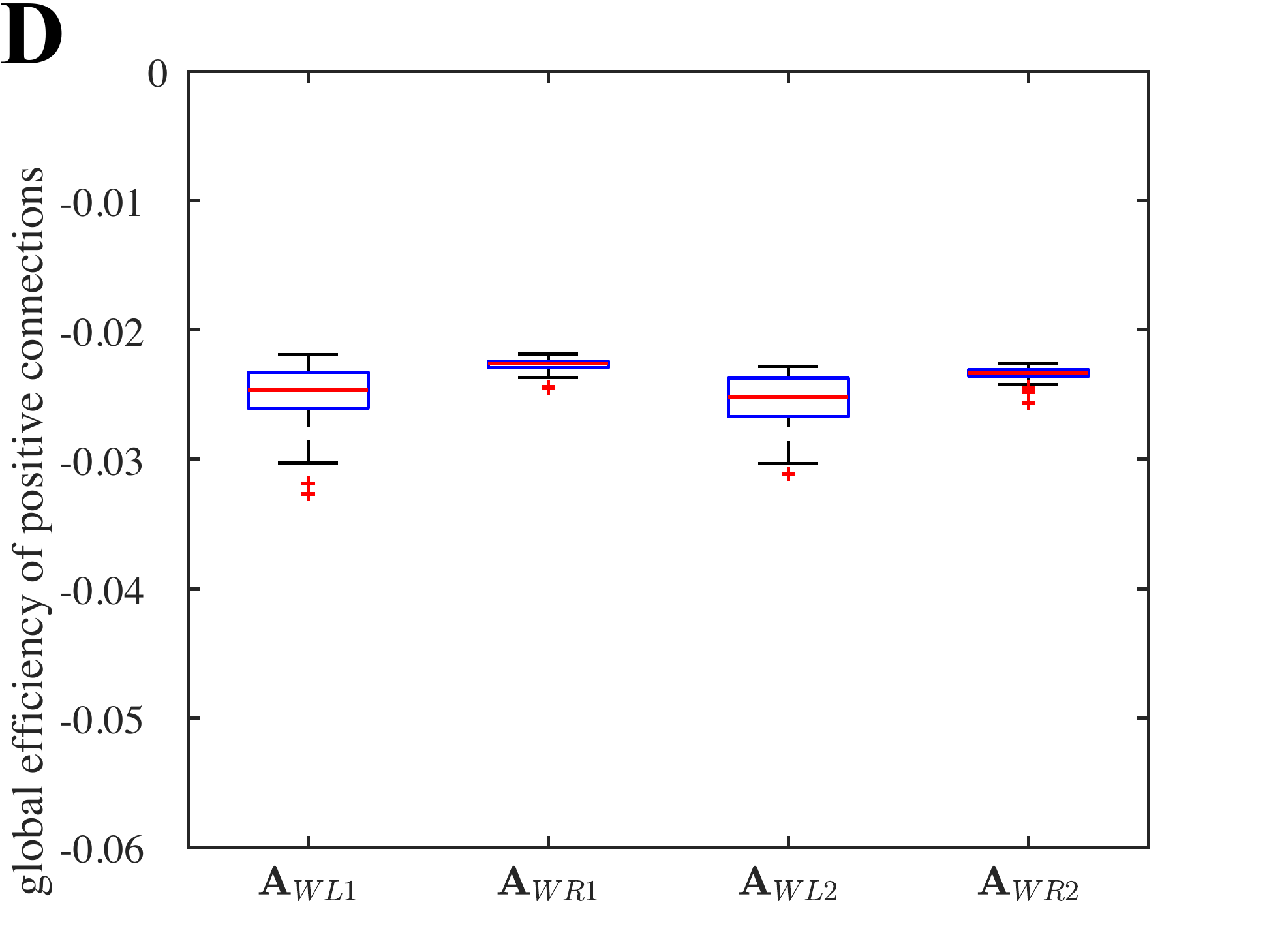}
  \includegraphics[width=5.5cm]{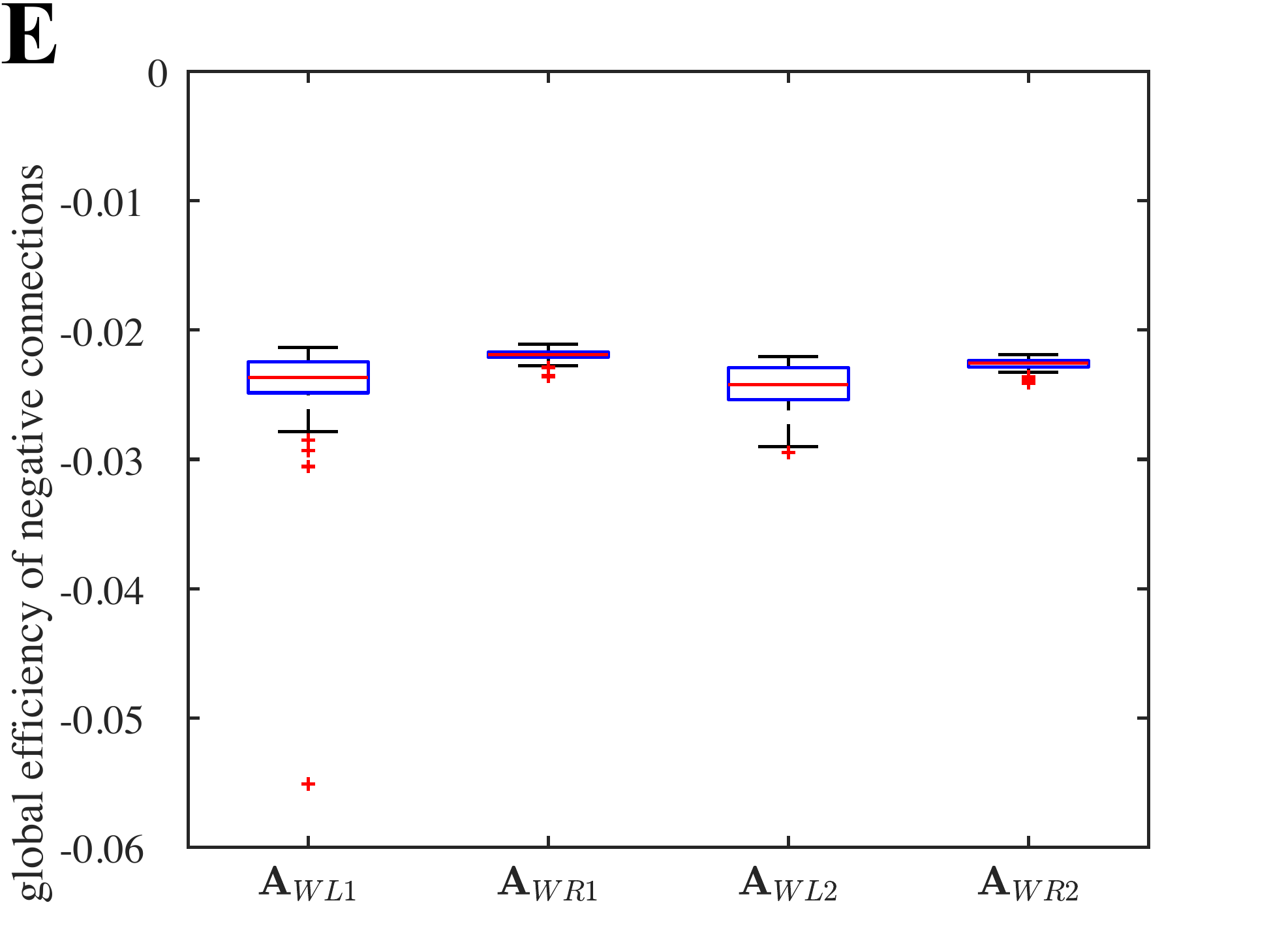}
  \includegraphics[width=5.5cm]{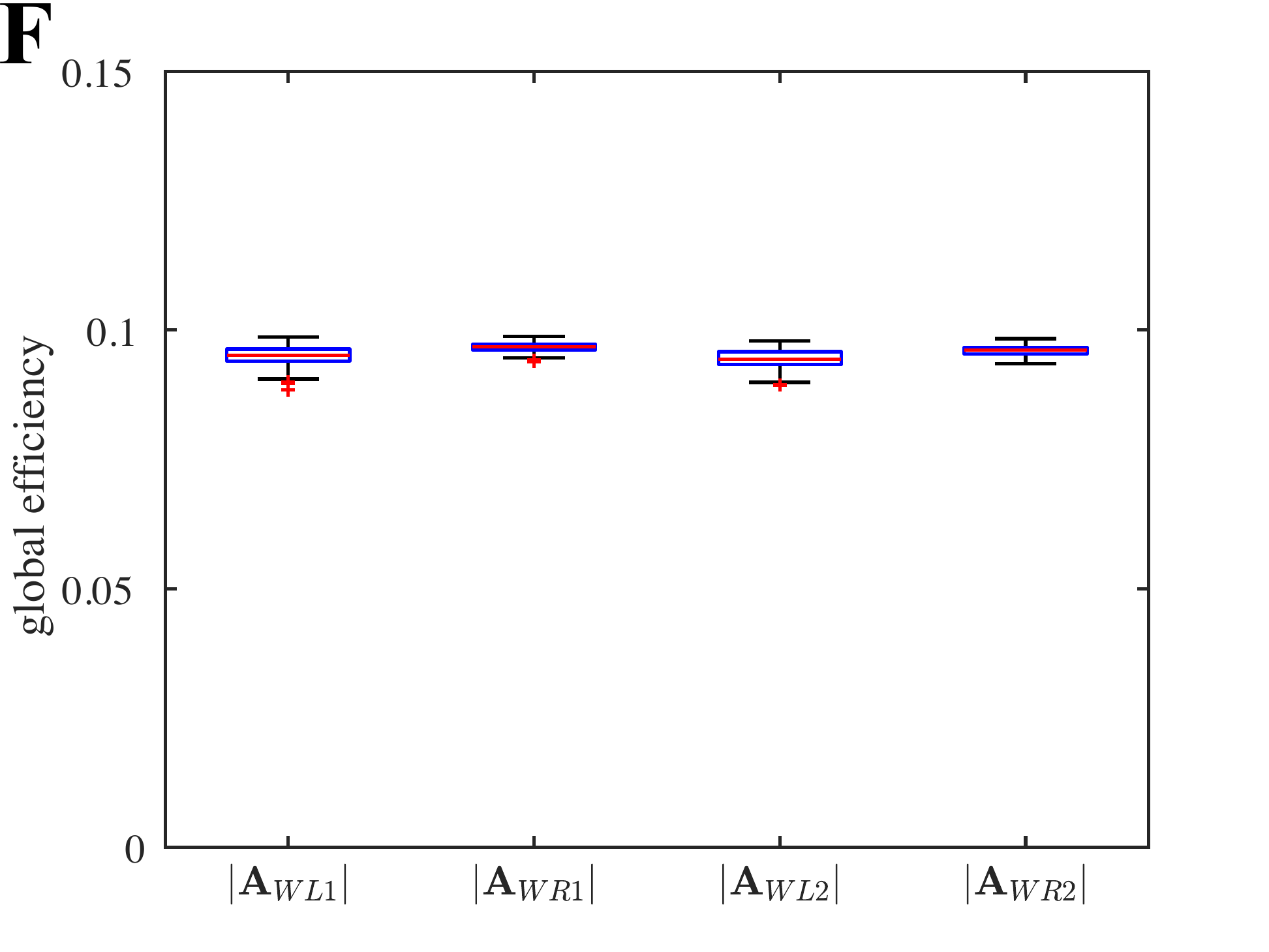}
  \caption{\textbf{Topological measures of subnetworks in the structural matrix that comprise the clusters detected from the functional matrices.} \emph{(A)} Strength of positive connections in the structural subnetwork of functional community one and two elicited by the \Rossler and Lorentz tasks. \emph{(B)} Strength of negative connections in the structural subnetwork of functional community one and two elicited by the \Rossler and Lorentz tasks. \emph{(C)} Signed clustering coefficient in the structural subnetwork of functional community one and two elicited by the \Rossler and Lorentz tasks. \emph{(D)} Global efficiency of positive connections in the structural subnetwork of functional community one and two elicited by the \Rossler and Lorentz tasks. \emph{(E)} Global efficiency of negative connections in the structural subnetwork of functional community one and two elicited by the \Rossler and Lorentz tasks. \emph{(F)} Global efficiency of the magnitude of connections in the structural subnetwork of functional community one and two elicited by the \Rossler and Lorentz tasks.}
  \label{fig:structural_matrix_topology_measure}
\end{figure*}

We therefore turn to statistical analyses. Specifically, we begin by evaluating the strength of structural connectivity within the functional communities elicited by the \Rossler task, and we compare it to the strength of structural connections within the functional communities elicited by the Lorentz task (Fig.~\ref{fig:structural_matrix_topology_measure}A--B). We consider the positive and negative structural connections separately. We find that the strength of structural connectivity is consistently greater within communities elicited by the \Rossler task than within communities elicited by the Lorentz task (two-sample $t$-tests for positive connections in community one $t=-11.57$, $p = 5.53\times 10^{-24}$ and in community two $t=-10.05$, $p=1.88\times 10^{-19}$, and for negative connections in community one $t=-12.02$, $p=2.37\times 10^{-25}$ and in community two $t=-10.58$, $p=5.22\times 10^{-21}$). We also note visually that the variance of structural connection strength in the functional communities elicited by the \Rossler task is smaller than the variance of structural connection strength in the functional communities elicited by the Lorentz task. 

To further assess potential structural differences in the functional communities elicited by each task, we calculate two different graph statistics: the clustering coefficient and the global efficiency (Fig.~\ref{fig:structural_matrix_topology_measure}C--F). We begin with a signed clustering coefficient metric, and find no significant differences between (i) the clustering coefficient of the structural subnetwork composed of nodes assigned to the first functional community elicited by the \Rossler task and (ii) the clustering coefficient of the structural subnetwork composed of nodes assigned to the first functional community elicited by the Lorentz task (two-sample $t$-test $t=-0.39$, $p=0.70$). We observe similar non-significant results for the second functional community ($t=0.04$, $p = 0.96$). Finally, we consider the global efficiency, and we assess this metric in the structural subnetwork of positive connections and negative connections, separately, as well as in the full network constructed by taking the absolute value of edge strengths in $\mathbf{A}$. We find consistently that global efficiency of the structural network is higher in functional communities elicited by the \Rossler task than in functional communities elicited by the Lorentz task. Specifically, when considering the subnetwork of positive structural connections a two-sample $t$-test gives $t =-9.86$, $p = 6.66 \times 10^{-19}$ for the first community and $t=-9.83$, $p = 8.3 \times 10^{-19}$ for the second community. Similarly, when considering the subnetwork of negative structural connections, a two-sample $t$-test gives $t=-6.31$, $p = 1.75 \times 10^{-9}$ for the first community and $t=-9.57$, $p = 4.71\times 10^{-18}$ for the second community. Finally, when considering the full network constructed by taking the absolute value of edge strengths in $\mathbf{A}$, a two-sample $t$-test gives $t=-8.25$, $p = 2.14\times 10^{-14}$ for the first community and $t=-7.70$, $p=6.31 \times 10^{-13}$  for the second community.

Collectively, these results demonstrate that the functional communities elicited by the two tasks (\Rossler and Lorentz) differ in their structural support. This fact is perhaps intuitive given that the functional communities are driven by the input dynamics, which differ appreciably across the two attractors. Yet, further work is needed to understand whether there might be a way of predicting exactly what sort of structural topology is expected in the functional communities elicited by distinct attractors. Finally, we note that these data support the notion that attractor dynamics are not learned by distinct subnetworks of neurons, but instead by all neurons in specific regions in the phase space.

\section*{Appendix F: Missing variable inference with neurons receiving multiple input signals}
\label{sec:new_missing_infer}

\begin{figure*}
  \center
  \includegraphics[width=16cm]{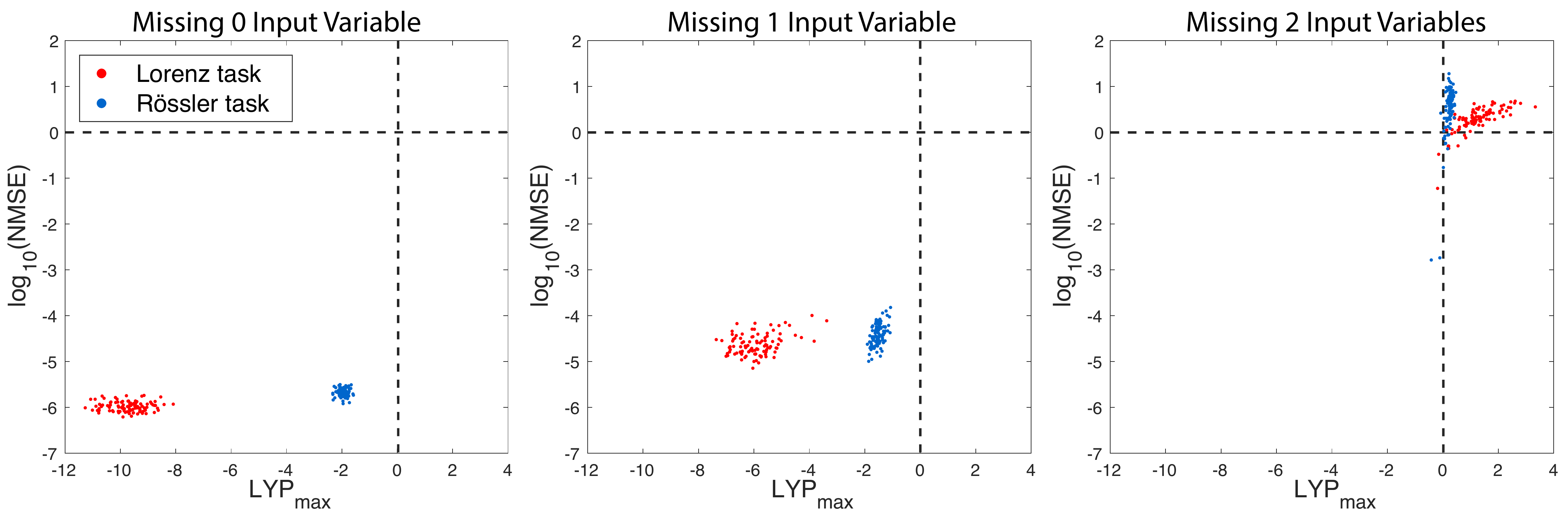}
  \linespread{1.0}
  \caption{\textbf{Supplementary results related to the inference of missing input variables.} In contrast to our simulations in the main text where we let each neuron connect to $1$ out of $3$ input channels, here instead we let each neuron connect to all $3$ input channels. Specifically, each of the elements in the input weight matrix $\mathbf{W}_{\text{in}}\in\mathbb{R}^{N\times 3}$ is randomly and uniformly drawn from $[-0.05,0.05]$. Here we show the performance of this new system on the task of inferring missing input variables, which we note is comparable to that observed for the original system (see Fig.~\ref{fig:observer}(d) in the main text).}
  \label{fig:input_overlaps_inference}
\end{figure*}

In the demonstration given in the main text, the missing variable inference task is accomplished by the central system where each of the $N$ neurons is connected to $1$ of the $3$ input channels. To better understand this behavior, it is useful to consider whether the system could still perform the inference task if all $N$ neurons in the central system are connected to all $3$ input channels. In this section, we perform additional numerical simulations to directly address this question. Specifically, in contrast to our simulations in the main text, each of the elements in the input weight matrix $\mathbf{W}_{\text{in}}\in\mathbb{R}^{N\times 3}$ is now randomly and uniformly drawn from $[-0.05,0.05]$. In Fig.~\ref{fig:input_overlaps_inference}, we show that this new system's performance on the missing-variable inference problem is similar to the performance that we observed for the original system where every neuron receives signals from only one input channel (compare to Fig.~5(D) in the main text). We conclude that the robustness that we observe in Fig.~5 cannot be attributed solely to the fact that each neuron only receives one of the three input variables.

\bibliography{citation}


\end{document}